\pdfoutput=1  

\documentclass[aps,twocolumn,groupedaddress,floatfix,preprintnumbers,nofootinbib,eqsecnum]{revtex4}

\usepackage{amsmath, float, graphicx,placeins,amssymb,multirow,pgfplots,pgfplotstable}
\usepackage{tikz,hyperref}
\usepackage{CJKutf8}
\usetikzlibrary{shapes.geometric,arrows}

\begin{document}
\title{
Partially Composite Dynamical Dark Matter  
}

\author{Yusuf Buyukdag$^{1}$\footnote{E-mail address:  {\tt buyuk007@umn.edu}},
  Keith R.\ Dienes$^{2,3}$\footnote{E-mail address:  {\tt dienes@email.arizona.edu}},
  Tony Gherghetta$^{1}$\footnote{E-mail address:  {\tt tgher@umn.edu}},
  Brooks Thomas$^{4}$\footnote{E-mail address:  {\tt thomasbd@lafayette.edu}}\\ ~}
\affiliation{
  $^1\,$School of Physics and Astronomy, University of Minnesota, Minneapolis, MN  55455  USA\\
  $^2\,$Department of Physics, University of Arizona, Tucson, AZ  85721  USA\\
  $^3\,$Department of Physics, University of Maryland, College Park, MD 20742 USA\\
  $^4\,$Department of Physics, Lafayette College, Easton, PA  18042  USA}

\preprint{UMN-TH-3910/20}

\begin{abstract}
In this paper, we consider a novel realization of the Dynamical Dark Matter (DDM) framework in 
which the ensemble of particles which collectively constitute the dark matter are the composite 
states of a strongly-coupled conformal field theory.  Cosmological abundances for
these states are then generated through mixing with an additional, elementary state.  As a
result, the physical fields of the DDM dark sector at low energies are partially 
composite --- {\it i.e.}\/, admixtures of elementary and composite states.  Interestingly, 
we find that the degree of compositeness exhibited by these states varies across the DDM
ensemble.  We calculate the masses, lifetimes, and abundances of these states --- along 
with the effective equation of state of the entire ensemble --- by considering the gravity
dual of this scenario in which the ensemble constituents are realized  as the Kaluza-Klein
states associated with a scalar propagating within a slice of five-dimensional 
anti-de Sitter (AdS) space.  Surprisingly, we find that the warping of the AdS space 
gives rise to parameter-space regions in which the decay widths of the dark-sector
constituents vary non-monotonically with their masses.  We also find that there exists a
maximum degree of AdS warping for which a phenomenologically consistent dark-sector 
ensemble can emerge.  Our results therefore suggest the existence of a potentially rich
cosmology associated with partially composite DDM.
\end{abstract}

\maketitle


\newcommand{\PRE}[1]{{#1}} 
\newcommand{\ul}{\underline}
\newcommand{\del}{\partial}
\newcommand{\nbox}{{\,\lower0.9pt\vbox{\hrule \hbox{\vrule height 0.2 cm
\hskip 0.2 cm \vrule height 0.2 cm}\hrule}\,}}

\newcommand{\postscript}[2]{\setlength{\epsfxsize}{#2\hsize}
   \centerline{\epsfbox{#1}}}
\newcommand{\gweak}{g_{\text{weak}}}
\newcommand{\mweak}{m_{\text{weak}}}
\newcommand{\mplanck}{M_{\text{Pl}}}
\newcommand{\mstar}{M_{*}}
\newcommand{\sigmaan}{\sigma_{\text{an}}}
\newcommand{\sigmatot}{\sigma_{\text{tot}}}
\newcommand{\sigmaSI}{\sigma_{\rm SI}}
\newcommand{\sigmaSD}{\sigma_{\rm SD}}
\newcommand{\OmegaM}{\Omega_{\text{M}}}
\newcommand{\OmegaDM}{\Omega_{\text{DM}}}
\newcommand{\ipb}{\text{pb}^{-1}}
\newcommand{\ifb}{\text{fb}^{-1}}
\newcommand{\iab}{\text{ab}^{-1}}
\newcommand{\ev}{\text{eV}}
\newcommand{\kev}{\text{keV}}
\newcommand{\mev}{\text{MeV}}
\newcommand{\gev}{\text{GeV}}
\newcommand{\tev}{\text{TeV}}
\newcommand{\pb}{\text{pb}}
\newcommand{\mb}{\text{mb}}
\newcommand{\cm}{\text{cm}}
\newcommand{\m}{\text{m}}
\newcommand{\km}{\text{km}}
\newcommand{\kg}{\text{kg}}
\newcommand{\g}{\text{g}}
\newcommand{\s}{\text{s}}
\newcommand{\yr}{\text{yr}}
\newcommand{\Mpc}{\text{Mpc}}
\newcommand{\etal}{{\em et al.}}
\newcommand{\eg}{{\em e.g.}}
\newcommand{\ie}{{\em i.e.}}
\newcommand{\ibid}{{\em ibid.}}
\newcommand{\Eqref}[1]{Equation~(\ref{#1})}
\newcommand{\secref}[1]{Sec.~\ref{sec:#1}}
\newcommand{\secsref}[2]{Secs.~\ref{sec:#1} and \ref{sec:#2}}
\newcommand{\Secref}[1]{Section~\ref{sec:#1}}
\newcommand{\appref}[1]{App.~\ref{sec:#1}}
\newcommand{\figref}[1]{Fig.~\ref{fig:#1}}
\newcommand{\figsref}[2]{Figs.~\ref{fig:#1} and \ref{fig:#2}}
\newcommand{\Figref}[1]{Figure~\ref{fig:#1}}
\newcommand{\tableref}[1]{Table~\ref{table:#1}}
\newcommand{\tablesref}[2]{Tables~\ref{table:#1} and \ref{table:#2}}
\newcommand{\Dsle}[1]{\slash\hskip -0.28 cm #1}
\newcommand{\met}{{\Dsle E_T}}
\newcommand{\mpt}{\not{\! p_T}}
\newcommand{\Dslp}[1]{\slash\hskip -0.23 cm #1}
\newcommand{\Dsl}[1]{\slash\hskip -0.20 cm #1}

\newcommand{\mB}{m_{B^1}}
\newcommand{\mq}{m_{q^1}}
\newcommand{\mf}{m_{f^1}}
\newcommand{\mKK}{m_{\rm KK}}
\newcommand{\WIMP}{\text{WIMP}}
\newcommand{\SWIMP}{\text{SWIMP}}
\newcommand{\NLSP}{\text{NLSP}}
\newcommand{\LSP}{\text{LSP}}
\newcommand{\mWIMP}{m_{\WIMP}}
\newcommand{\mSWIMP}{m_{\SWIMP}}
\newcommand{\mNLSP}{m_{\NLSP}}
\newcommand{\mchi}{m_{\chi}}
\newcommand{\mgravitino}{m_{\gravitino}}
\newcommand{\mmed}{M_{\text{med}}}
\newcommand{\gravitino}{\tilde{G}}
\newcommand{\Bino}{\tilde{B}}
\newcommand{\photino}{\tilde{\gamma}}
\newcommand{\stau}{\tilde{\tau}}
\newcommand{\slepton}{\tilde{l}}
\newcommand{\snu}{\tilde{\nu}}
\newcommand{\squark}{\tilde{q}}
\newcommand{\mgaugino}{M_{1/2}}
\newcommand{\epsEM}{\varepsilon_{\text{EM}}}
\newcommand{\mmess}{M_{\text{mess}}}
\newcommand{\lmess}{\Lambda}
\newcommand{\nmess}{N_{\text{m}}}
\newcommand{\signmu}{\text{sign}(\mu)}
\newcommand{\Omegachi}{\Omega_{\chi}}
\newcommand{\lambdafs}{\lambda_{\text{FS}}}
\newcommand{\be}{\begin{equation}}
\newcommand{\ee}{\end{equation}}
\newcommand{\bea}{\begin{eqnarray}}
\newcommand{\eea}{\end{eqnarray}}
\newcommand{\beq}{\begin{equation}}
\newcommand{\eeq}{\end{equation}}
\newcommand{\beqn}{\begin{eqnarray}}
\newcommand{\eeqn}{\end{eqnarray}}
\newcommand{\baln}{\begin{align}}
\newcommand{\ealn}{\end{align}}
\newcommand{\lsim}{\lower.7ex\hbox{$\;\stackrel{\textstyle<}{\sim}\;$}}
\newcommand{\gsim}{\lower.7ex\hbox{$\;\stackrel{\textstyle>}{\sim}\;$}}

\newcommand{\ssection}[1]{{\em #1.\ }}
\newcommand{\rem}[1]{\textbf{#1}}

\def\ie{{\it i.e.}\/}
\def\eg{{\it e.g.}\/}
\def\etc{{\it etc}.\/}
\def\calN{{\cal N}}

\def\mptwo{{m_{\pi^0}^2}}
\def\mp{{m_{\pi^0}}}
\def\sqtsn{\sqrt{s_n}}
\def\sqtsn{\sqrt{s_n}}
\def\sqtsn{\sqrt{s_n}}
\def\sqts0{\sqrt{s_0}}
\def\Dsqts{\Delta(\sqrt{s})}
\def\Omegatot{\Omega_{\mathrm{tot}}}
\def\rhotot{\rho_{\mathrm{tot}}}
\def\rhocrit{\rho_{\mathrm{crit}}}
\def\OmegaDM{\Omega_{\mathrm{DM}}}
\def\OmegaDMbar{\overline{\Omega}_{\mathrm{DM}}}
\def\weff{w_{\mathrm{eff}}}
\def\tLS{t_{\mathrm{LS}}}
\def\aLS{a_{\mathrm{LS}}}
\def\zLS{z_{\mathrm{LS}}}
\def\tnow{t_{\mathrm{now}}}
\def\znow{z_{\mathrm{now}}}
\def\tMRE{t_{\mathrm{MRE}}}
\def\tast{t_{\ast}}
\def\Ndof{N_{\mathrm{d.o.f.}}\/}
\def\LambdaIR{\Lambda_{\rm IR}}
\def\LambdaUV{\Lambda_{\rm UV}}


\section{Introduction}


Dynamical Dark Matter~\cite{DDM1,DDM2} (DDM) provides an alternative framework for
dark-matter physics in which the notion of dark-matter stability is replaced by something
more general and powerful:  a balancing of decay widths against cosmological abundances
across an ensemble of individual dark-matter constituents.  Within this framework, those
dark-sector states with larger decay widths (shorter lifetimes) must have smaller 
abundances, while those with smaller decay widths (longer lifetimes) can have larger
abundances.  This balancing allows the ensemble to
exhibit a variety of lifetimes that stretch across all cosmological periods, leading to an
extremely ``dynamic'' universe in which quantities such as the total dark-matter abundance
$\Omega_{\rm CDM}$ evolve non-trivially throughout all periods of cosmological history ---
all while remaining consistent with experimental and observational constraints.

If such a balancing could only be arranged by adjusting the masses and couplings 
associated with the individual constituent particles of the ensemble by hand, such a 
dark-matter scenario would clearly require an unacceptable degree of fine-tuning. 
However, it turns out that large collections of particles with the appropriate balancing 
between decay widths and abundances arise in a number of top-down scenarios for new physics.
In such realizations of the DDM framework, the properties of all the constituent particles
within the ensemble are completely specified by only a small number of parameters.
The masses, lifetimes, abundances, \etc, of these particles scale across the ensemble
according to a set of scaling relations.  Examples of scenarios which yield a DDM-appropriate 
set of scaling relations include higher-dimensional theories of an axion or axion-like 
particle propagating in the bulk~\cite{DDGAxion} in which the Kaluza-Klein (KK) resonances
collectively constitute the DDM ensemble~\cite{DDM2,DDMAxion}; theories with additional fields 
which transform non-trivially under large, spontaneously-broken symmetry groups, in which the
ensemble constituents are the physical degrees of freedom within the corresponding symmetry
multiplets~\cite{DDMRandom}; and theories with strongly-coupled hidden-sector gauge groups, 
in which the ensemble constituents are identified with the ``hadrons'' which emerge in the 
confining phase of the theory at low energies~\cite{DDMHagedorn,DDMHagedornProc}. 
      
In this paper, we consider another possible top-down realization of the DDM framework in 
the context of a conformal field theory (CFT).  In particular, we consider a strongly-coupled 
theory which exhibits conformal invariance at high energies, but in which this invariance is 
spontaneously broken at low energies.  Below the corresponding symmetry-breaking scale,  
a spectrum of particle-like composite states emerges.  As we shall show, these 
composite states can acquire a spectrum of decay widths and abundances by mixing with 
an additional, elementary degree of freedom external to the CFT.  However,
since the theory is strongly coupled, it is in general not possible to calculate the 
masses, couplings, \etc, of the physical fields of the the low-energy theory directly 
from first principles.  It is therefore not {\it a priori}\/ obvious whether these fields
can collectively exhibit an appropriate balancing of decay widths against abundances 
for DDM.  

Fortunately, the AdS/CFT correspondence~\cite{AdSCFT,GubserHolography,WittenHolography} 
provides us with a tool for overcoming this obstacle.  By studying the gravity dual 
of our partially composite DDM scenario we can infer information about the values of these 
parameters and ultimately determine how the lifetimes, abundances, \etc, of the individual 
constituents scale across the ensemble.  This dual theory involves a scalar propagating in 
the bulk of a spacetime orbifold which is tantamount to a slice of five-dimensional anti-de
Sitter (AdS) space.  A spectrum of decay widths and abundances for the physical fields 
in the dual theory, which are admixtures of the KK modes of this bulk scalar, arises as a 
result of physics localized on the boundaries of this slice of AdS$_5$.     

Moreover, the gravity dual of our partially composite DDM scenario is not only
useful as a tool for gleaning information about this scenario, but is also interesting
in its own right.  It has been shown~\cite{DDM2,DDMAxion} that the KK modes of an 
axion-like particle propagating in the bulk of a theory with a single, flat extra dimension 
constitute a phenomenologically viable DDM ensemble with a particular set of scaling 
relations.  The dual of our partially composite DDM scenario can be viewed as a 
generalization of these flat-space bulk-scalar DDM scenarios to warped space, and thus
can allow us to address a variety of questions related to such DDM scenarios. 
To what extent do these scenarios survive in warped space?  How much warping of the 
space can be tolerated?  As we shall see, the warping has a profound effect on the 
phenomenology of the ensemble.  Indeed, constraints on warped-space bulk-scalar DDM 
scenarios become increasingly stringent as the AdS curvature scale increases.
Moreover, there exist interesting qualitative differences between these warped-space 
scenarios and their flat-space analogues.  One such difference is that, in the case of
a warped extra dimension, there exist regions of parameter space within which the decay 
widths of the ensemble constituents scale non-monotonically with their masses
across the ensemble.  Another difference arises due to the fact that, as a consequence
of the warp factor, the effect of brane-localized dynamics on one of the boundaries of 
the AdS$_5$ slice is generically different from the effect of identical dynamics 
on the other boundary.  As a result, a variety of different possible scaling behaviors
can arise within the basic scenario, depending on which of the boundaries the 
operators responsible for establishing the abundances and decay widths of the 
ensemble constituents reside.  
  
This paper is organized as follows.  In Sect.~\ref{sec:PartiallyCompositeAxion}, we 
present our partially composite DDM scenario and show how a spectrum of 
abundances for the mass-eigenstate fields in this scenario can be generated via 
misalignment production.  In Sect.~\ref{sec:DualMisalignmentMechanism}, we construct
the gravity dual of this scenario.  In Sect.~\ref{sec:DynamicalDarkMatter},  
we calculate the total abundance and equation of state for the ensemble in this dual 
as functions of time and use this information to constrain the parameter space
of our scenario.  We also show that there exist substantial regions of that parameter space 
in the decay widths and abundances exhibit the appropriate scaling relations for a DDM 
ensemble.  In Sect.~\ref{sec:WarpedvsFlat}, we complete the dictionary which relates 
the parameters of the partially composite 4D theory to those of the 5D dual theory  
and investigate to what the flat-space limit of the dual theory corresponds  
in the partially composite theory.  In Sect.~\ref{sec:Conclusions}, we conclude with a 
summary of our findings and a discussion of some possible implications for future work.  
In Appendix~\ref{app:FlatLimit}, we show how the results obtained in the flat-space DDM 
scenario in Ref.~\cite{DDM1} are recovered from the warped case in the limit of vanishing 
curvature.  In Appendix~\ref{app:DifferentScenarios}, we generalize the results obtained in 
Sect.~\ref{sec:DualMisalignmentMechanism} by considering different possible 
locations for the relevant boundary terms which give rise to the decay widths 
and abundances for the ensemble constituents.

\section{Partially Composite Scalar Ensembles and Misalignment Production\label{sec:PartiallyCompositeAxion}}

Partially composite scalars arise in a variety of extensions of the Standard Model (SM).
The QCD axion, for example, is an elementary scalar which mixes with with composite states 
such as the $\pi^0$ and $\eta'$.  Models involving composite invisible axions have also been 
posited to explain why the allowed window for the axion decay constant lies between the 
grand-unification scale and the electroweak scale~\cite{KimComposite}.  In this paper,
we consider a scenario in which a single elementary scalar mixes with a large --- and
potentially even infinite --- number of composite states.  As we shall see, scenarios of 
this sort can be fertile ground for DDM model-building.

In constructing the elementary sector of our theory, we consider a complex scalar field 
$\Phi$ which is charged under a global $U(1)$ symmetry.  We shall assume that the 
potential for $\Phi$ is such that it receives a non-zero vacuum expectation value (VEV)
$\langle \Phi \rangle = \hat{f}_X/\sqrt{2}$, thereby spontaneously breaking this symmetry at 
the scale $\hat{f}_X$.  At scales well below $\hat{f}_X$, this complex scalar may be 
parametrized as 
\begin{equation}
  \Phi ~\approx~ \frac{\hat{f}_X}{\sqrt{2}} e^{i\phi_0/\hat{f}_X}~,
\end{equation}
where $\phi_0$ is a real ($CP$-odd) scalar field which can be viewed as the 
Nambu-Goldstone boson associated with the breaking of this symmetry.  This field 
$\phi_0$, which could in principle be identified with the QCD axion, but could also be 
some additional axion-like particle, shall effectively constitute the elementary sector 
of our theory in and of itself.

Since $\phi_0$ is a Nambu-Goldstone boson, the manner in which it interacts with any 
other fields present in the theory is in this case dictated in part by a global shift 
symmetry under which $\phi_0 \rightarrow \phi_0 + C$, where $C$ is an arbitrary real 
constant. For example, in the presence of an additional non-Abelian gauge group $G$, 
the action for $\phi_0$ takes the form  
\begin{equation}
  \mathcal{S}_{\phi} ~=~ \int d^4x \left[
     \frac{1}{2} \partial_{\mu} \phi_0 \partial^{\mu} \phi_0
    +\frac{g_G^2 c_g\phi_0}{32\pi^2 \hat{f}_X} G_{\mu \nu} \tilde{G}^{\mu \nu} \right]~, 
  \label{eq:AxionLag}
\end{equation}
where $g_G$ is the gauge coupling associated with the gauge group $G$, where 
$G_{\mu \nu}$ is the corresponding field-strength tensor, where 
$\widetilde{G}^{\mu \nu} \equiv \frac{1}{2}\epsilon^{\mu\nu\rho\sigma}G_{\rho\sigma}$ is 
the corresponding dual field-strength tensor, and where $c_g$ is a model-dependent 
coefficient that parametrizes the interaction between $\phi_0$ and the gauge fields.

Strict invariance under the classical shift symmetry of Eq.~(\ref{eq:AxionLag}) would 
imply that the potential for $\phi_0$ vanishes.  However, this classical symmetry is 
broken dynamically at the quantum level by non-perturbative instanton effects
associated with the gauge group $G$ which become significant at scales around or 
below the scale $\Lambda_G$ at which $G$ becomes confining.  Thus, $\phi_0$ is 
effectively massless at scales above $\Lambda_G$, while at lower scales it
generically acquires a mass as a consequence of these instanton effects.  The
implications of this dynamically generated mass term shall be discussed in
greater detail below.

We now turn to discuss the composite sector of the theory.  We take the fields 
$\varphi_n$ of this sector to be the composite states of a $SU(N)$ gauge theory with 
$N\gg1$ which appear in the spectrum of the infrared theory at scales below the 
confinement scale $\LambdaIR$.  We emphasize that this $SU(N)$ group is distinct from 
the non-Abelian gauge group $G$ discussed above.  At scales above $\LambdaIR$, the 
unconfined theory rapidly approaches an ultraviolet fixed point and effectively behaves 
as a CFT up to some ultraviolet scale $\LambdaUV$.  At higher scales, the approximate
conformal invariance of the theory is explicitly broken by the presence of additional 
fields $\Psi$ with masses of order $M_\Psi \sim \LambdaUV$ which transform non-trivially 
under the same $SU(N)$ gauge group --- fields which are integrated out of the effective 
theory below $\LambdaUV$.  We shall also assume that this $SU(N)$ gauge theory is 
vector-like and therefore yields no contribution to the chiral anomaly.

We shall assume that the quantum numbers of the $\varphi_n$ are such that they can mix with 
$\phi_0$.  Moreover, the shift symmetry once again dictates that this mixing
occurs as the result of Lagrangian terms linear in $\phi_0$.  For concreteness,
we shall consider the simple case in which this mixing arises as the result of a coupling
between $\phi_0$ and an operator $\mathcal{O}_c$ of mass dimension
$d_{\mathcal{O}_c} = 4$ constructed from the fundamental degrees of freedom of the 
unconfined $SU(N)$ theory.  At the scale $\LambdaUV$, the action for $\phi_0 $ therefore 
takes the form 
\begin{eqnarray}
  \mathcal{S}_{\phi} &=& \int d^4x\bigg[
     \frac{1}{2} \partial_{\mu} \phi_0 \partial^{\mu} \phi_0 
    + \left( \frac{\Phi}{\LambdaUV} \mathcal{O}_c + h.c. \right) \nonumber \\ & & 
    ~~~~~~~~~
    + \frac{g_G^2 c_g\phi_0}{32\pi^2 \hat{f}_X} G_{\mu \nu} \tilde{G}^{\mu \nu}
    +\ldots\bigg]~.
  \label{eq:axionL}
\end{eqnarray}
We shall assume that the operator $\mathcal{O}_c$ transforms non-trivially under the global 
$U(1)$ symmetry in such a way that the action is invariant under this symmetry. 
At scales $\LambdaIR < \mu \leq \LambdaUV$, radiative corrections to the kinetic term for
$\phi_0$ arise as a result of the interaction in Eq.~(\ref{eq:axionL}).  
The effect of these corrections can be interpreted as a renormalization 
of the kinetic term for $\phi_0$. Thus, at an arbitrary scale 
$\LambdaIR < \mu \leq \LambdaUV$, the kinetic term in Eq.~(\ref{eq:axionL}) takes the
form~\cite{WittenHolography,ContinoPomarol}
\begin{equation}
  \mathcal{L}_{\phi} ~\ni~ \frac{Z(\mu)}{2} \partial_{\mu} \phi_0 \partial^{\mu} \phi_0~,
  \label{eq:RunningLag}
\end{equation}
where $Z(\LambdaUV) = 1$.  The renormalization-group equation for $Z(\mu)$ in the presence of 
the $SU(N)$ operator $\mathcal{O}_c$, where 
$\langle \mathcal{O}_c \mathcal{O}_c \rangle \propto N/16\pi^2$ for 
large $N$, takes the form
\begin{equation}
  \frac{\partial Z(\mu)}{\partial \log \mu} ~\approx~ -2 
    \gamma \frac{N}{16\pi^2} \left( \frac{\mu}{\LambdaUV} \right)^2~,
  \label{eq:RGeqn}
\end{equation}
where $\gamma$ is an $\mathcal{O}(1)$ constant.  In the
large-$N$ limit, the solution to this equation at low scales $\mu \ll \LambdaUV$ is
approximately
\begin{equation}
  Z \left( \mu \right) ~\approx~ \gamma \frac{N}{16\pi^2}~.
\end{equation}
In the confined phase of the theory at scales $\mu < \LambdaIR$, there exists a tower of 
composite states $\varphi_n$ with the masses, $\widetilde{m}_n \sim n \LambdaIR$.  The 
precise mass spectrum of these states and the extent to which each of them mixes with the
elementary field $\phi_0$ cannot in general be determined in a straightforward manner from 
the properties of the theory in the unconfined phase, due to the strong dynamics involved.
Thus, for the moment, we simply seek to parametrize the Lagrangian for the fields of the
confined phase in a meaningful way, given certain reasonable assumptions about the symmetry
structure of the theory and certain results which are known to hold for $SU(N)$ gauge
theories in the large-$N$ limit.  As we shall see, however, whenever these assumptions 
hold, it will be possible for us to determine the properties of the physical fields of 
the theory using other means. 

In parametrizing the Lagrangian for the confined phase, we choose to work in a basis in 
which the kinetic terms for all physical fields are canonical, and mixing between these
fields occurs only via the mass matrix.  In this basis, it can be shown that in the 
large-$N$ limit, the matrix element of the operator $\mathcal{O}_c$ between the vacuum 
and each scalar $\varphi_n$ takes the 
form~\cite{WittenBaryons}  
\begin{equation}
  \langle 0 | \mathcal{O}_c | \varphi_n \rangle ~\propto~  
      \frac{\sqrt{N}}{4\pi}~.
\end{equation} 
The corresponding operator-field identity takes the form
\begin{equation}
  \mathcal{O}_c ~=~ \frac{N}{16\pi^2}
     \frac{\LambdaIR^4}{\sqrt{2}}  \sum_{n = 1}^\infty 
    \tilde{\xi}_n^2 e^{ i \frac{4\pi}{\sqrt{N}}
    \frac{\varphi_n}{\LambdaIR}} ~,
  \label{eq:OperatorFieldID}
\end{equation}
where $\tilde{\xi}_n$ is a dimensionless $\mathcal{O}(1)$ coefficient.  

We now turn to consider what the action for the theory looks like in the confined phase.
Given that the $SU(N)$ gauge group in our scenario is assumed to be vector-like,
no coupling between the Chern-Simons term and $\phi_0$ is generated.
We therefore expect that the global shift symmetry of the original action in
Eq.~(\ref{eq:AxionLag}) is not disturbed by the confining phase transition at 
$\mu \sim \LambdaIR$ and remains intact within the confined phase.  This implies that a
massless degree of freedom should likewise be present in the spectrum 
of the theory within the confined phase. To remove the constant potential that appears when we expand Eq.~(\ref{eq:axionL}), we add the appropriate terms at the IR scale. It therefore follows that the 
Lagrangian at scales $\mu \lesssim \LambdaIR$ takes the form  
\begin{eqnarray}
  \mathcal{L}_\phi &=& \frac{1}{2} \partial_{\mu} \phi_0 \partial^{\mu} \phi_0 + 
     \sum_{n=1}^\infty\frac{1}{2} \partial_{\mu} \varphi_n \partial^{\mu} \varphi_n 
     \nonumber \\ & &
     + \frac{g_G^2 c_g\phi_0}{32\pi^2 \hat{f}_X} G_{\mu \nu} \tilde{G}^{\mu \nu} 
     + \frac{1}{2} \Lambda_{\text{IR}}^2 \sum_{n=1}^\infty
    \big( \epsilon_n \phi_0 + g_n \varphi_n \big)^2~, \nonumber \\
  \label{eq:lambdascale}
\end{eqnarray}
where the $g_n$ and $\epsilon_n$ are dimensionless parameters which cannot, in general, be
calculated from first principles.  Indeed, we observe that the corresponding action is 
invariant under the combined transformations
\begin{eqnarray}
  \phi_0 &\rightarrow& \phi_0 + C \nonumber \\ 
    \varphi_n &\rightarrow& \varphi_n - \frac{\epsilon_n}{g_n} C~.
  \label{eq:shiftbyCunmixed}
\end{eqnarray}   
The parameters $g_n \equiv \widetilde{m}_n/\LambdaIR$ in Eq.~(\ref{eq:lambdascale}) can be 
viewed as a convenient parametrization for the mass $\widetilde{m}_n$ that the field
$\varphi_n$ would have had in the absence of mixing.  By contrast, the $\epsilon_n$, each 
of which determines the degree of mixing between $\phi_0$ and the corresponding 
$\varphi_n$, arise as a consequence of the operator $\mathcal{O}_c$ and may be viewed 
as a convenient reparametrization of the corresponding coefficients $\tilde{\xi}_n$ in
Eq.~(\ref{eq:OperatorFieldID}).  Indeed, through use of this operator-field identity, 
we see that  
\begin{equation}
  \epsilon_n ~=~ \frac{\xi_n}{\sqrt{\gamma}} \frac{\LambdaIR}{\LambdaUV}~,
\end{equation}
where $\xi_n \equiv \tilde{\xi}_n \sqrt{\LambdaUV/\hat{f}_X}$.
Of course, if $\epsilon_n \neq 0$ for one or more of the $\varphi_n$, the mass eigenstates 
of the theory are not $\phi_0$ and the $\varphi_n$, but rather linear combinations of 
these fields.  The mass-squared matrix which follows from the Lagrangian in
Eq.~(\ref{eq:lambdascale}) is   
\begin{equation}
  \mathcal{M}^2 ~=~ \left( \begin{matrix}
    \sum_{n=1}^\infty\epsilon_n^2 & \epsilon_1 g_1 & \epsilon_2 g_2 & \ldots \\ 
    \epsilon_1 g_1         & g_1^2          & 0              & \ldots \\ 
    \epsilon_2 g_2         & 0              & g_2^2          & \ldots \\
    \vdots                 & \vdots         & \vdots         & \ddots
    \end{matrix} \right) \LambdaIR^2~.
  \label{eq:MassMatCompositeHighEnergy}
\end{equation}  
Within the regime in which $\LambdaIR \ll \LambdaUV$, a hierarchy among the parameters
develops in which $\epsilon_n \ll 1 \lesssim g_n$ for each of the $\varphi_n$.  Within
this regime, the eigenvalues and eigenvectors of $\mathcal{M}^2$ can be reliably 
calculated using a perturbation expansion in the $\epsilon_n$.  In particular, to 
$\mathcal{O}(\epsilon_n^2)$, the squared masses are 
\begin{equation}
  m_n^2 ~\approx~ 
    \begin{cases}
      0 & n = 0 \\
      (\epsilon_n^2 + g_n^2)\LambdaIR^2 & n > 0
    \end{cases}
\end{equation}
and the corresponding mass-eigenstate fields are approximately
\begin{widetext}
\begin{equation}
  | \chi_n \rangle ~\approx~ 
    \begin{cases} 
      \displaystyle \left(1 - \sum_{m=1}^\infty\frac{\epsilon_m^2}{2g_m^2} \right) 
      | \phi_0 \rangle - \sum_{m=1}^\infty\frac{\epsilon_m}{g_m} 
      | \varphi_m \rangle 
        & n = 0 \\ 
      \displaystyle \left(1 - \frac{\epsilon_n^2}{2g_n^2}\right) | \varphi_n \rangle + 
        \frac{\epsilon_n}{g_n} | \phi_0 \rangle 
        + \sum_{m\neq 0,n}^\infty\frac{\epsilon_n\epsilon_m g_m}{g_n(g_n^2-g_m^2)} 
        |\varphi_m\rangle  
       & n > 0~.
    \end{cases}
  \label{eq:EvecsCompositeHighEnergy} 
\end{equation}
\end{widetext}
The presence of a massless physical degree of freedom is a direct consequence of the global 
shift symmetry.  Indeed, $\chi_0$ transforms under the corresponding symmetry transformation 
according to the relation 
\begin{equation}
    \chi_0 ~\rightarrow~ \chi_0 + C\left(1 + \sum_{n=1}^\infty\frac{\epsilon_n^2}{g_n^2}
    \right)^{1/2} ~\approx~ \chi_0 + C~,
\end{equation}
while $\chi_n \rightarrow \chi_n$ for all $n > 0$.  Since $\chi_0$ transforms non-trivially 
under this symmetry transformation, a mass term for this field is forbidden as long as the 
shift symmetry remains intact.  

While we have assumed that the shift symmetry is preserved, at least approximately, during 
the confining phase transition at $\LambdaIR$, this classical symmetry is in general broken 
at the quantum level by instanton effects associated with the gauge group $G$, as discussed
above.  At early times, when the temperature $T$ of the thermal bath greatly exceeds the scale 
$\Lambda_G$ at which $G$ becomes confining --- a scale which we shall assume is much smaller 
than $\LambdaIR$ --- these effects are negligible.  However, when the temperature of 
the universe falls to around $T \sim \Lambda_G$ these effects dynamically generate a
potential for $\phi_0$, which generically includes a temperature-dependent mass term
$m_{\rm dyn}(T)$.  Exactly how $m_{\rm dyn}(T)$ behaves as a function of $T$ at temperatures 
$T \sim \Lambda_G$ depends on the details of the instanton dynamics.  Nevertheless, we
generically expect that $m_{\rm dyn}(T) \approx 0$ at temperatures $T \gg \Lambda_G$, while
$m_{\rm dyn}(T)$ asymptotically approaches a constant value 
$m_\phi \equiv \lim_{T\rightarrow 0} m_{\rm dyn}(T)$ at temperatures 
$T \ll \Lambda_G$.  Provided that the phase transition is sufficiently rapid, it is
reasonable to work in the ``rapid-turn-on'' approximation in which we approximate the 
phase transition as infinitely rapid and model $m_{\rm dyn}(T)$ with a 
step function of the form  
\begin{equation}
  m_{\rm dyn}(T) ~\approx~ 
   \begin{cases}
     0 & T > \Lambda_G \\ 
     m_\phi & T \leq \Lambda_G~.
  \end{cases}
\end{equation}
In this approximation, the mass matrix in Eq.~(\ref{eq:MassMatCompositeHighEnergy}) is 
modified at temperatures $T \leq \Lambda_G$ to
\begin{equation}
  \mathcal{M}^2 ~=~ \left( \begin{array}{cccc} 
    \frac{m_\phi^2}{\LambdaIR^2} + \sum_{n=1}^\infty\epsilon_n^2  
                           & \epsilon_1 g_1 & \epsilon_2 g_2 & \ldots \\ 
    \epsilon_1 g_1         & g_1^2          & 0              & \ldots \\ 
    \epsilon_2 g_2         & 0              & g_2^2          & \ldots \\
    \vdots                 & \vdots         & \vdots         & \ddots
    \end{array} \right) \LambdaIR^2~.
  \label{eq:warpedmass}
\end{equation}

Since we are assuming $\Lambda_G \ll \LambdaIR$, we are primarily interested in the 
regime within which $m_\phi^2 \ll \epsilon_n^2 \LambdaIR^2$.  Within this 
regime, the additional dynamical contribution to the mass matrix in Eq.~(\ref{eq:warpedmass})
represents a small perturbation to the original mass matrix in 
Eq.~(\ref{eq:MassMatCompositeHighEnergy}).  Within the regime in which $\LambdaIR \ll \LambdaUV$,
the squared masses $\hat{m}_n^2$ of the theory at temperatures $T \leq \Lambda_G$ are to 
$\mathcal{O}(\epsilon_n^2)$ given by 
\begin{equation}
  \hat{m}_n^2 ~\approx~ 
  \begin{cases} 
    m_\phi^2 & n = 0 \\  
    \displaystyle \left( g_n^2 + \epsilon_n^2 \right) 
    \Lambda_{\text{IR}}^2 + \frac{\epsilon_n^2}{g_n^2} m_\phi^2 
    & n > 0~,
  \end{cases}
  \label{eq:MassesCompositeLowEnergy}
\end{equation}
while the corresponding mass-eigenstate fields $\hat{\chi}_n$ are 
\begin{widetext}
\begin{equation}
  | \hat{\chi}_n \rangle ~\approx~ 
    \begin{cases} 
      \displaystyle \left(1 - \sum_{m=1}^\infty\frac{\epsilon_m^2}{2g_m^2}
      - \sum_{m=1}^\infty\frac{\epsilon_m^2}{g_m^4}
      \frac{m_\phi^2}{\LambdaIR^2}
      \right) 
      | \phi_0 \rangle - \sum_{m=1}^\infty\left(\frac{\epsilon_m}{g_m} 
         + \frac{\epsilon_m}{g_m^3}
        \frac{m_\phi^2}{\LambdaIR^2} \right) | \varphi_m \rangle 
        & n = 0 \\ 
      \displaystyle \left(1 - \frac{\epsilon_n^2}{2g_n^2}
        - \frac{\epsilon_n^2}{g_n^4}\frac{m_\phi^2}{\LambdaIR^2}
        \right) | \varphi_n \rangle + 
        \left(\frac{\epsilon_n}{g_n}
        + \frac{\epsilon_n}{g_n^3}\frac{m_\phi^2}{\LambdaIR^2} \right)
        | \phi_0 \rangle 
        + \sum_{m\neq 0,n}^\infty\frac{\epsilon_n\epsilon_m g_m}{g_n(g_n^2-g_m^2)} 
        \left(1 + \frac{1}{g_n^2}\frac{m_\phi^2}{\LambdaIR^2}
        \right)|\varphi_m\rangle  
       & n > 0~.
    \end{cases}
  \label{eq:EvecsCompositeLowEnergy}
\end{equation}
\end{widetext} 

We now turn to assess whether the partially composite states $\hat{\chi}_n$ which emerge 
in this scenario at $T \leq \Lambda_G$ can collectively play the role of a DDM ensemble.  
In order for this to be the case, these states must exhibit an appropriate balancing of 
decay widths against abundances across the ensemble as a whole.  On the other hand, without 
additional information about the values of the constants $\xi_n$ and $g_n$, we cannot at 
this point make any more rigorous assessment as to whether such a balancing in fact arises. 
On the other hand, there are many qualitative features of this partially composite theory
which are auspicious from a DDM perspective.   The theory includes a potentially vast 
number of particle species with a broad spectrum of masses, all of which are neutral under 
the SM gauge group.  Moreover, as we shall discuss in further detail below, there exists a
natural mechanism --- namely, misalignment production --- for generating a spectrum of
abundances for the $\hat{\chi}_n$ in this scenario.       

The consequences of a bulk axion acquiring a misaligned vacuum value were investigated 
in Ref.~\cite{DDGAxion}.  Since $\chi_0$ is forbidden from acquiring a potential at 
$T \gtrsim \Lambda_G$ by the shift symmetry, the VEV $\langle \chi_0 \rangle$ of this 
field at such temperatures is arbitrary.  We may parametrize this VEV in terms of a
misalignment angle $\theta$ as
\begin{equation}
  \langle \chi_0 \rangle ~=~ \theta \hat{f}_X~. 
  \label{eq:VEV4D}
\end{equation}
By contrast, $\langle \chi_n\rangle = 0$ for all $\chi_n$ with $n > 0$ at 
$T \gtrsim \Lambda_G$, since these fields already have non-zero masses 
$m_n \sim \mathcal{O}(\LambdaIR)$.

After the mass-generating phase transition occurs, however, the mass eigenstates of the 
theory are no longer the $\chi_n$, but rather the $\hat{\chi}_n$.  These latter fields can 
be expressed as linear combinations of the $\chi_n$.  In general, we may write 
\begin{equation}
  |\hat{\chi}_n\rangle ~=~ \sum_{n=0}^\infty U_{n\ell} |\chi_\ell \rangle~,
  \label{eq:MixingStatesIn4D}
\end{equation}
where the $U_{n\ell} \equiv \langle \chi_\ell | \hat{\chi}_n \rangle$ are the elements of the
mixing matrix between these two sets of basis states.  Of particular significance for the 
phenomenology of the $\hat{\chi}_n$ are the {\it mixing coefficients}\/ $A_n ~\equiv~ U_{n 0}$ 
between these states and the massless state $\chi_0$.  Indeed, since $\chi_0$ is the only one 
of the $\chi_n$ which acquires a non-zero VEV from the misalignment mechanism, the mixing 
coefficient $A_n$ determines the VEV $\langle\hat{\chi}_n\rangle$ of each $\hat{\chi}_n$.  
In particular, in the rapid-turn-on approximation, Eq.~(\ref{eq:VEV4D}) implies that at 
the time $t_G$ at which the phase transition occurs~\cite{DDGAxion}, we have
\begin{equation}
  \langle \hat{\chi}_n(t_G) \rangle ~=~ \theta A_n \hat{f}_X~.
  \label{eq:MixedHatPhiVEV4D}
\end{equation}
As a result, each $\hat{\chi}_n$ acquires an energy density at $t = t_G$, given by 
\begin{equation}
  \rho_n(t_G) ~=~ \frac{1}{2} \hat{m}_n^2 \langle \hat{\chi}_n(t_G)\rangle^2~,
  \label{eq:RhonMisaligned}
\end{equation}
and hence also a cosmological abundance.  

Similarly, in order to assess whether our ensemble of $\hat{\chi}_n$ constitute a viable
DDM ensemble, we must also evaluate the corresponding decay widths $\Gamma_n$ of these
particles.  One way in which the $\hat{\chi}_n$ can decay is through interactions with
fields outside the composite sector --- interactions which these fields inherit from 
the elementary field $\phi_0$.  Such interactions are typically suppressed by 
powers of the scale $\hat{f}_X$.  Since these interactions are a consequence of 
mixing with $\phi_0$, the matrix element for any process by which one of the 
$\hat{\chi}_n$ decays necessarily includes one or more factors of the 
{\it projection coefficient}\/ $A_n' \equiv \langle \phi_0 | \hat{\chi}_n \rangle$
which quantifies the extent of this mixing.  

Another way in which contributions to the $\Gamma_n$ might arise is through intra-ensemble
decays --- processes in which one of the $\hat{\chi}_n$ decays to a final state 
involving one or more other, lighter ensemble constituents.  However, given that our
composite sector consists of the meson-like bound states of a large-$N$ $SU(N)$ 
gauge theory, we expect the collective contribution to each $\Gamma_n$ from such processes 
to be suppressed relative to the contribution from decays inherited from $\phi_0$ 
into final states consisting solely of particles external to the ensemble.  In a 
large-$N$ gauge theory of this sort, the three-point functions for meson-like 
states scale as $\sim 1/\sqrt{N}$, while correlation functions with larger numbers 
of external lines are suppressed by higher powers of 
$N$~\cite{WittenBaryons}.  The amplitudes for two-body decay processes in which 
one such state decays to a pair of other, lighter meson-like states therefore also scale as  
$\sim 1/\sqrt{N}$.  Thus, in the $N\rightarrow \infty$ limit, these meson-like states become
free particles and their decay widths vanish, while for large but finite values of $N$ they 
are heavily suppressed.  An alternative way of understanding this suppression is to note that
if we were to model the flux tubes of our $SU(N)$ theory as strings, as was done in 
the ``dark-hadron'' DDM model presented Refs.~\cite{DDMHagedorn,DDMHagedornProc},
the string coupling which governs the interactions of these flux tubes with each other 
scales as $g_s \sim 1/N$.  For these reasons, we shall assume that decays to states 
external to the ensemble dominate the decay width of each $\hat{\chi}_n$ and neglect 
the effect of intra-ensemble decays in what follows.

For concreteness, we shall focus on the case in which the dominant 
contribution to each $\Gamma_n$ arises due to two-body decay processes associated with 
Lagrangian operators of mass dimension $d=5$.  Such an assumption is well motivated, 
given that $\phi_0$ is an axion-like particle and therefore naturally couples 
to fermion and gauge fields through such operators.  In the regime in which the decay 
products of $\hat{\chi}_n$ decay are much lighter than $\hat{\chi}_n$ itself for all 
ensemble constituents, the decay width of each constituent is
\begin{equation}
  \Gamma_n ~\sim~ \frac{\hat{m}_n^3}{\hat{f}_X^2} A_n^{\prime 2}~.
  \label{eq:decaywidth4D}
\end{equation} 

Within the $\LambdaIR \ll \LambdaUV$ regime, Eqs.~(\ref{eq:EvecsCompositeHighEnergy})
and~(\ref{eq:EvecsCompositeLowEnergy}) together imply that
\begin{equation}
  A_n ~\approx~ 
    \begin{cases} 
      1 & n = 0 \\ 
     \frac{\xi_n}{g_n^3 \sqrt{\gamma}} 
     \frac{m_\phi^2}{\LambdaIR \LambdaUV} & n > 0~.
  \end{cases}
\end{equation}
Likewise, in this same regime, the projection coefficients are well approximated by
\begin{equation}
  A_n' ~\approx~ 
    \begin{cases} 1 & n = 0 \\ 
    \frac{\xi_n}{g_n\sqrt{\gamma}} 
    \frac{\Lambda_{\text{IR}}}{\Lambda_{\text{UV}}} & n > 0~.
  \end{cases}
\end{equation}
However, without additional information about the constants $\xi_n$ and $g_n$,
we cannot determine how the $A_n$, and by extension the cosmological abundances of
the $\hat{\chi}_n$, scale across the ensemble.  Nevertheless, as we shall see in
the next section, we can glean the information we require in order to determine whether 
or not this partially composite DDM scenario is phenomenologically viable by exploiting 
certain aspects of the AdS/CFT correspondence.


\section{The Gravity Dual: Misalignment Production in Warped Space\label{sec:DualMisalignmentMechanism}}


Our ignorance of strong dynamics prevents us from being able to determine directly 
the manner in which the decay widths and cosmological abundances of our partially composite 
scalars scale across the ensemble.  Nevertheless, 
inspired by AdS/CFT correspondence~\cite{AdSCFT}, we 
may hope to glean additional information about these scaling exponents by examining the 
gravity dual of our partially composite DDM scenario.  As discussed in the Introduction, 
this dual theory involves a higher-dimensional scalar $\chi$ which propagates throughout 
the bulk of a five-dimensional spacetime orbifold which is tantamount to a slice of AdS$_5$.  
The spacetime metric on this orbifold is 
\begin{equation}
  ds^2 ~=~ e^{-2ky} \eta_{\mu \nu} dx^{\mu} dx^{\nu} + dy^2~,
\end{equation}   
where $\eta_{\mu\nu}$ is the Minkowski metric, where $y$ is the coordinate in the fifth
dimension, and where $k$ is the AdS curvature scale.  This fifth dimension is compactified 
on an $S^1/\mathbb{Z}_2$ orbifold of radius $R$, and a pair of $3$-branes, to which we shall 
refer as the UV and IR branes, are assumed to reside at the orbifold fixed points at $y = 0$ 
and $y = \pi R$, respectively~\cite{RS1}.  While $\chi$ propagates through the entirety of 
the bulk, the fields of the SM are assumed to be localized on the UV brane.  Consistency also 
requires that an additional non-Abelian gauge group $G$ is also assumed to be present in the dual 
theory, the gauge fields of which are likewise localized on the UV brane.  Like the 
corresponding gauge group in the 4D theory, this gauge group is assumed to become confining 
at temperatures $T \lesssim \Lambda_G$, or equivalently, at times $t \gtrsim t_G$.   
  
The bulk scalar which appears in the gravity dual of the theory presented in 
Sect.~\ref{sec:PartiallyCompositeAxion} is the axion or axion-like particle associated
with a global $U(1)$ symmetry which is broken by some bulk dynamics at the scale $f_X$.  
The action for the dual theory is therefore invariant under a global shift symmetry under 
which $\chi \rightarrow \chi + C$, where $C$ is an arbitrary real constant.  In particular, 
this action takes the form 
\begin{eqnarray}
  \mathcal{S}_\chi &=& - \int d^5 x \sqrt{-g} \Bigg[ 
    \frac{1}{2} \partial_M \chi \partial^M \chi \nonumber \\ & & ~~~~~~~~~~~~~~
    + \frac{g_G^2 c_g\chi}{32\pi^2 f_X^{3/2}} G_{\mu \nu} \tilde{G}^{\mu \nu}
    \delta(y)\Bigg]~, ~~~~
  \label{eq:lambdascale2}
\end{eqnarray}
where $g$ is the metric determinant, and where $g_G$, $G_{\mu\nu}$, $\widetilde{G}^{\mu\nu}$,
and $c_g$ are defined as in Eq.~(\ref{eq:AxionLag}).  We note that according to the AdS/CFT 
dictionary, the 5D scalar $\chi$ corresponds to an operator of mass dimension $d=4$ in the 
4D CFT~\cite{Klebanov}.  We also note that since a potential for $\chi$ is forbidden by the 
shift symmetry, the VEV $\langle \chi \rangle$ of this field at times $t\ll t_G$ is arbitrary.  
We parametrize this VEV in terms of a misalignment angle $\theta$ as follows:
\begin{equation}
  \langle \chi(x,y) \rangle ~=~ \theta f_X^{3/2}~.
  \label{eq:VEVin5D}
\end{equation}

In analyzing the implications of this setup, we begin by performing a
KK decomposition of our bulk scalar.  In particular, we write  
\begin{equation}
  \chi(x, y) ~=~ \sum_{n=0}^{\infty} \chi_n (x) \zeta_n (y)~,
  \label{eq:KKDecomp5D}
\end{equation}
where $\chi_n(x)$ is the four-dimensional KK mode of $\chi(x,y)$ with KK 
number $n$ and where $\zeta_n(y)$ is the bulk profile of the corresponding 
KK mode.  We note that since the potential for our bulk scalar $\chi$ vanishes at
times $t \lesssim t_G$, the only contribution to the mass matrix for the $\chi_n$ at
such times is the contribution from the KK masses.  Thus, the 
$\chi_n$ are also mass eigenstates of the theory at such times.  
The masses $m_n$ and profiles $\zeta_n(y)$ of these
fields can be determined by solving the equation of motion for $\chi(x,y)$ which 
follows from the action in Eq.~(\ref{eq:lambdascale2}) with the boundary conditions 
\begin{equation}
 \partial_y  \chi(x, y) \big|_{y = 0, \pi R} ~=~ 0~.
  \label{eq:HigEnergyBCs}
\end{equation}
In particular, one finds that the KK spectrum contains one massless mode $\chi_0$
with a flat profile~\cite{GherghettaPomarol} 
\begin{equation}
  \zeta_0(y) ~=~ \sqrt{\frac{2k}{1-e^{-2\pi kR}}}~, 
  \label{eq:profilezeromode}
\end{equation}
as well as a tower of massive modes with masses which are solutions to the 
transcendental equation
\begin{equation}
  J_1\left(\frac{m_n}{\mKK}\right)Y_1\left(\frac{m_n}{k}\right) ~=~
    Y_1\left(\frac{m_n}{\mKK}\right)J_1\left(\frac{m_n}{k}\right)~,
  \label{eq:MassSpecEqnNomphi}
\end{equation}
where $J_\alpha(x)$ and $Y_\alpha(x)$ respectively denote the Bessel functions of 
the first and second kind and where we have defined 
\begin{equation}
  m_{\text{KK}} ~\equiv~ ke^{-\pi kR}~.
\end{equation}
The corresponding bulk profiles of these massive modes are given by 
\begin{equation}
  \zeta_n(y) ~=~ \mathcal{N}_n e^{2ky} 
    \left[ J_2 \left(\frac{m_n}{ke^{-ky}}  \right) 
  + b_n Y_2 \left(\frac{m_n}{ke^{-ky}} \right) \right]~,
  \label{eq:profile}
\end{equation}
where $b_n$ is a constant whose value is specified by the boundary conditions for 
$\chi(x,y)$ at $y=0$ and $y= \pi R$ and where the normalization constant $\mathcal{N}_n$ 
is determined by the orthogonality relation
\begin{equation}
  \int_0^{\pi R} e^{-2ky}\zeta_m(y)\zeta_n(y) dy ~=~ \delta_{mn}~.
\end{equation}
For the boundary conditions given in Eq.~(\ref{eq:HigEnergyBCs}), we have 
\begin{equation}
  b_n ~=~ \frac{J_1\left(\frac{m_n}{\mKK}\right)}{Y_1\left(\frac{m_n}{\mKK}\right)}~.
  \label{eq:bn}
\end{equation}

The massless mode $\chi_0$, which has a flat profile in the extra dimension, inherits 
the misaligned VEV in Eq.~(\ref{eq:VEVin5D}) from the bulk scalar.  Thus, we have
\begin{equation}
  \langle \chi_0 \rangle ~=~ \theta \sqrt{\frac{1-e^{-2\pi k R}}{2k}} f_X^{3/2}
     ~\equiv~ \theta \hat{f}_X~, 
\label{eq:VEV}
\end{equation}
while $\langle \chi_n \rangle = 0$ for all of the $\chi_n$ with $n > 0$.

At times $t \sim t_G$, instanton effects associated with the gauge group $G$ give   
rise to a potential for $\chi$ on the UV brane.  We focus here on the consequences 
of the brane-localized mass term $m_B$ for $\chi$ which generically appears in
this potential.  In the presence of such a mass term, the action in Eq.~(\ref{eq:lambdascale2}) 
is modified at times $t \gtrsim t_G$ to  
\begin{equation}
  \mathcal{S}_{\chi} ~=~ - \int d^5 x \sqrt{-g} 
    \Bigg[ \frac{1}{2} \partial_M \chi \partial^M \chi 
    + m_B \chi^2 \delta (y) \Bigg]~.
  \label{eq:SwithaUVBraneMass}
\end{equation}
The corresponding boundary condition for $\chi$ on the UV brane at late times is
\begin{equation}
  \left( \partial_y - m_B \right) \chi(x, y) \big|_{y = 0} ~=~ 0~,
  \label{eq:LowEnergyBCsUVMass}
\end{equation}
while the boundary condition on the IR brane remains unchanged.
As a result of this modification, the mass eigenstates $\hat{\chi}_n$ of the 
four-dimensional theory at $t \gtrsim t_G$ are no longer the KK-number 
eignenstates $\chi_n$, but rather admixtures of these fields.  The masses 
$\hat{m}_n$ of these fields are the solutions to the equation
\begin{multline}
  J_1\left(\frac{\hat{m}_n}{\mKK}\right)\left[ 
    \frac{m_B}{ \hat{m}_n} 
    Y_2\left(\frac{\hat{m}_n}{k}\right) 
    - Y_1\left(\frac{\hat{m}_n}{k}\right)\right] \\ ~=~
  Y_1\left(\frac{\hat{m}_n}{\mKK}\right)
    \left[ \frac{m_B}{\hat{m}_n} 
    J_2\left(\frac{\hat{m}_n}{k}\right) 
    - J_1\left(\frac{\hat{m}_n}{k}\right)\right]~.
  \label{eq:MassSpecEqnmphi}
\end{multline}   
The bulk profiles $\hat{\zeta}_n(y)$ of the $\hat{\chi}_n$ are given by an expression 
identical in form to the expression appearing in Eq.~(\ref{eq:profile}), but with 
$\hat{m}_n$ in place of $m_n$ and a constant $\hat{b}_n$ which reflects the modified 
boundary condition on the UV brane in place of $b_n$.  In particular, $\hat{b}_n$ 
turns out to have the same form as in Eq.~(\ref{eq:bn}), but with $\hat{m}_n$ in place 
of $m_n$.  We note that in the presence of a non-zero mass term $m_B$, all of the 
$\hat{\chi}_n$ --- including even the lightest such state $\hat{\chi}_0$ --- are massive.

We now turn to examine how the brane-localized mass term $m_B$ affects the physics
of these mass-eigenstate fields.  In doing so, we shall find it convenient to adopt
an alternative parametrization for this mass term.  In particular, without loss of 
generality, we choose to parametrize the brane-localized mass term $m_B$ in terms of a 
``brane-mass parameter'' $m_\phi$, which we define such that   
\begin{equation}
  m_B ~=~ \frac{m_\phi^2}{2k}\left(1-e^{-2\pi k R}\right)~.
\end{equation}
We note that parameter $m_\phi$ has a straightforward physical interpretation.
In particular, given the normalization for the KK zero mode in 
Eq.~(\ref{eq:profilezeromode}), we observe that $m_\phi^2$ represents the element
$\mathcal{M}_{00}^2$ of the squared-mass matrix $\mathcal{M}^2$ in the basis of 
the unmixed KK modes $\chi_n$.  In this way, the parameter $m_\phi$ can be viewed 
as the warped-space analogue of the similarly-named parameter in Ref.~\cite{DDM1}. 

As discussed above, the late-time mass eigenstates $\hat{\chi}_n$ of the theory 
can be represented as linear combinations of the KK-number eigenstates $\chi_\ell$.  
In particular, one finds that~\cite{BatellGherghetta}      
\begin{equation}
  |\hat{\chi}_n \rangle ~=~ \sum_{\ell = 0}^\infty U_{n\ell} |\chi_\ell \rangle~,
\end{equation}
where the elements $U_{n\ell}$ of the mixing matrix which relates these two sets of
states are given by  
\begin{equation}
    U_{n\ell} ~\equiv~ \langle \chi_\ell | \hat{\chi}_n \rangle ~=~
      \int_{0}^{\pi R} e^{-2ky} \zeta_\ell(y) \hat{\zeta}_n(y) dy~.
   \label{eq:Unl}
\end{equation}
We shall once again find it useful here, as we did when analyzing our partially composite 
theory in Sect.~\ref{sec:PartiallyCompositeAxion}, to define a set of mixing coefficients 
$A_n ~\equiv~ U_{n 0}$, which in the dual theory represent the mixing between these mass 
eigenstates and the KK zero-mode $\chi_0$.  Indeed, these mixing coefficients once again
play an important role in the phenomenology of the $\hat{\chi}_n$.  Since $\chi_0$ is 
the only one of the KK-number eigenstates which acquires a non-zero VEV from the 
misalignment mechanism, $A_n$ determines the VEV $\langle\hat{\chi}_n\rangle$ 
of $\hat{\chi}_n$.  In particular, in the rapid-turn-on approximation, 
Eq.~(\ref{eq:VEV}) implies that~\cite{DDGAxion}  
\begin{equation}
  \langle \hat{\chi}_n(t_G) \rangle ~=~ \theta A_n \hat{f}_X~.
  \label{eq:MixedHatPhiVEV}
\end{equation} 

The mixing coefficients $A_n$ can be obtained from the general expression for 
$U_{n \ell}$ in Eq.~(\ref{eq:Unl}), which holds regardless of the relationship between 
$m_\phi$, $k$, and $R$.  However, a simple analytic approximation for $A_n$ may also be 
obtained within one of the regimes of greatest phenomenological interest, which is 
the regime in which $m_\phi$ is small compared to the other relevant scales 
in the theory.  In particular, in the regime in which $m_\phi \ll \mKK$, the 
mixing coefficient $A_0$ for the lightest mass eigenstate $\hat{\chi}_0$ is 
approximately unity.  Moreover, within this same regime, the mixing coefficients 
for all $\hat{\chi}_n$ with masses in the regime $k \gg \hat{m}_n \gg \mKK$ are 
approximately given by
\begin{equation}
  A_n ~\approx ~ \sqrt{\frac{\pi}{2}} e^{-\pi kR} 
    \left( \frac{m_\phi}{\mKK} \right)^2 
    \left( \frac{\mKK}{\hat{m}_n} \right)^{3/2}~,
  \label{eq:mixing}
\end{equation}
while the masses themselves are well approximated by
\begin{equation}
  \hat{m}_n ~ \approx ~ \left( n + \frac{1}{4} \right) \pi \mKK~.
 \label{eq:mnApproxLowmphi}
\end{equation}
We note that since this analytic approximation is valid in the regime in which 
$k \gg \hat{m}_n \gg \mKK$, the greatest degree of agreement between the values of $A_n$ 
obtained from this approximation and the exact result obtained from Eq.~(\ref{eq:Unl})
occurs for intermediate values of $n$.

In Sect.~\ref{sec:PartiallyCompositeAxion}, we saw that a second set of coefficients,
namely the projection coefficients $A_n'$, also played a crucial role in the phenomenology 
of our partially composite DDM scenario.  The analogous quantity in the dual theory for
each $\hat{\chi}_n$ is the coefficient 
$A_n' \equiv \sqrt{1-e^{-2\pi k R}} 
\langle \chi(x, 0) | \hat{\chi}_n \rangle / \sqrt{2k}$ which 
describes the projection of this state onto the UV brane at $y=0$.  Indeed, since the fields 
of the SM are also assumed to be localized on the UV brane, all interactions between
the $\hat{\chi}_n$ and any SM field necessarily include one or more factors of $A'_n$.
In general, these projection coefficients are given by
\begin{eqnarray}
  A'_n &\equiv & \sqrt{\frac{1-e^{-2\pi k R}}{2k}} \sum_{\ell=0}^{\infty} \zeta_\ell(0) 
    \int_{0}^{\pi R} e^{-2ky} \zeta_\ell(y) \hat{\zeta}_n(y) dy \nonumber \\
    & = & \sqrt{\frac{1-e^{-2\pi k R}}{2k}}\hat{\zeta}_n(0)~.
  \label{eq:projectionUV}
\end{eqnarray}
where in going from the first to the second line, we have used the completeness relation
\begin{equation}
  \sum_{n=0}^\infty \zeta_n (y) \zeta_n (y') ~=~ e^{2 k y} \delta(y-y')~.
  \label{eq:CompletenessRel}
\end{equation} 

Once again, while the expression in Eq.~(\ref{eq:projectionUV}) is completely general, 
simple analytic approximations for the $A'_n$ may also be obtained within our regime of 
phenomenological interest --- \ie, the regime in which $m_\phi$ is much smaller than the 
other relevant scales in the theory.  Indeed, we find that within this regime, the $A'_n$ 
for those $\hat{\chi}_n$ with masses which satisfy $k \gg \hat{m}_n \gg \mKK$ are well 
approximated by
\begin{equation}
  A'_n ~\approx~ 
    \sqrt{\frac{\pi}{2}} e^{-\pi kR} \left( \frac{\hat{m}_n}{\mKK} \right)^{1/2}~. 
  \label{eq:projectionUVApprox}
\end{equation}


\section{Dynamical Dark Matter from a Warped Extra Dimension\label{sec:DynamicalDarkMatter}}


Thus far, we have analyzed the properties of the mass-eigenstate fields 
$\hat{\chi}_n$ which emerge in the gravity dual of our partially composite
DDM scenario.  We shall now show that an appropriate balancing of decay widths against 
abundances can emerge across this collection of fields such that the $\hat{\chi}_n$ 
collectively constitute a viable DDM ensemble.  

Cosmological constraints on dark-matter decays arise primarily as a consequence of two
considerations.  First, such decays lead to a modification of the total dark-matter
abundance and the effective dark-matter equation of state, and thus to a departure 
from the standard cosmology.  Second, observational limits constrain the 
production rate of SM particles which might appear in the final states into which the
dark-matter particles decay.  Since the corresponding constraints on DDM scenarios 
depend sensitively on the mass scales involved and on the particular channels through 
which the different dark-matter species decay, we focus here on the constraints on 
the total abundance and equation of state for our ensemble of $\hat{\chi}_n$. 

\subsection{Total Abundance and Effective Equation of State}
  
In order to determine how the total abundance and effective equation of state for our 
ensemble evolve in time, we begin by assessing how the cosmological abundances 
$\Omega_n$ of the individual $\hat{\chi}_n$ scale across the ensemble as a function of 
$\hat{m}_n$ immediately
after these abundances are established.  In general, $\Omega_n = \rho_n /\rhocrit$ represents 
the ratio of the energy density $\rho_n$ of $\hat{\chi}_n$ to the critical density 
$\rhocrit \equiv 3M_P^2H^2$ of the universe, where $M_P$ is the reduced Planck mass and 
$H$ is the Hubble parameter.  We focus here on the contribution to each of the $\Omega_n$ from 
misalignment production, which arises as a consequence of dynamics associated with the 
mass-generating phase transition described in Sect.~\ref{sec:DualMisalignmentMechanism}.  
We have seen that each of the $\hat{\chi}_n$ acquires a misaligned VEV as a consequence of this 
phase transition.  As a result, each of these fields acquires an energy density $\rho_n(t_G)$ 
given by Eq.~(\ref{eq:RhonMisaligned}).
In the rapid-turn-on approximation, $\langle \hat{\chi}_n(t_G) \rangle$ is given by
Eq.~(\ref{eq:MixedHatPhiVEV}) and the corresponding initial abundance $\Omega_n(t_G)$ 
of each $\hat{\chi}_n$ at $t=t_G$ is
\begin{equation}
  \Omega_n(t_G) ~=~ \frac{\theta^2 A_n^2 \hat{m}_n^2\hat{f}_X^2}{6M_P^2H^2(t_G)}~.
\end{equation}

It is also important to note that the $\Omega_n$ do not necessarily all evolve with 
$t$ in the same way for all $t > t_G$.  Indeed, at any particular $t$, only those 
$\hat{\chi}_n$ for which $2\hat{m}_n \gtrsim 3H(t)$ experience underdamped oscillations, 
whereas the $\hat{\chi}_n$ for which $2\hat{m}_n \lesssim 3H(t)$ remain overdamped.  We may 
therefore associate an oscillation-onset time $t_n$ with each such field.  At any given 
time $t$, the energy densities of those fields for which $t_n < t $ evolve in time like massive 
matter, whereas, the energy densities of those $\hat{\chi}_n$ with $t_n > t$ scale like vacuum 
energy.  Since successively lighter fields begin oscillating at successively later times, we may 
consider the time $t_0$ at which the lightest ensemble constituent $\hat{\chi}_0$ begins 
oscillating as the time at which the initial abundance for the DDM ensemble is effectively 
established, since at all subsequent times $t \geq t_0$ all of the ensemble constituents behave 
like massive matter.  Of course, the manner in which the initial abundances 
$\Omega_n^0 \equiv \Omega_n(t_0)$ at this time scale with $\hat{m}_n$ over some range of $n$ 
depends on whether the $\hat{\chi}_n$ all begin oscillating instantaneously at $t = t_G$, 
or whether the $t_n$ are staggered in time.  As a result, the overall scaling behavior of 
$\Omega_n^0$ with $\hat{m}_n$ turns out to be~\cite{DDM1}
\begin{equation}
  \Omega_n^0 ~\propto~ \begin{cases} 
    \hat{m}_n^2 A_n^2 & \text{instantaneous} \\ 
    \hat{m}_n^{1/2} A_n^2 & \text{staggered (RD era)} \\ 
    A_n^2 & \text{staggered (MD era)}~,
  \end{cases}
  \label{eq:abundances}
\end{equation} 
where in cases in which the oscillation-onset times are staggered, the manner in which 
$\Omega_n^0$ scales with $\hat{m}_n$ depends on whether these oscillation-onset times occur 
during a radiation-dominated (RD) or matter-dominated (MD) epoch.  While the expressions 
in Eq.~(\ref{eq:abundances}) do not account for the decays of shorter-lived ensemble 
constituents at times $t < t_0$, we note that these expressions are nevertheless valid  
either if $t_0 \ll \tau_n$ for all of the $\hat{\chi}_n$ in an ensemble with a 
finite number of constituents, or else if the $\hat{\chi}_n$ with $\tau_n \lesssim t_0$ 
collectively contribute only a negligible fraction of the total abundance of the ensemble 
at $t_0$.

At times $t \geq t_0$, all of the $\hat{\chi}_n$ behave like massive matter.  Thus, their
energy densities are all affected by Hubble expansion in exactly the same way.  In 
particular, each $\rho_n$ evolves according to an equation of the form
\begin{equation}
  \frac{d\rho_n}{dt} ~=~ -(3H + \Gamma_n) \rho_n~,
  \label{eq:rhonDiffEq}  
\end{equation}   
where $\Gamma_n$ is the decay width of $\hat{\chi}_n$.  Within either a radiation-dominated (RD) 
or matter-dominated (MD) era, $H$ is well approximated by
\begin{equation}
  H ~ \approx ~ \frac{\kappa}{3t}~,
\end{equation}
where $\kappa$ is constant and given by
\begin{equation}
  \kappa ~=~ \begin{cases}
  3/2 & {\rm RD}\\
  2 & {\rm MD}~.
  \end{cases}
\end{equation} 
Solving Eq.~(\ref{eq:rhonDiffEq}) for $H$ of this form and using the fact that 
the critical density scales with the scale factor $a$ like 
$\rhocrit \propto a^{-\frac{6}{\kappa}}$ within a RD or MD era, we find that    
\begin{equation}
  \Omega_n(t) ~=~ \Omega_n^\ast 
    \left(\frac{a}{a_\ast}\right)^{\frac{6}{\kappa}-3}
    e^{-\Gamma_n(t-\tast)}~,
  \label{eq:OmeganIndivid}
\end{equation}
where $t_\ast$ is an arbitrary fiducial time within the same era and where 
$\Omega_n^\ast =\Omega_n(\tast)$ and $a_\ast = a(\tast)$ respectively denote the values of 
$\Omega_n$ and $a$ at this fiducial time.  The total abundance $\Omegatot$ of the ensemble,
which is simply the sum of the individual $\Omega_n$, is therefore given by
\begin{equation}
  \Omegatot(t) ~=~ \sum_{n=0}^\infty \Omega_n^\ast 
    \left(\frac{a}{a_\ast}\right)^{\frac{6}{\kappa}-3}
    e^{-\Gamma_n(t-\tast)}~.
 \label{eq:Omegatot}
\end{equation}

The effective dark-matter equation of state for a DDM ensemble can be characterized 
by a time-dependent parameter $\weff(t)$, which is defined by the relation 
$p_{\rm tot}(t) = \weff(t) \rho_{\rm tot}(t)$, where $p_{\rm tot}(t)$ is the total momentum 
density of the ensemble as a whole at time $t$ and where $\rhotot(t)$ is the 
corresponding energy density.  This equation-of-state parameter can be written in the
general form~\cite{DDM1}
\begin{equation}
  \weff ~=~ \frac{1}{3H}\frac{d\log\rhotot}{dt}~.
\end{equation}
Within a RD or MD era, this expression reduces to
\begin{equation}
  \weff ~=~ -\frac{t}{\kappa\Omegatot}\frac{d\Omegatot}{dt} + \frac{2}{\kappa} - 1~.
  \label{eq:weffOmega}
\end{equation}
The time derivative of the expression for $\Omegatot$ in Eq.~(\ref{eq:OmeganIndivid}) is simply 
\begin{equation}
  \frac{d\Omegatot}{dt} ~=~ \sum_{n=0}^\infty\left[\frac{2-\kappa}{t} - \Gamma_n\right] 
     \Omega_n^\ast \left(\frac{a}{a_\ast}\right)^{\frac{6}{\kappa}-3}
    e^{-\Gamma_n(t-\tast)}~.    
\end{equation}  
Thus, we find that with the assumptions outlined above, the expression for $\weff$ in
Eq.~(\ref{eq:weffOmega}) simplifies to 
\begin{eqnarray}
  \weff &=& \frac{\sum_{n=0}^\infty\Omega_n^\ast [\Gamma_n t - (2-\kappa)] 
    e^{-\Gamma_n(t-t_\ast)}}{\kappa \sum_{n=0}^\infty\Omega_n^\ast e^{-\Gamma_n(t-t_\ast)}} 
    +\frac{2}{\kappa}-1 \nonumber \\
    &=& \frac{\sum_{n=0}^\infty\Omega_n^\ast \Gamma_n t 
    e^{-\Gamma_n(t-t_\ast)}}{\kappa \sum_{n=0}^\infty\Omega_n^\ast e^{-\Gamma_n(t-t_\ast)}}~.
  \label{eq:weffOmegaArbt}
\end{eqnarray}

We now turn to assess how the $\Gamma_n$ scale with $\hat{m}_n$ across the ensemble.
Since the fields of the SM are assumed to be localized on the UV brane, the partial width 
for any tree-level process in which one of the $\hat{\chi}_n$ decays directly into a final 
state involving these fields necessarily involves the projection coefficients $A_n'$.
In particular, in situations in which two-body decays directly to a pair of much lighter 
SM particles dominate the width of $\hat{\chi}_n$, one finds that~\cite{DDM1} 
\begin{equation}
  \Gamma_n ~\sim~ \frac{\hat{m}_n^3}{\hat{f}_X^2} A_n^{\prime 2}~.
  \label{eq:decaywidth}
\end{equation} 

In principle an additional contribution to $\Gamma_n$ for each of the $\hat{\chi}_n$ can
arise as a result of intra-ensemble decays.  In the case of a flat extra 
dimension~\cite{DDM1,DDM2}, KK-number conservation serves to suppress such contributions,
which arise in this case only through brane-localized operators.  In the case of 
a warped extra dimension, no such conservation principle holds.  
Nevertheless, we expect any bulk interactions which could give rise to intra-ensemble 
decays to be suppressed, based on the general arguments advanced in
Sect.~\ref{sec:PartiallyCompositeAxion} concerning the scaling properties of the 
decay amplitudes of the $\hat{\chi}_n$ in the 4D dual picture.  
We shall therefore neglect the contribution from intra-ensemble decays
in what follows.

\subsection{Constraining Deviations from the Standard Cosmology}

Having derived general expressions for $\Omegatot$ and $\weff$ for our 
ensemble within a RD or MD era, we now turn to consider how these quantities
are constrained by data.  First of all, consistency with observation requires that 
$\Omegatot$ not differ significantly from the abundance of a stable, cold 
dark-matter (CDM) candidate over the range of timescales extending from the time 
$t_{\rm BBN}$ at which Big-Bang nucleosynthesis (BBN) begins until the present
time $\tnow$.  Motivated by this consideration, at all times 
$t_0 \leq t \leq \tnow$, we shall impose the bound
\begin{equation}
  \frac{\Omegatot(t)}{\widetilde{\Omega}_{\rm tot}(t)} ~ > ~ 0.95~,
  \label{eq:Omegatotbound} 
\end{equation}    
where $\widetilde{\Omega}_{\rm tot}(t)$ represents the total abundance that the ensemble
{\it would}\/ have had at a given time $t$ if all of the ensemble 
constituents had been absolutely stable.  The value $0.95$ has been chosen in accord
with the value adopted in Ref.~\cite{DDMHagedorn} in order to ensure that the total abundance 
of the ensemble does not deviate significantly from the case of a stable CDM candidate.

In addition to imposing this constraint on $\Omegatot$, we must also ensure that $\weff$ 
does not deviate significantly from that of a stable CDM candidate at any 
time during the recent cosmological past.  For the case of a flat extra dimension,
it was shown in Ref.~\cite{DDM1} that this bound on $\weff$ could be phrased primarily
as a constraint on the scaling relations which govern how the abundances, decay widths, \etc,
of the individual ensemble constituents scale in relation to one another across the ensemble.
Indeed, in the flat-space limit, one finds that the decay widths $\Gamma_n$ of the ensemble 
constituents increase monotonically with $\hat{m}_n$.  As a result, within the regime which 
the spectrum of decay widths $\Gamma_n$ within the ensemble is reasonably dense, one may 
sensibly approximate the spectrum of abundances $\Omega(\Gamma)$ and the density of states 
per unit decay width $n_\Gamma(\Gamma)$ within the ensemble as functions of a continuous 
variable $\Gamma$.  Without loss of generality, one may parametrize these functions as  
\begin{eqnarray}
  \Omega(\Gamma) & =  & A\Gamma^{\alpha(\Gamma)} \nonumber \\
  n_\Gamma(\Gamma) & = & B\Gamma^{\beta(\Gamma)}~, 
  \label{eq:OmegaGammaScaling}
\end{eqnarray}
where $A$ and $B$ are constants and where the scaling exponents $\alpha(\Gamma)$ and 
$\beta(\Gamma)$ are functions of $\Gamma$.  Moreover, for the DDM ensembles considered in 
Ref.~\cite{DDM1}, $\alpha(\Gamma) \approx \alpha$ and $\beta(\Gamma)\approx \beta$ 
are typically roughly constant either across the entire ensemble or else across a large range 
of $\Gamma$.  Under these assumptions, it was shown that at times $t \gtrsim \tMRE$, where 
$\tMRE$ denotes the time of matter-radiation equality, $\weff$ is well approximated by   
\begin{equation}
  \weff(t) ~\approx~ \weff(\tnow) \left(\frac{t}{\tnow}\right)^{-1-x}~,
\end{equation} 
where $x \equiv \alpha + \beta$.  Thus, constraints on $\weff$ for the case of a flat extra
dimension can be phrased as bound on $x$ and $\weff(\tnow)$.  In particular, ensembles which 
are likely to be phenomenologically viable are those for which $\weff(\tnow)$ is fairly small
and $x  \leq -1$.  The former criterion ensures that the equation-of-state parameter for the 
ensemble does not differ significantly from the constant value $w = 0$ associated with a stable 
CDM candidate at present time, while the latter criterion ensures that 
$0 \leq \weff(t) \leq \weff(\tnow)$ for all $t \leq \tnow$.
    
By contrast, for the case of a warped extra dimension, constraints on $\weff$ cannot always be
characterized in this way.  The reason is that within certain regions of the parameter space of 
our scenario, $\Gamma_n$ is not a monotonic function of $\hat{m}_n$.  A non-monotonicity of
this sort implies that ensemble constituents with significantly different $\hat{m}_n$ --- and 
hence, in general, significantly different individual abundances $\Omega_n$ --- can have similar 
or identical values of $\Gamma_n$.  When this is the case, the function $\Omega(\Gamma)$ in
Eq.~(\ref{eq:OmegaGammaScaling}) cannot be sensibly defined and indeed may not even be 
single-valued.  Thus, within any region of parameter space in which such non-monotonicities 
in the spectrum of decay widths develop, there is no meaning to the parameter $x$.     

\begin{figure} 
\centering 
  \includegraphics[width=0.95\linewidth]{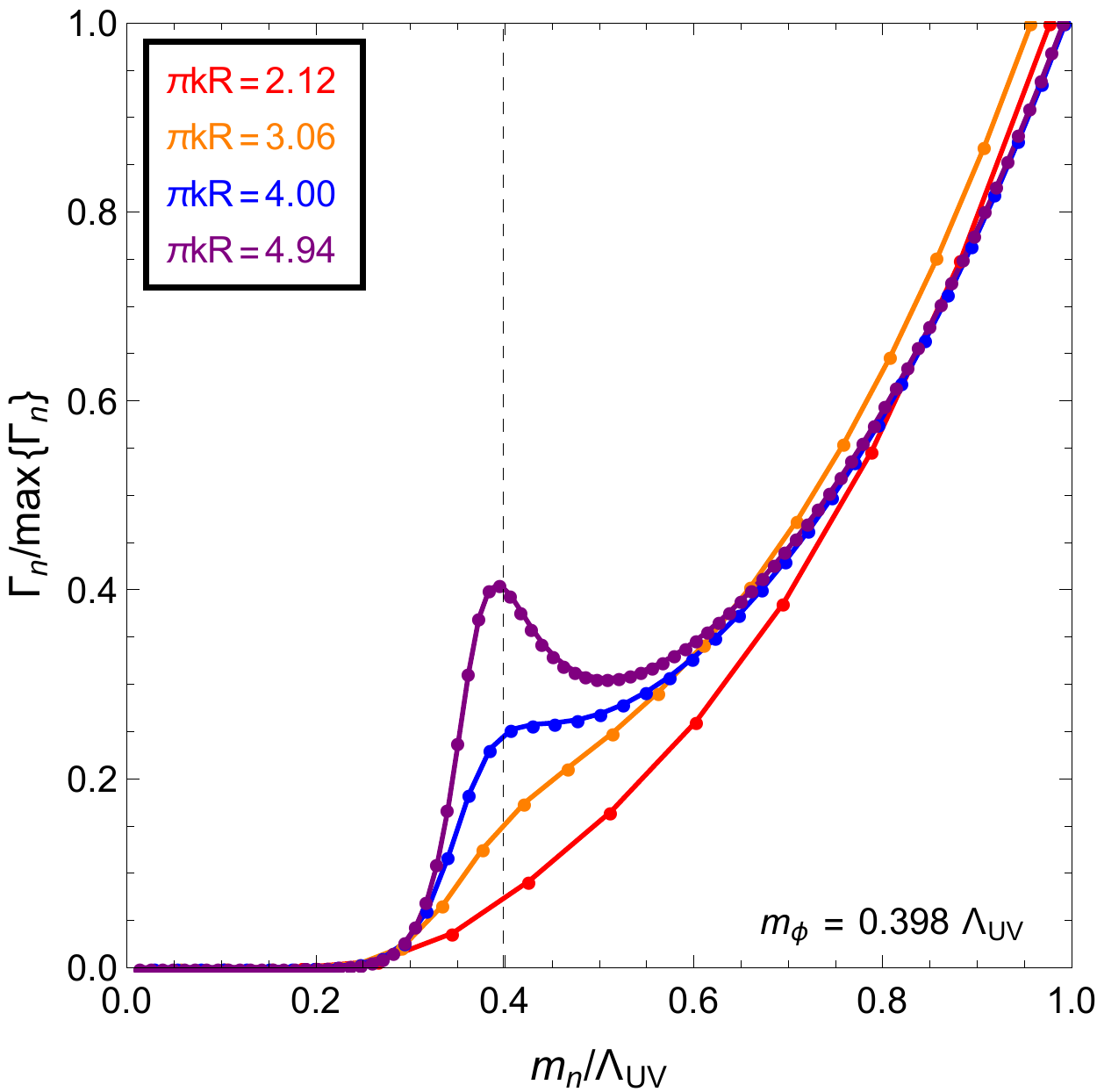}
  \includegraphics[width=0.95\linewidth]{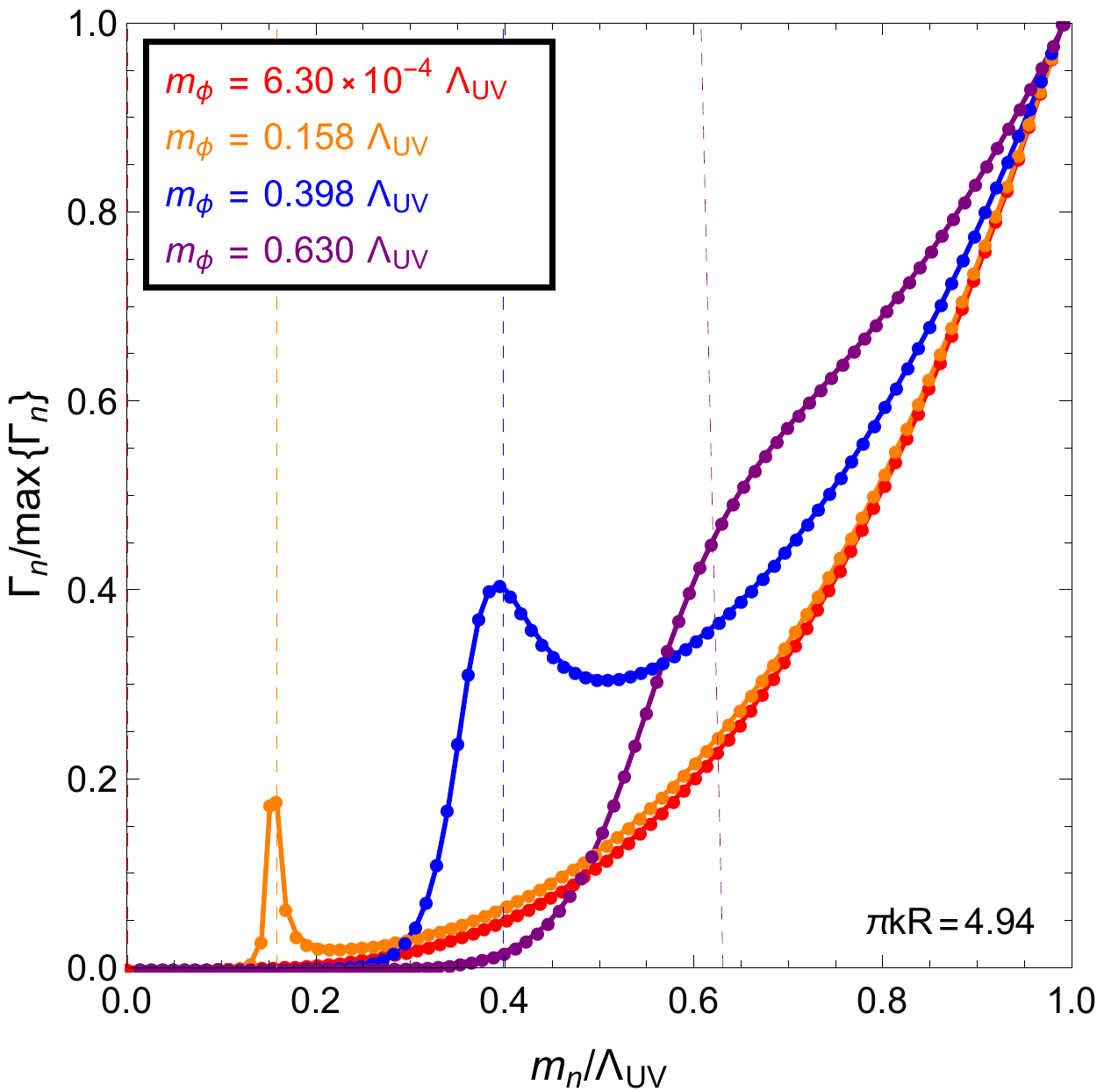}
\caption{The decay-width spectra obtained for several different choices of our model 
  parameters.  The dots of each color indicate the decay widths $\Gamma_n$ of the 
  ensemble constituents, normalized to the maximum width $\Gamma_{\rm max}$ obtained for 
  any ensemble constituent with a mass in the range $\hat{m}_n \leq \LambdaUV$.
  The continuous solid curve which connects each set of dots is included simply to guide the 
  eye.  The four decay-width spectra shown in the top panel illustrate the effect of varying 
  the AdS curvature scale in the regime in which $m_\phi$ is large with $\LambdaUV R = 3$  
  and $m_\phi/\LambdaUV = 0.398$ held fixed.  We observe that as $\pi k R$ increases, a 
  non-monotonicity emerges in the spectrum wherein a local maximum in $\Gamma_n$ occurs 
  around $\hat{m}_n \sim m_\phi$.  The four decay-width spectra shown in the 
  bottom panel illustrate the effect of varying $m_\phi$ with $\pi k R = 4.94$ and 
  $\LambdaUV R = 3$ held fixed.  
  \label{fig:GammaSpectrumvsmn}}
\end{figure}

In Fig.~\ref{fig:GammaSpectrumvsmn}, in order to show how and where such non-monotonicities can 
arise within the parameter space of our scenario, we display the decay-width spectra obtained for 
several different choices of model parameters.  The dots of each color indicate the actual 
$\Gamma_n$ values of the $\hat{\chi}_n$, and the continuous solid curve connecting these dots 
is included simply to guide the eye.  In order to facilitate comparison between the 
different spectra, we normalize the decay width of each state in a given ensemble to the maximum 
decay width 
\begin{equation}
  \Gamma_{\rm max} ~\equiv~ \max_{\hat{m}_n \leq \LambdaUV}\{\Gamma_n\}
\end{equation}
obtained for any ensemble constituent with a mass in the range $\hat{m}_n \leq \LambdaUV$.  The 
four decay-width spectra shown in the top panel illustrate the effect of varying the AdS$_5$ 
curvature scale in the regime in which $m_\phi$ is large.  For all spectra shown in the panel, 
we have fixed $m_\phi/\LambdaUV = 0.398$ (indicated by the black dashed vertical line) and 
$\LambdaUV R =3$.  The four decay-width spectra shown in the bottom panel illustrate the effect 
of varying $m_\phi$ with $\pi k R = 4.94$ and $\LambdaUV R = 3$ held fixed.  For each of these
four spectra, the dashed vertical line of the same color indicates the corresponding value of 
$m_\phi/\LambdaUV$.

We observe from the top panel of Fig.~\ref{fig:GammaSpectrumvsmn} that for small values of 
$\pi k R$, the decay-width spectrum of the ensemble rises monotonically with $\hat{m}_n$, just 
as it does in the flat-space limit.  However, as $\pi k R$ increases, a local maximum in 
$\Gamma_n$ develops around $\hat{m}_n \sim m_\phi$.  
This non-monotonicity, which is a consequence of the warping of the extra dimension, is 
an example of a qualitative feature which does not arise in flat-space DDM scenarios.  
This behavior is a consequence of the manner in which the projection coefficient $A'_n$, 
which is proportional to the value 
$\hat{\zeta}_n(0) = \mathcal{\hat{N}}_n[J_2(\hat{m}_n/k) + \hat{b}_nY_2(\hat{m}_n/k)]$ 
of the bulk profile of the corresponding field on the UV brane, varies across the ensemble.  
Since $|J_2(\hat{m}_n/k)| \ll |Y_2(\hat{m}_n/k)|$ in the regime in which $\hat{m}_n \ll k$, the 
magnitude of $A_n'$ will be maximized in this regime when the constant $\hat{b}_n$ is 
large.  While it is not immediately obvious from the form of the expression in 
Eq.~(\ref{eq:profile}) that the value of $\hat{b}_n$ is enhanced for ensemble constituents
with masses $\hat{m}_n \sim m_\phi$, we note that this expression can also be recast in the 
alternative form
\begin{equation}
  \hat{b}_n ~=~ - \left[
    \frac{m_B J_2\left(\frac{\hat{m}_n}{k}\right) 
      - \hat{m}_nJ_1\left(\frac{\hat{m}_n}{k}\right)}
    {m_B Y_2\left(\frac{\hat{m}_n}{k}\right) 
      - \hat{m}_nY_1\left(\frac{\hat{m}_n}{k}\right)}\right]~,
  \label{eq:bnAlt}
\end{equation}
which is obtained applying the boundary condition at $y=0$ in 
Eq.~(\ref{eq:LowEnergyBCsUVMass}) to $\chi(x,y)$ instead of the boundary condition at 
$y = \pi R$ in Eq.~(\ref{eq:HigEnergyBCs}).  For $\hat{m}_n \ll k$, we may 
approximate $J_\alpha(x)$ and $Y_\alpha(x)$ using the standard asymptotic expansions for 
$x \ll 1$.  After some algebra, we find that 
\begin{equation}
    \hat{b}_n ~\approx~ \frac{\pi \hat{m}_n^4}{32k^4} \left[
      \frac{m_\phi^2(1-e^{-2\pi k R}) - 8k^2}{m_\phi^2(1-e^{-2\pi k R})-\hat{m}_n^2}\right]~.
  \label{eq:bnAltApprox}
\end{equation}
We see that within the regime in which $\pi k R$ is large, the denominator in this expression 
is quite small for $\hat{m}_n \sim m_\phi$.  Indeed, the expression for $\hat{b}_n$ in 
Eq.~(\ref{eq:bnAltApprox}) exhibits a singularity at 
$\hat{m}_n = m_\phi (1-e^{-2\pi k R})^{1/2}$, though 
none of the physical masses for the ensemble constituents ever takes precisely this singular 
value.  As a result, $\hat{b}_n$ --- and therefore also $A_n'$ --- is sharply peaked for 
$\hat{\chi}_n$ with masses near $m_\phi$ in this regime. 
By contrast, within the regime in which $\pi k R$ is small or vanishing, the 
asymptotic expansions which led from Eq.~(\ref{eq:bnAlt}) to Eq.~(\ref{eq:bnAltApprox}) are
not valid.  In Appendix~\ref{app:FlatLimit}, we derive a general expression for 
$A_n'$ in the $\hat{m}_n \gg k$ regime using the asymptotic expansions for 
$J_\alpha(x)$ and $Y_\alpha(x)$ valid for $x \gg 1$ and show that this expression 
contains no such singularities.

As $m_\phi$ is further increased, the peak in $\Gamma_n$ around $\hat{m}_n \sim m_\phi$ becomes 
higher and broader.  However, for sufficiently large $m_\phi$, the decrease in $A'_n$ with 
$\hat{m}_n$ beyond this peak is more than compensated for by the $\hat{m}_n^2$ factor in 
Eq.~(\ref{eq:decaywidth}).  As a result, the decay-width spectrum once again becomes monotonic in 
$\hat{m}_n$.  Thus, we see that both in the regime in which $m_\phi \ll \mKK$ and in the regime 
in which $m_\phi \sim \LambdaUV$, the scaling relation $\Omega(\Gamma)$ in 
Eq.~(\ref{eq:OmegaGammaScaling}) can still be meaningfully defined, even for large $\pi k R$.  
Rather, it is for intermediate values of $m_\phi$ that this description breaks down when the
warping of the space becomes significant.  We note that while $\Gamma_n$ scales 
non-monotonically with $\hat{m}_n$, the abundances $\Omega_n$ nevertheless scale 
monotonically.  Interestingly, this is the converse of the situation in Ref.~\cite{DDMThermal},
where it is the abundances which scale non-monotonically with mass while the decay widths are
monotonic.

Since the parameter $x$ is not well defined across the entire parameter space of our 
warped-space DDM scenario, we must establish a different method for constraining deviations 
of the effective equation-of-state parameter $\weff$ of the ensemble from the constant
value $w = 0$ which characterizes a stable CDM candidate during the recent cosmological past.
In particular, at all times $t_0 \leq t \leq \tnow$, we shall impose the bound
condition   
\begin{equation}
  \weff(t) ~<~ 0.05~.
  \label{eq:weffbound} 
\end{equation}
Once again, the value $0.05$ has been chosen in accord with the value adopted in
Ref.~\cite{DDMHagedorn} in order to ensure that the equation of state for the ensemble 
does not deviate significantly from that of a stable CDM candidate.

\subsection{Case Study: Small Brane Mass, Strong Warping}

Before embarking on a general exploration of the parameter space of our 
5D scenario, we begin by focusing on a particular region of interest within 
that parameter space.  In particular, we consider the region in which 
the AdS curvature scale is large, in the sense that $\pi k R \gg 1$, while $m_\phi$ is 
small in comparison with all other relevant scales in the theory --- a criterion 
which, in the highly-warped regime, is tantamount to requiring that 
$m_\phi \ll \mKK$.  This region is interesting for several reasons.  On the one hand, 
the region in which $\pi k R \gg 1$ represents the greatest degree of departure from the 
flat-space limit investigated as a context for DDM model-building in 
Refs.~\cite{DDM1,DDM2,DDMAxion}.  Moreover, this highly-warped regime corresponds to the 
regime in the 4D dual theory within which $\LambdaUV/\LambdaIR = e^{\pi k R}$ is large and 
a significant hierarchy exists between the UV and IR scales.  On the other hand as discussed 
above, the scaling relation $\Omega(\Gamma)$ is nevertheless sensibly 
defined within the regime in which $m_\phi \ll \mKK$.  Thus, within this region we
may compare our results to those obtained in these previous studies in a straightforward
manner.  Indeed, as we shall demonstrate, the scaling exponents $\alpha(\Gamma)$ and 
$\beta(\Gamma)$ in Eq.~(\ref{eq:OmegaGammaScaling}) are roughly constant across the 
range of $\Gamma$ values associated with the lighter constituents in the ensemble which
carry the majority of the abundance.  Thus, within this region, we can meaningfully define 
a single value of $x$ with the ensemble.  

Within this parameter-space region of interest, the low-lying states within the ensemble include 
a single extremely light state $\hat{\chi}_0$ with a mass $\hat{m}_0 \sim m_\phi$, as well as a 
large number of additional $\hat{\chi}_n$ with masses $k \gg \hat{m}_n \gtrsim \mKK$. 
While of course heavier states with $\hat{m}_n \geq k$ are also present within the ensemble, 
the collective abundance of these states is typically so small that the phenomenology of the 
ensemble is not terribly sensitive to how $\Omega_n$ and $\Gamma_n$ scale with $\hat{m}_n$ 
across this set of states.  Thus, we shall focus on the lighter $\hat{\chi}_n$ in deriving
a value of $x$ for the ensemble.  The expressions for $A_n$ and $A'_n$ for the light states 
with $n > 0$ are given by Eqs.~(\ref{eq:mixing}) and~(\ref{eq:projectionUV}), respectively.  
Each of these expressions scales with $\hat{m}_n$ according to a simple power law.  Thus,
we find that the abundances of the $\hat{\chi}_n$ in our 5D dual theory 
scale with $\hat{m}_n$ according to the relation
\begin{equation}
  \Omega_n ~\propto~ \begin{cases} 
    \hat{m}_n^{-1} & \text{instantaneous} \\ 
    \hat{m}_n^{-5/2} & \text{staggered (RD era)} \\ 
    \hat{m}_n^{-3} & \text{staggered (MD era)}~,
  \end{cases}
  \label{eq:abundance}
\end{equation} 
while the decay widths of these states scale with $\hat{m}_n$ according to the relation
\begin{equation}
  \Gamma_n ~\propto~ \hat{m}_n^4~.
  \label{eq:DecayWidthScaling}
\end{equation}
Given the results in Eq.~(\ref{eq:abundance}) and~(\ref{eq:DecayWidthScaling}), we
find that the functional form for $\Omega(\Gamma)$ in this case is
\begin{equation}
  \Omega(\Gamma) ~\propto~ \begin{cases} 
    \Gamma^{-1/4} & \text{instantaneous} \\ 
    \Gamma^{-5/8} & \text{staggered (RD era)} \\ 
    \Gamma^{-3/4} & \text{staggered (MD era)}~.
  \end{cases}
  \label{eq:balancingequation}
\end{equation} 

We emphasize that the scaling relation in Eq.~(\ref{eq:balancingequation}) was derived 
from asymptotic expressions for $A_n$ and $A_n'$ valid only for $n > 0$.  Within the region of 
parameter space in which $m_\phi$ is much smaller than all other relevant scales in the theory, 
the abundance $\Omega_0$ and decay width $\Gamma_0$ of $\hat{\chi}_0$ do not accord with this 
scaling relation.  Moreover, within this region of parameter space, $\Omega_0$ typically dominates 
the abundance of the ensemble, while $\Gamma_0$ is typically significantly smaller than the decay 
widths of all of the remaining $\Gamma_n$.
Indeed, this behavior arises not only in the case of a warped extra dimension, but in the 
corresponding $m R \ll 1$ regime in the case of a flat extra dimension as well~\cite{DDM1}.  
Nevertheless, since $\Omega_0$ represents a significant fraction of $\Omegatot$ within this
region, $\hat{\chi}_0$ is typically required to be sufficiently long-lived that its decays
at $t < \tnow$ have a negligible effect on the phenomenology of the ensemble.  Rather, it is 
primarily the $\hat{\chi}_n$ with $n > 0$ which dictate that phenomenology.  Thus, in what 
follows, we shall focus on the $\hat{\chi}_n$ with $n > 0$ in deriving an effective value of $x$ 
for our warped-space DDM ensembles --- as was done in the analysis in Ref.~\cite{DDM1}.  

In order to determine the scaling relation for $n_\Gamma(\Gamma)$, we 
begin by noting that the splitting $\hat{m}_{n+1} - \hat{m}_n$ between the masses of any two adjacent 
states $\hat{\chi}_{n+1}$ and $\hat{\chi}_n$ is approximately uniform across the ensemble 
for $n > 0$.  We are once again primarily interested in the regime in which the mass 
spectrum is sufficiently dense that we may approximate the density of states per unit mass 
$n_m(m)$ within the ensemble as a function of the continuous variable $m$.  Within this
regime, a uniform mass splitting implies that $n_m(m)$ is approximately constant
across the ensemble.  The corresponding density of states per unit $\Gamma$ is therefore  
\begin{equation}
  n_{\Gamma}(\Gamma) ~=~ n_m(\Gamma) 
    \left( \frac{d\Gamma}{dm} \right)^{-1} ~\sim~ \Gamma^{-3/4}~. 
  \label{eq:densityofstates}
\end{equation} 

Combining the results in Eqs.~(\ref{eq:balancingequation}) and~(\ref{eq:densityofstates}), 
we find that within the parameter-space region in which $\pi k R \gg 1$ and $m_\phi \ll \mKK$, 
the value of $x$ obtained for our ensemble of $\hat{\chi}_n$ is 
\begin{equation}
  x ~\approx~ \begin{cases} 
    -1 & \text{instantaneous}  \\ 
    -11/8 & \text{staggered (RD era)} \\ 
    -3/2  & \text{staggered (MD era)}~. \\
  \end{cases}
  \label{eq:parameterx1UV}
\end{equation}
These results indicate that within this region of parameter space, our ensemble satisfies the
the rough consistency criterion $x \lesssim -1$ independent of the details of when the 
the individual constituents begin oscillating.  Thus, we find that ensembles of this 
sort indeed exhibit an appropriate balancing of decay widths against abundances for DDM.
The values of $x$ appearing in Eq.~(\ref{eq:parameterx1UV}), along with the corresponding values 
of $\alpha$ and $\beta$ obtained in each case, are collected in Table~\ref{tab:scalingtable} for
ease of reference.

\subsection{Generalizing the Scenario}

It is possible to generalize the results of the previous section in several ways,
even if we wish to restrict our focus to the region of parameter space within which
$m_\phi$ is much smaller than all other relevant scales in the problem and $x$ is 
well defined.   

\begin{table*}[t]
  \centering
  \begin{minipage}{1.0\textwidth}
    \centering
    \begin{tabular}{||c|c||c|c|c||c|c|c||c|c|c||}
      \hline \hline
      \multicolumn{2}{||c||}{Model} &
      \multicolumn{3}{c||}{~~Instantaneous~~} & 
      \multicolumn{3}{c||}{~~Staggered (RD Era)~~} & 
      \multicolumn{3}{c||}{~~Staggered (MD Era)~~ } \\
      \hline
      ~Brane Mass~ & ~SM Fields~ & 
      ~~~~~$\alpha$~~~~~ & ~~~~~$\beta$~~~~~ & ~~~~~$x$~~~~~ & 
      ~~~~~$\alpha$~~~~~ & ~~~~~$\beta$~~~~~ & ~~~~~$x$~~~~~ & 
      ~~~~~$\alpha$~~~~~ & ~~~~~$\beta$~~~~~ & ~~~~~$x$~~~~~ \\
      \hline\hline
      UV & UV         & $-1/4$ & $-3/4$ & $-1$
                      & $-5/8$ & $-3/4$ & $-11/8$ 
                      & $-3/4$ & $-3/4$ & $-3/2$ \\
      \hline
      UV & IR         & $-1/3$ & $-2/3$ & $-1$
                      & $-5/6$ & $-2/3$ & $-3/2$
                      & $-1$   & $-2/3$ & $-5/3$ \\
      \hline
      IR & UV         & $-1/2$ & $-3/4$ & $-5/4$
                      & $-7/8$ & $-3/4$ & $-13/8$ 
                      & $-1$   & $-3/4$ & $-7/4$ \\
      \hline
      IR & IR         & $-2/3$ & $-2/3$ & $-4/3$ 
                      & $-7/6$ & $-2/3$ & $-11/6$ 
                      & $-4/3$ & $-2/3$ & $-2$ \\
      \hline\hline
    \end{tabular}
  \end{minipage}
  \caption{The scaling exponents $\alpha$ and $\beta$ and the parameter $x = \alpha + \beta$
    obtained for the four different possible combinations of locations for the brane mass and 
    the SM fields in our 5D scenario within the regime in which $\pi k R \gg 1$ and 
    $m_\phi \ll \mKK$.  Within this regime, $x$ is well defined and approximately constant 
    across a large number of the lower-lying $\hat{\chi}_n$ with $n > 0$ within the ensemble.  
    Results are shown for three different possible scenarios
    depending on whether all of the ensemble constituents begin oscillating (and thus 
    behaving as matter rather than as vacuum energy) instantaneously at the time of
    the mass-generating phase transition, or whether different constituents begin 
    oscillating at different times in staggered fashion after the phase transition has 
    occurred, during either a RD or MD epoch.    
    \label{tab:scalingtable}}
\end{table*}

\begin{figure*}[t]
  \centering
  \includegraphics[clip, width=0.33\linewidth]{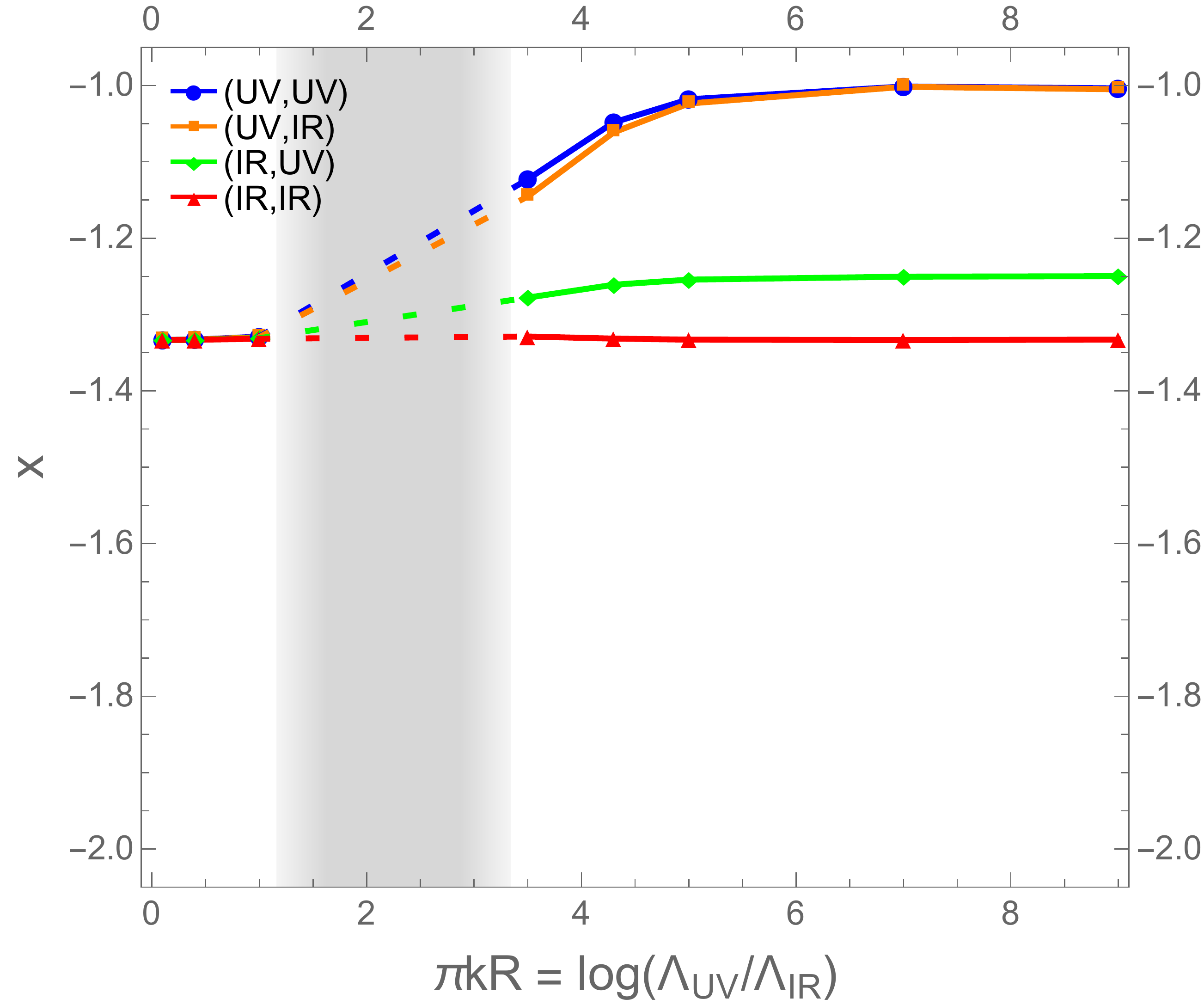}~
  \includegraphics[clip, width=0.33\linewidth]{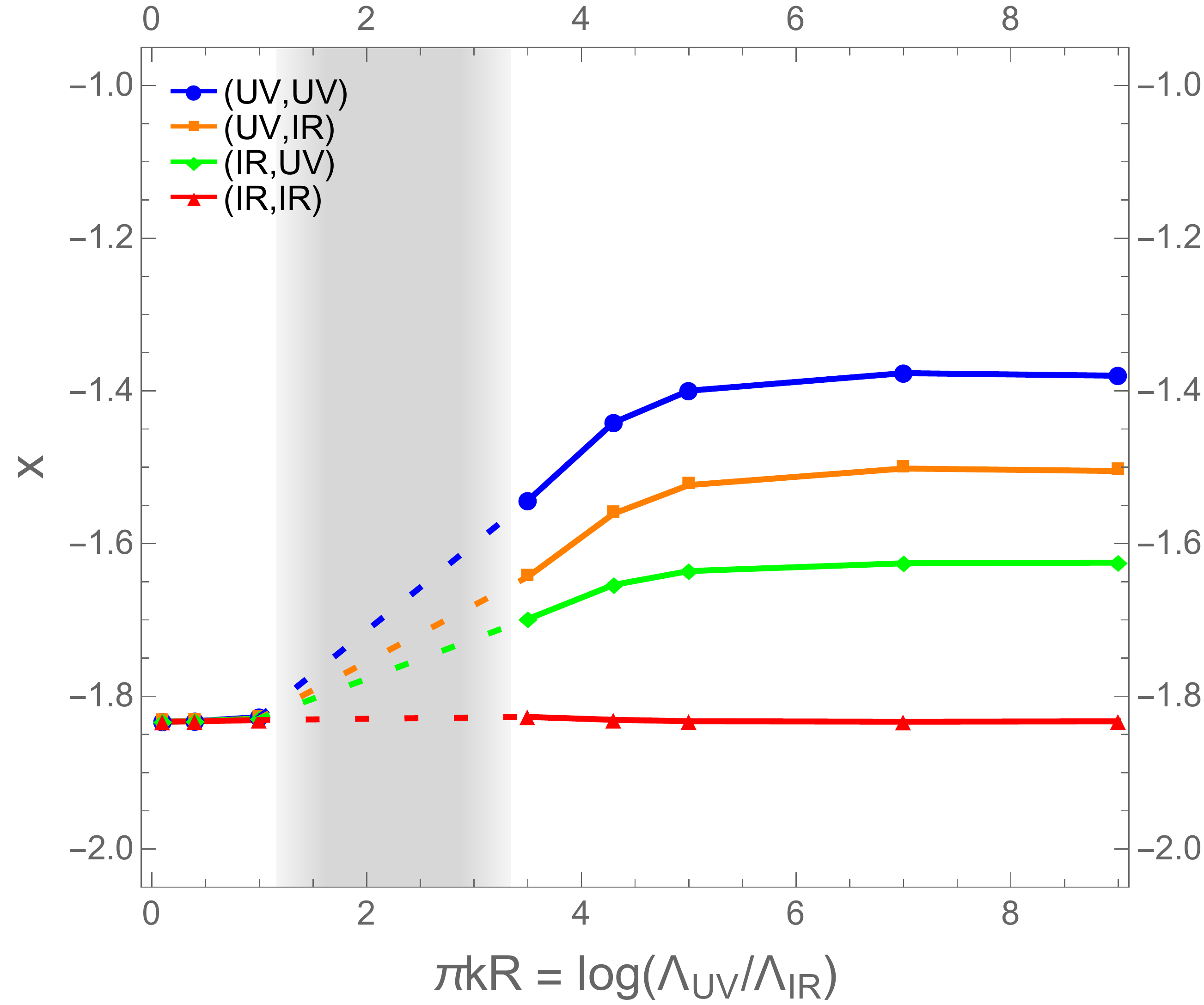}~
  \includegraphics[clip, width=0.33\linewidth]{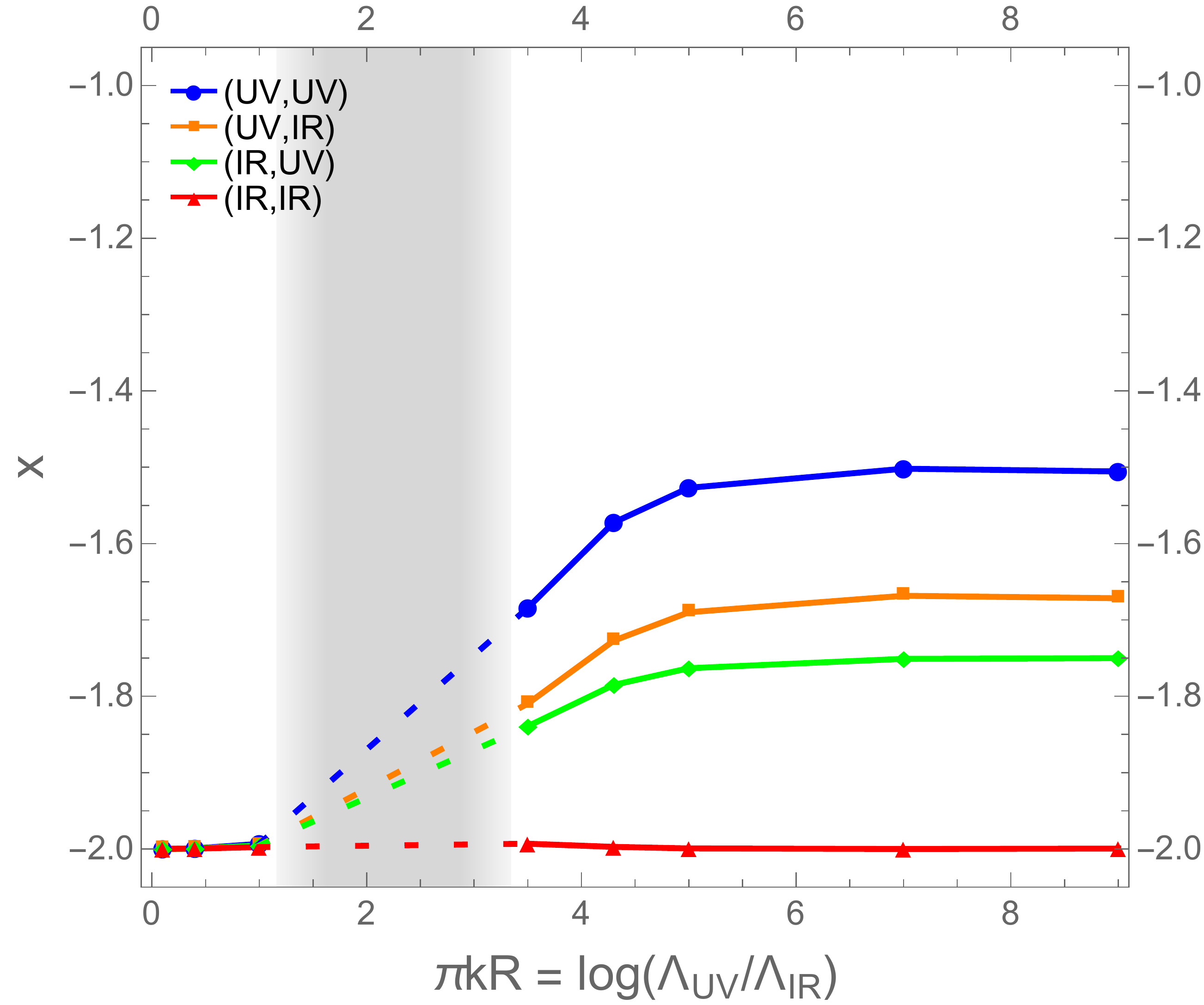}
  \caption{The scaling exponent $x = \alpha + \beta$, plotted as a function of the 
    ratio $\pi k R = \log(\LambdaUV/\LambdaIR)$, for the four different possible 
    combinations of locations for the brane mass and the SM fields.
    All curves shown in all panels of the figure correspond to the same value for the 
    dimensionless product $m_\phi R \approx 3.5 \times 10^{-4}$.  The left, middle, and 
    right panels of the figure correspond respectively to the case in which the 
    $\hat{\chi}_n$ all begin oscillating instantaneously at the time of
    the mass-generating phase transition, the case in which the $t_n$ are staggered in 
    time during a RD epoch, and the case in which the $t_n$ are staggered in time during 
    a MD epoch.  We observe that all of the curves shown in each panel approach 
    a common $x$ value in the flat-space limit, which corresponds to taking 
    $\pi k R \rightarrow 0$.
  \label{fig:xvsLambdaRatio}}
\end{figure*} 

Thus far, we have focused on the case in which the fields of the SM and the 
dynamics which generates $m_\phi$ are both localized on the UV brane.  However,
we are also free to consider alternative possibilities in which this dynamics,
the SM fields, or both are instead localized on the IR brane.  Such modifications
of our scenario can have a significant impact on $\Omega(\Gamma)$ and $n_\Gamma(\Gamma)$.  
For example, if the SM fields are localized on the IR brane, 
it is not the projection coefficients $A'_n$ which determine the decay widths of our 
ensemble constituents, but rather a different set of coefficients 
$A_n'' \equiv e^{-4\pi k R} \sqrt{1-e^{-2\pi k R}} /\sqrt{2k}
\langle \chi(x, \pi R) | \hat{\chi}_n \rangle$
which represent the projection of the $\hat{\chi}_n$ onto the IR brane at $y=\pi R$.
A detailed derivation of the values of $\alpha$ and $\beta$ for each of the 
four possible combinations of locations for the brane mass and the SM
fields is provided in Appendix~\ref{app:DifferentScenarios}.  Once again, in deriving
these scaling exponents, we focus on the regime in which $\pi k R \gg 1$ and 
$m_\phi \ll \mKK$.  The main results are summarized in Table~\ref{tab:scalingtable}.

It is also interesting to consider how the results in Table~\ref{tab:scalingtable}
are modified when we depart from the $\pi k R \gg 1$ regime.  However, while the parameter 
$x$ is always well defined within the region of parameter space wherein $m_\phi$ is much 
smaller than all other relevant scales in the problem, regardless of the value of $\pi k R$, it
is not always constant.  Thus, in assessing how our results for $x$ generalize for arbitrary 
values of $\pi k R$, we must first identify the regions of parameter space within which $x$ is 
constant across a large number of the lower-lying $\hat{\chi}_n$ with $n > 0$ within the ensemble,
since it is only within these regions where we can meaningfully associate a single value of 
$x$ with the ensemble.  We have already seen that this is the case within the regime wherein 
$\pi k R \gg 1$.  For $\pi k R$ outside this regime, however, the number of states 
with masses $k \gg \hat{m}_n \gtrsim \mKK$ is far smaller.  When this is the case, $x$ is not 
necessarily constant even across the lightest several $\hat{\chi}_n$ with $n > 0$ 
in the ensemble.  That said, we also note that for $\pi k R \lesssim 1$, all of the 
low-lying $\hat{\chi}_n$ with $n > 0$ have $\hat{m}_n \gtrsim k$.  As a result, $x$ is 
approximately constant across this portion of the ensemble within this regime.  Thus, 
it is once again sensible from a DDM perspective to identify this value of $x$ as the 
effective value of $x$ for the ensemble.

Given these considerations, we adopt the following procedure in analyzing how $x$
varies as a function of $\pi k R$.  We calculate a value of $x$ only for those 
ensembles for which the masses of the $\hat{\chi}_n$ with $1 \leq n \leq 10$ either
all satisfy the condition $\hat{m}_n \leq k$ or else all satisfy the condition 
$\hat{m}_n \geq k$.  We then calculate $x$ by performing linear fits of both 
$\log(A_n)$ and $\log(A_n')$ to $\log(\hat{m}_n)$ for the set of ensemble constituents 
$\hat{\chi}_n$ with $2 \leq n \leq 9$.  In this way, we may define an effective value of 
$x$ for all $\pi k R$ either above or below the rough range $1 \lesssim \pi k R \lesssim 3$.   

In Fig.~\ref{fig:xvsLambdaRatio}, we plot this effective value of $x$ as a function of 
$\pi k R = \log(\LambdaUV/\LambdaIR)$ for all four possible combinations of locations 
for the brane mass and the SM fields.  The results shown in
the left, middle, and right panels of the figure correspond respectively to the case in 
which the $\hat{\chi}_n$ all begin oscillating instantaneously at $t_n = t_G$,
the case in which the $t_n$ are staggered in time during a RD epoch, and the case in which 
the $t_n$ are staggered in time during a MD epoch.  
All points displayed in all panels of the figure correspond to the same value for the 
dimensionless product $m_\phi R \approx 3.5 \times 10^{-4}$ --- a value chosen such that 
$m_\phi = \mKK$ for the largest value of $\pi k R$ within the range $0 \leq \pi k R \leq 9$ 
included in each plot.  This parameter choice ensures that $m_\phi \ll \hat{m}_n$ for all $n > 1$ 
across this entire range of $\pi k R$.  While we have connected these points in order to
guide the eye, we emphasize that we have only included $x$ values for $\pi k R$ within the 
ranges $\pi k R \lesssim 1$ and $\pi k R \gtrsim 3$ wherein this quantity is sensibly defined. 

For all of the curves shown in Fig.~\ref{fig:xvsLambdaRatio}, we observe that  
the value of $x$ rapidly approaches the corresponding asymptotic value quoted in 
Table~\ref{tab:scalingtable} as $\pi k R \gtrsim 1$.  Moreover, we see that the values of 
$x$ obtained for $\pi k R = 9$ accord well with this asymptotic value in all cases.
By contrast, we see in each panel that as $\pi k R \rightarrow 0$, the values of $x$ 
obtained for all possible combinations of brane-mass and SM-field locations
asymptote to a single, common value.  This common value is precisely the value of $x$
obtained in Ref.~\cite{DDM1} for the corresponding oscillation-onset behavior in
the flat-space limit: $x = -4/3$ for an instantaneous turn-on, $x = -11/10$ and 
for a staggered turn-on during a RD epoch, and $x = -2$ for a staggered turn-on during
a MD epoch.

\subsection{Surveying the Parameter Space}

We now turn to examine how the bounds in Eqs.~(\ref{eq:Omegatotbound}) 
and~(\ref{eq:weffbound}) constrain the full parameter space of our ensemble.
We shall assume that the lightest ensemble constituent begins oscillating well before 
the beginning of the BBN epoch --- \ie, that $t_0 \ll t_{\rm BBN}$.  When evaluating 
$\Omegatot$ and $\weff$ during the RD era prior to $\tMRE$, we take our fiducial time 
$t_\ast$ in Eqs.~(\ref{eq:Omegatot}) and~(\ref{eq:weffOmegaArbt}) to be some early time 
$t_0 \leq t_\ast \ll t_{\rm BBN}$.  Thus, for all $t_{\rm BBN} \leq t \leq \tMRE$, we
may approximate $t - t_\ast \approx t$.  For simplicity, at all times $t > \tMRE$, 
we ignore the effect of dark energy on $H$ at late times $t\sim \tnow$ and approximate 
the universe as strictly MD.  We also ignore any back-reaction on $H$ which results from the 
decay of the ensemble itself during this MD era, even though $\rhotot$ dominates the energy 
density of the universe at this time, given that we shall be imposing the bound in
Eq.~(\ref{eq:Omegatotbound}) and thereby mandating that $\rhotot$ does not differ 
significantly from the prediction of the $\Lambda$CDM cosmology.  With these approximations,
$\Omegatot$ and $\weff$ are given by Eqs.~(\ref{eq:Omegatot}) and~(\ref{eq:weffOmegaArbt})
at times $t < \tMRE$, but with $\kappa = 2$ rather than $\kappa = 3/2$.  When evaluating
$\Omegatot$ and $\weff$ during this MD era, we take $t_\ast = \tMRE$.  However, since
$\Omega_n(\tMRE) \propto \Omega_n^0 e^{-\Gamma_n (t - t_0)}$, where the constant of 
proportionality is the same for all $\hat{\chi}_n$ and is independent of the background
cosmology, we find that the ratio $\Omegatot/\widetilde{\Omega}_{\rm tot}$ at any time 
$t_{\rm BBN} < t \leq \tnow$, regardless of the relationship between $t$ and $\tMRE$, is 
given by
\begin{equation}
  \frac{\Omegatot}{\widetilde{\Omega}_{\rm tot}} ~\approx~ 
    \frac{\sum_{n=0}^\infty\Omega_n^0 e^{-\Gamma_n t}}
    {\sum_{n=0}^\infty\Omega_n^0}~.
  \label{eq:OmegatotoftRDMD}
\end{equation}
By contrast, the effective equation-of-state parameter for the ensemble is given by  
\begin{equation}
  \weff ~\approx~ \frac{\sum_{n=0}^\infty\Omega_n^0 \Gamma_n t 
    e^{-\Gamma_n t}}{\sum_{n=0}^\infty\Omega_n^0 e^{-\Gamma_n t}} 
    \times \begin{cases}
    2/3 & t < \tMRE \\
    1/2 & t > \tMRE~.  
    \end{cases}
  \label{eq:weffoftRDMD}
\end{equation} 
We note that while our expressions for $\weff$ before and after matter-radiation equality 
are not equal at $t=\tMRE$, this apparent discontinuity in $\weff$ is simply a reflection of 
the fact that we are approximating the transition from the RD era to the MD era as an 
instantaneous event occurring at time $t = \tMRE$, at which point the Hubble parameter leaps
discontinuously from $H=1/(2t)$ to $H=2/(3t)$.  In reality, of course, $H$ transitions continuously 
between these asymptotic values at $t \sim \tMRE$.  In order to describe the evolution of 
$\weff$ during this transition, one would need to treat the parameter $\kappa$ as a function 
of $t$.  However, since we are only interested in bounding $\weff$ and not its time derivatives,
approximating this transition as instantaneous is sufficient for our purposes.

In assessing how the constraints in Eqs.~(\ref{eq:Omegatotbound}) and~(\ref{eq:weffbound}) 
impact the parameter space of our scenario, we begin by noting that the expression for 
$\Omegatot/\widetilde{\Omega}_{\rm tot}$ in Eq.~(\ref{eq:OmegatotoftRDMD}) depends on the 
physical scales $\Gamma_0$ and $t$ only through the dimensionless quantity 
$\sigma \equiv \Gamma_0 t$.  Indeed, we observe that 
\begin{equation}
  \frac{\Omegatot}{\widetilde{\Omega}_{\rm tot}} ~\approx~ 
    \frac{\sum_{n=0}^\infty\Omega_n^0 e^{-\frac{\Gamma_n}{\Gamma_0} \sigma}}
    {\sum_{n=0}^\infty\Omega_n^0}~,
  \label{eq:OmegatotofsigmaRDMD}
\end{equation} 
which depends on $\sigma$ and on the ratios 
$\Gamma_n/\Gamma_0 = \hat{m}_n^3 A_n^{\prime 2}/(\hat{m}_0^3 A_0^{\prime 2})$ of the 
decay widths of the ensemble constituents, but not on the value of $\Gamma_0$ itself.
Likewise, our expression for $\weff$ in Eq.~(\ref{eq:weffoftRDMD}) can be written as
\begin{equation}
  \weff ~\approx~ \frac{\sum_{n=0}^\infty\Omega_n^0 \frac{\Gamma_n}{\Gamma_0} \sigma 
    e^{-\frac{\Gamma_n}{\Gamma_0} \sigma}}
    {\sum_{n=0}^\infty\Omega_n^0 e^{-\frac{\Gamma_n}{\Gamma_0} \sigma}} 
    \times \begin{cases}
    2/3 & \sigma < \Gamma_0\tMRE \\
    1/2 & \sigma > \Gamma_0\tMRE~,  
    \end{cases}
  \label{eq:weffofsigmaRDMD}
\end{equation} 
which depends on the value of $\Gamma_0$ only in that this parameter determines the 
value of $\sigma$ at the time of matter-radiation equality.  Moreover, we note that 
the expression for $\weff$ at times $t < \tMRE$ is always larger than the corresponding
expression at times $t > \tMRE$ by an overall multiplicative factor of precisely $4/3$.  
Given this, we shall hereafter adopt a conservative approach in establishing bounds on 
$\weff$ in which we always treat $\weff$ as being given by the expression valid during 
the RD era prior to matter-radiation equality, regardless of the actual relationship 
between $t$ and $\tMRE$.  With this modification, our expressions for both $\Omegatot$ 
and $\weff$ depend only on $\sigma$, and not on $\Gamma_0$ and $t$ independently.  In 
other words, these expressions are invariant under any simultaneous rescaling of 
$\Gamma_0$ and $t$ which leaves their product invariant.  
 
\begin{figure*}[t]
\centering
  \includegraphics[width=0.37\linewidth]{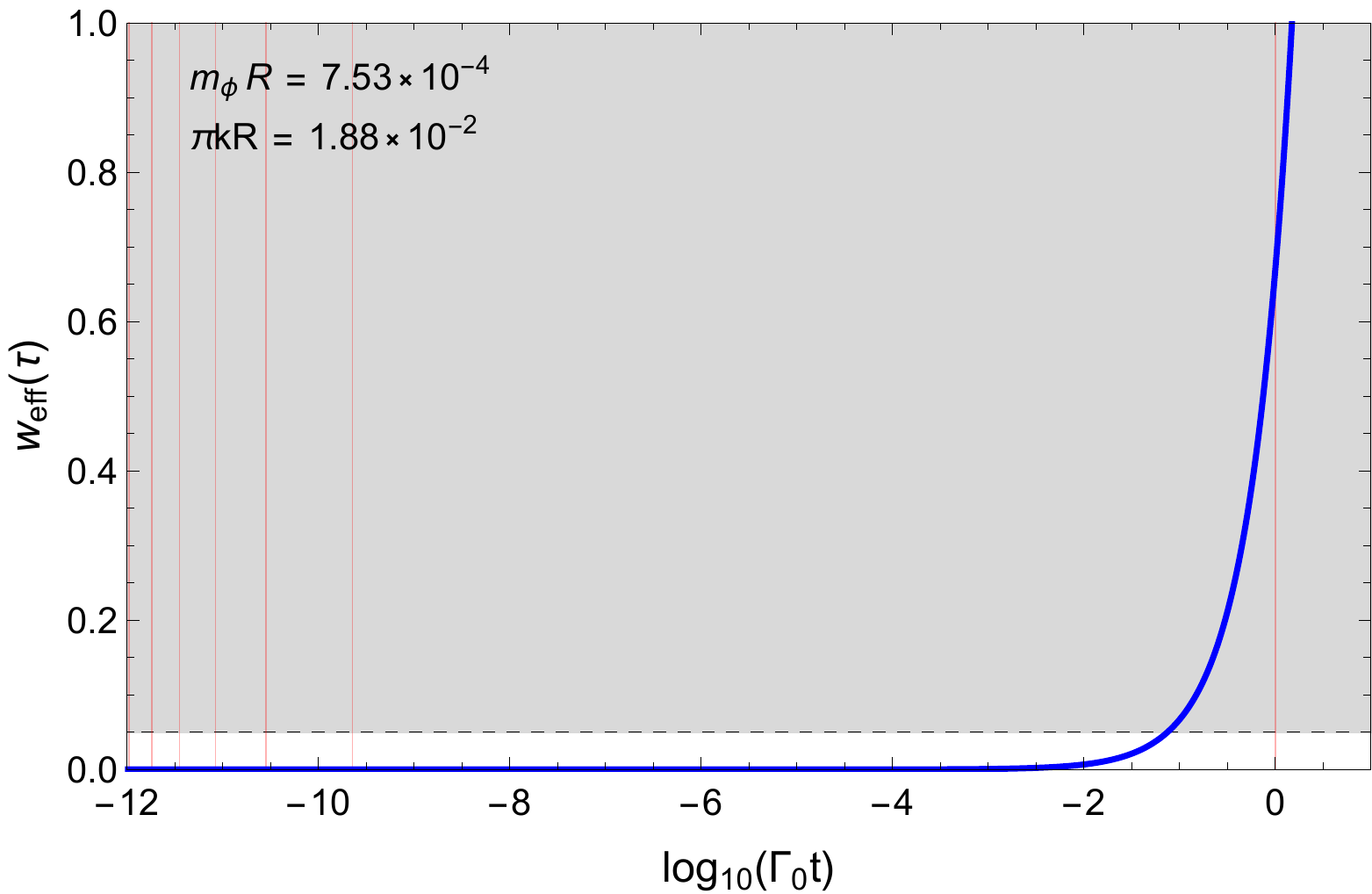}
    ~~~~~~~~~~~~~
  \includegraphics[width=0.37\linewidth]{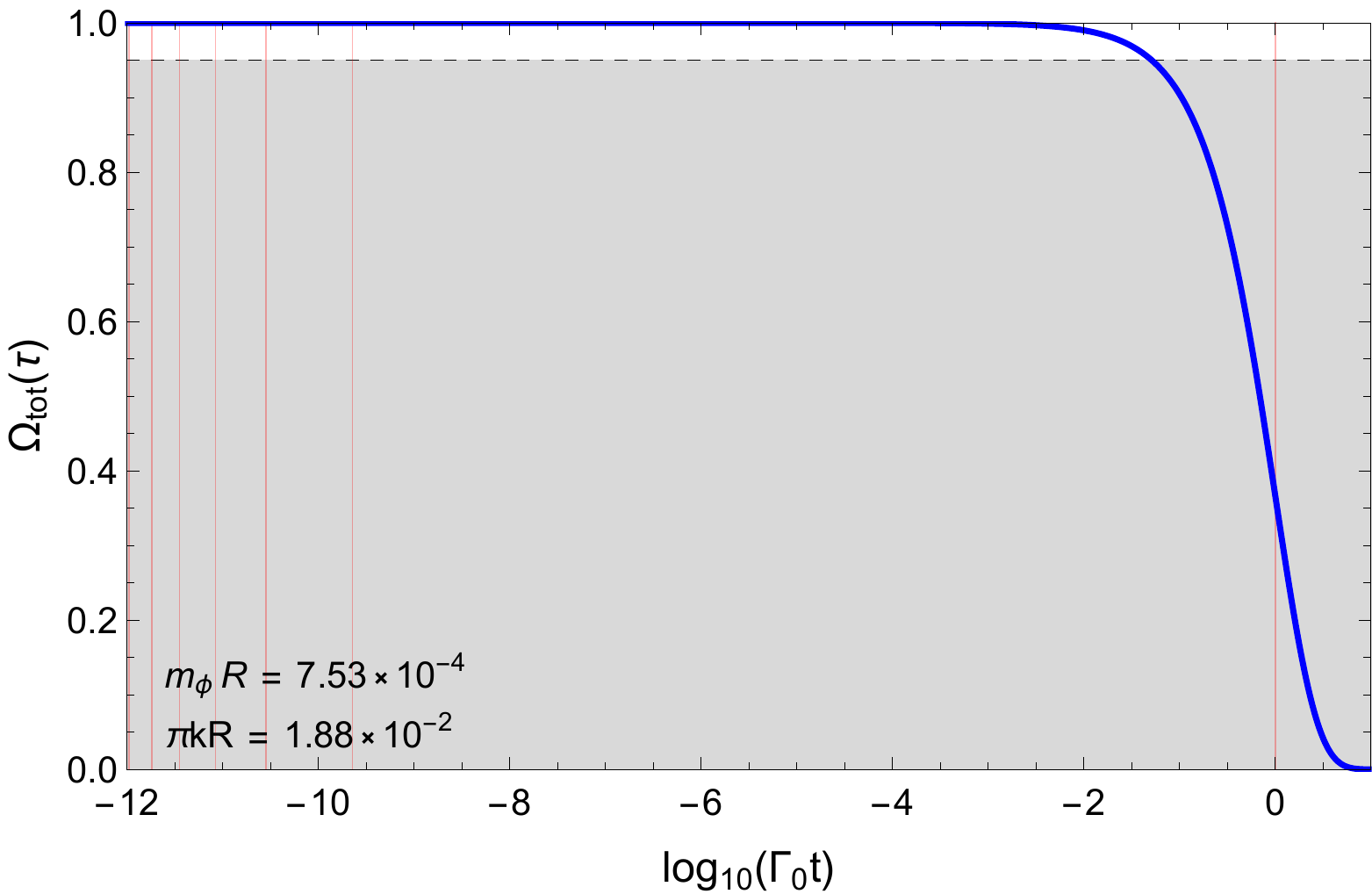}\\
  \includegraphics[width=0.37\linewidth]{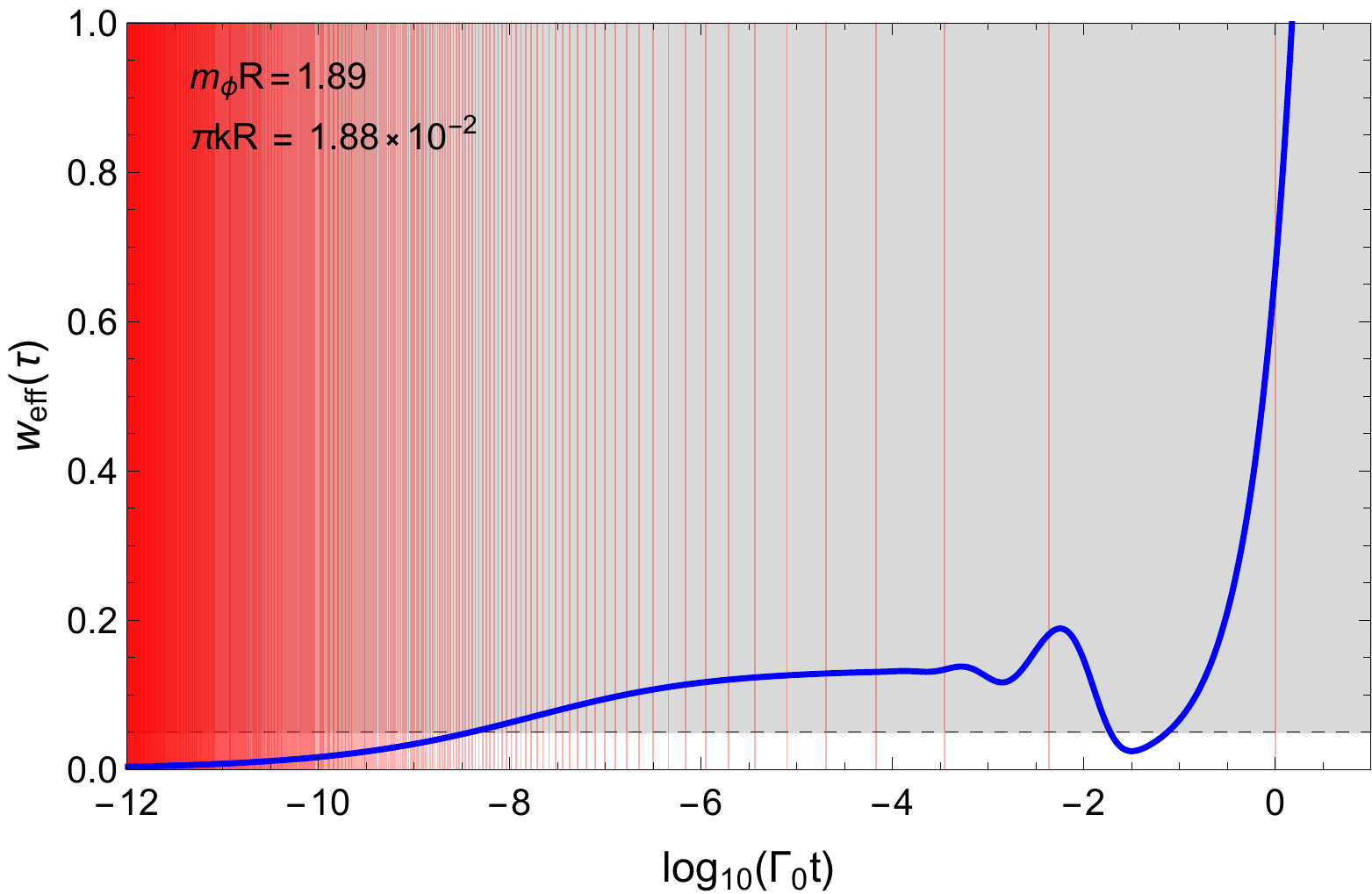}
    ~~~~~~~~~~~~~
  \includegraphics[width=0.37\linewidth]{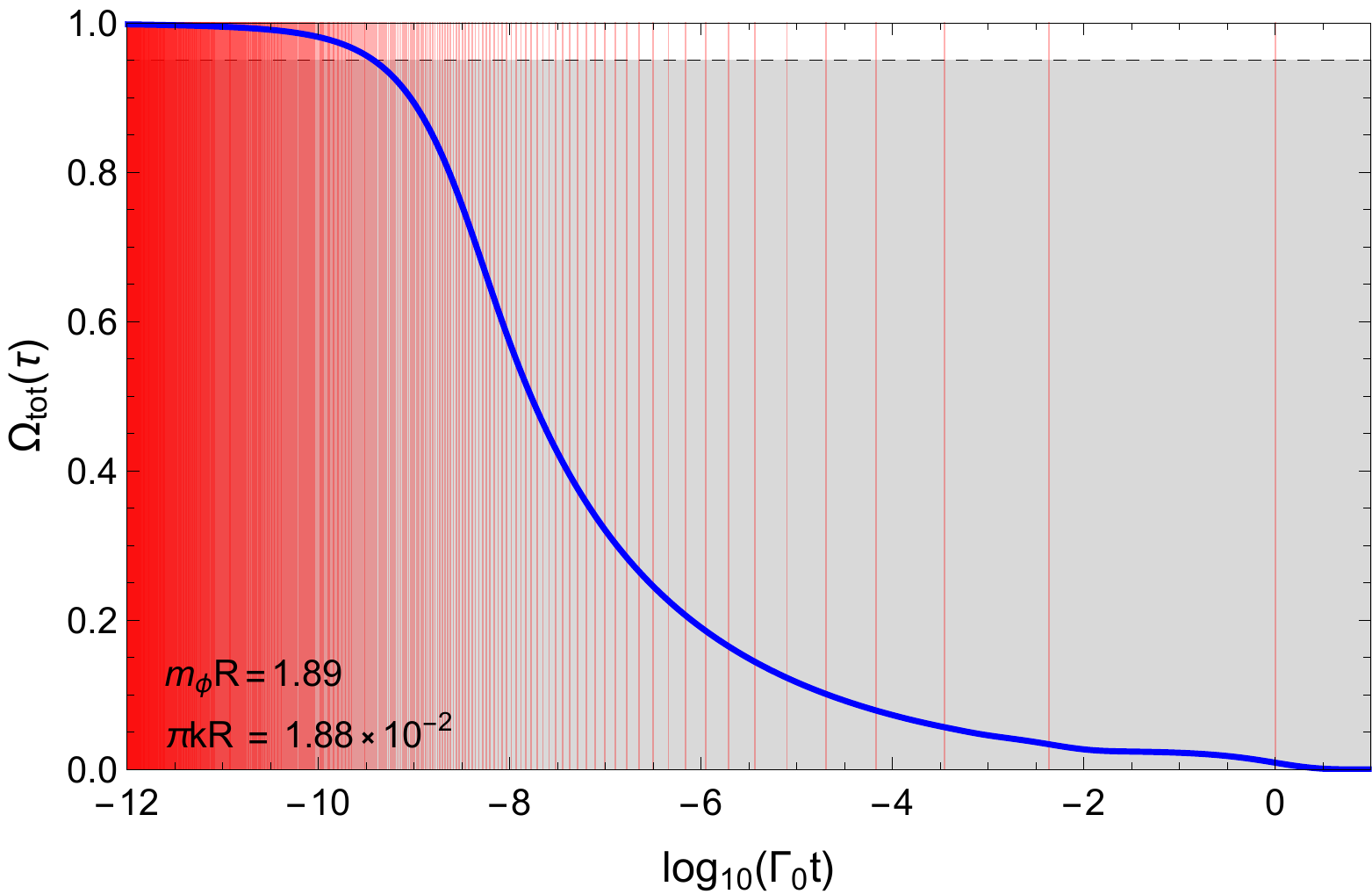}\\
  \includegraphics[width=0.37\linewidth]{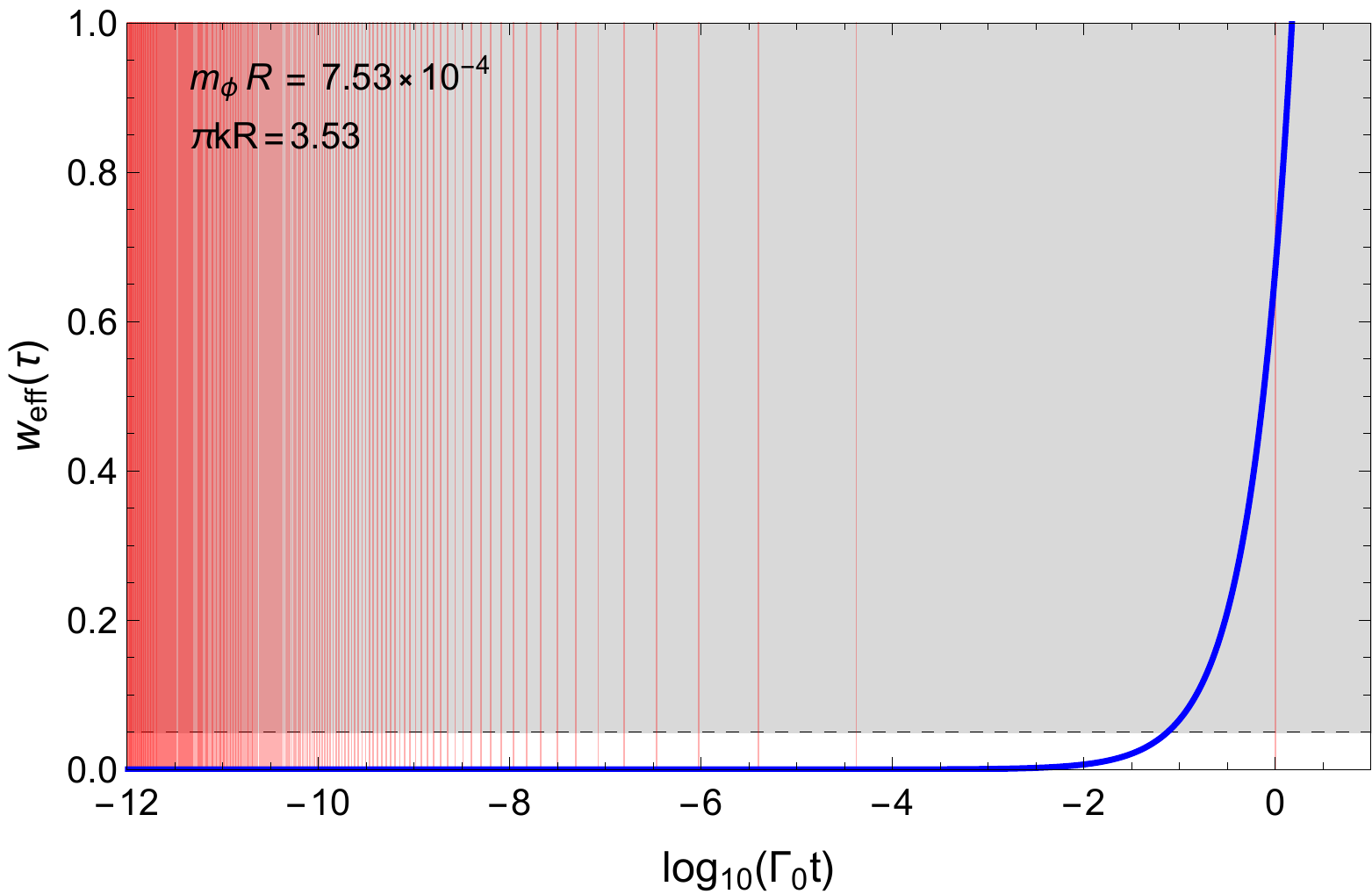}
    ~~~~~~~~~~~~~
  \includegraphics[width=0.37\linewidth]{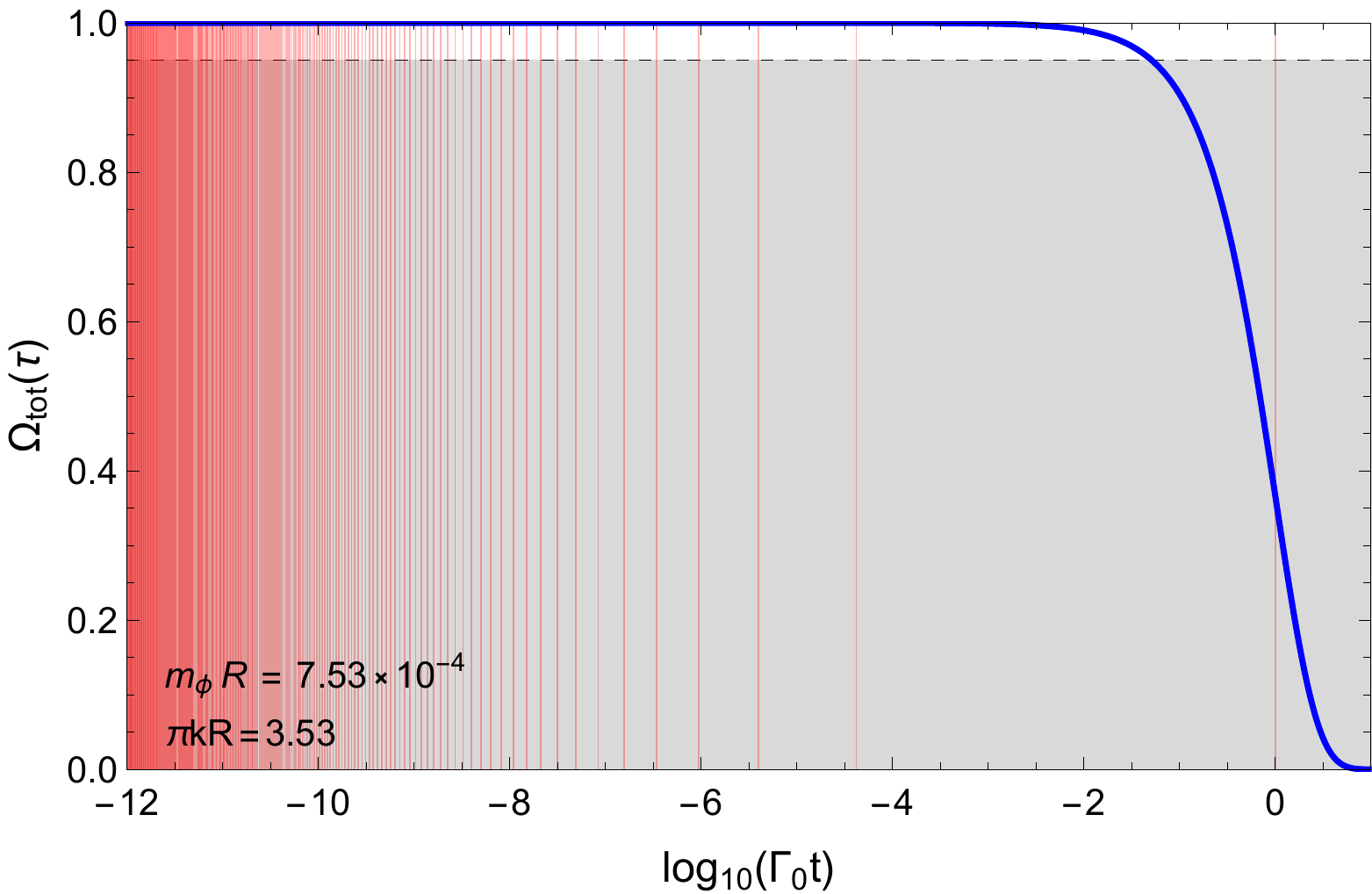}\\
  \includegraphics[width=0.37\linewidth]{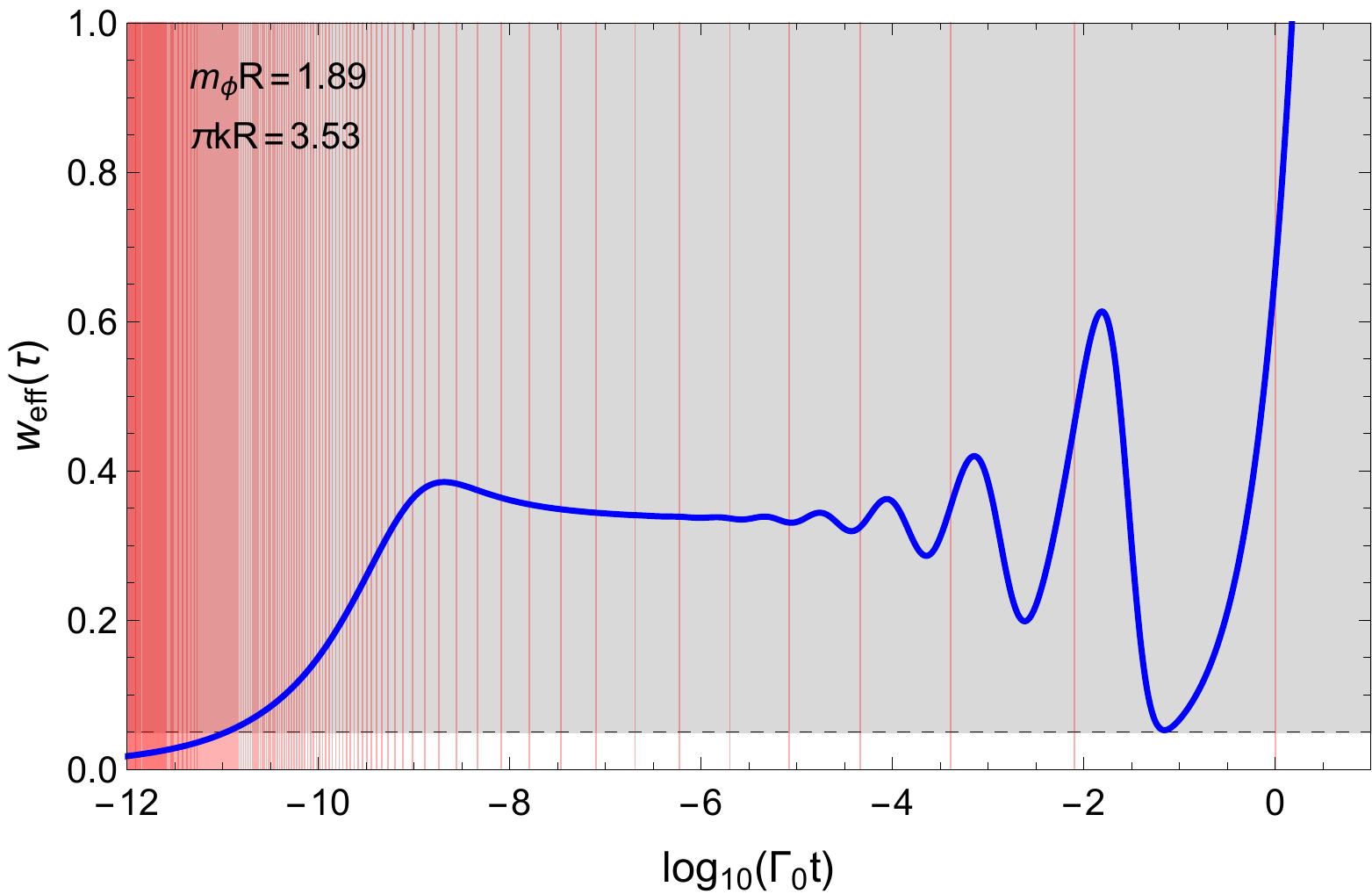}
    ~~~~~~~~~~~~~
  \includegraphics[width=0.37\linewidth]{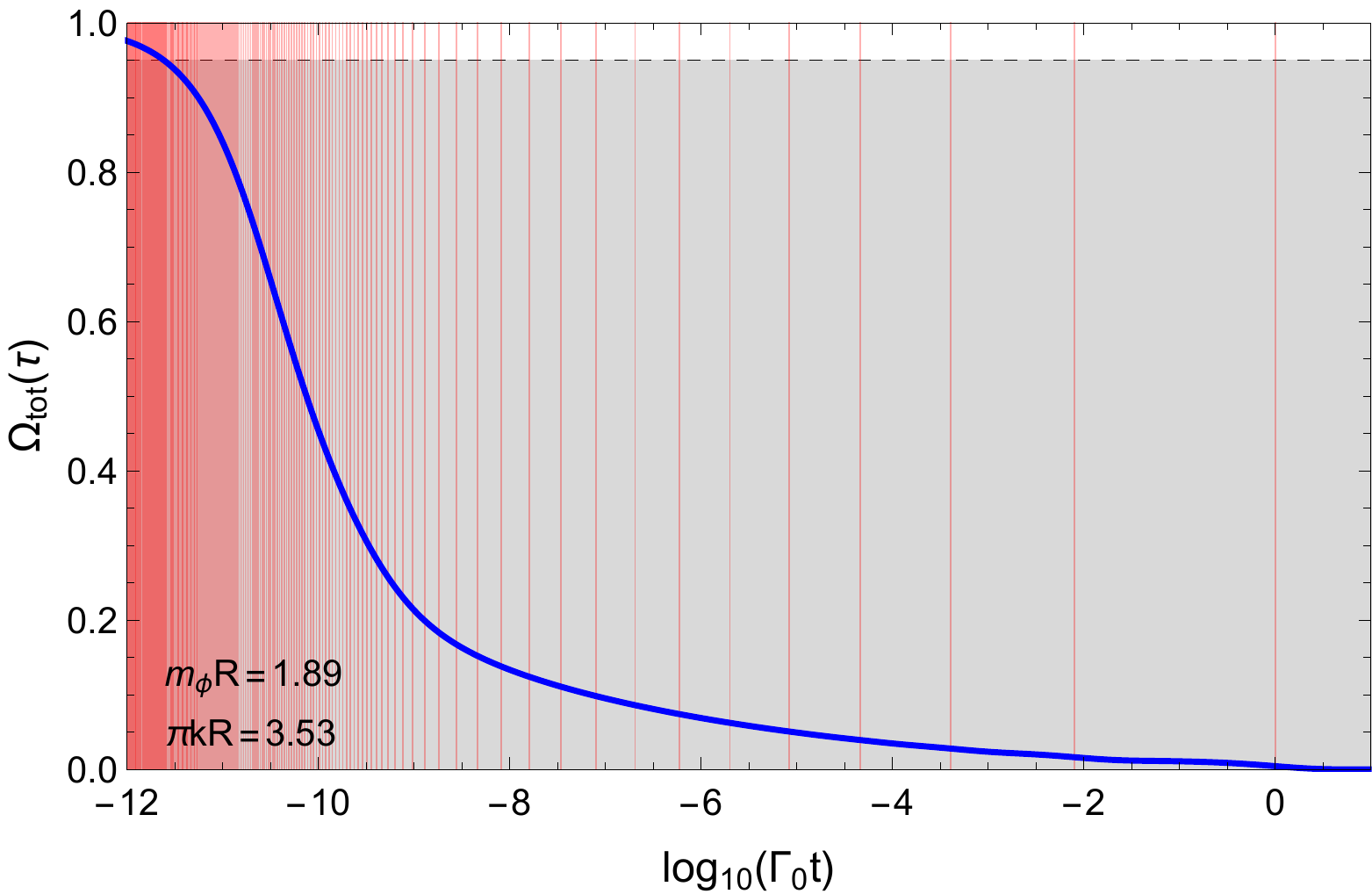}
\caption{The effective ensemble equation-of-state parameter $\weff(\sigma)$ (left panel 
  in each row) and total ensemble abundance $\Omegatot(\sigma)$ (right panel 
  in each row), plotted as functions of $\sigma \equiv \Gamma_0 t$.  Each row of the 
  figure corresponds to a 
  particular choice of $\pi k R$ and $m_\phi R$.  These choices are representative of the 
  regimes in which $\pi k R$ and $m_\phi R$ are both small (first row), in which $\pi k R$ is
  small but $\pi k R$ is large (second row), in which $\pi k R$ is large but $m_\phi R$ 
  is small (third row), and in which $\pi k R$ and $m_\phi R$ are both large (fourth row).    
  In all cases, we have taken $\LambdaUV R = 3$ and assumed that all of the $\hat{\chi}_n$ 
  begin oscillating instantaneously at $t = t_0$.  In each panel, the blue line is the 
  value of the quantity $\weff$ or $\Omegatot$ itself, while the black dashed line 
  indicates the corresponding constraint from either Eq.~(\protect\ref{eq:weffbound}) 
  or~(\protect\ref{eq:Omegatotbound}).  The red vertical lines indicate the values 
  $\sigma_n = \Gamma_0 \tau_n$ of $\sigma$ at which the various $\hat{\chi}_n$ decay.  
  The gray regions are excluded by the constraints.  In particular, for any given 
  ensemble, consistency with these constraints requires that $\Gamma_0$ be taken 
  sufficiently small that for all $\sigma$ within the range 
  $\sigma < \sigma_{\rm now} \equiv \Gamma_0 \tnow$, the blue curves
  for both $\weff$ and $\Omegatot$ do not enter the gray region. 
  \label{fig:TimeEvolweffandOmega}}
\end{figure*}

The utility of this invariance is perhaps best conveyed in the context of a graphical 
example.  In Fig.~\ref{fig:TimeEvolweffandOmega}, we show how $\weff(\sigma)$ and $\Omegatot(\sigma)$
actually evolve as functions of $\sigma$ for four different choices of $m_\phi R$ and $\pi k R$.
These four choices are intended to exemplify different possible regimes for these two 
parameters.  In particular, these choices are representative of the regimes in which $\pi k R$ 
and $m_\phi R$ are both small (first row), in which $\pi k R$ is small but $\pi k R$ is large 
(second row), in which $\pi k R$ is large but $m_\phi R$ is small (third row), and in which 
$\pi k R$ and $m_\phi R$ are both large (fourth row).  In all cases, we have taken 
$\LambdaUV R = 3$ and assumed that all of the $\hat{\chi}_n$ begin oscillating instantaneously
at $t = t_0$.  In each panel, the blue line indicates the value of the quantity 
$\weff(\sigma)$ or $\Omegatot(\sigma)$ itself, while the black dashed line indicates 
the corresponding constraint from either Eq.~(\protect\ref{eq:weffbound}) 
or Eq.~(\protect\ref{eq:Omegatotbound}).  The vertical red lines indicate the values 
$\sigma_n \equiv \Gamma_0 \tau_n$ of the dimensionless time variable $\sigma$ which correspond 
to the lifetimes of the $\hat{\chi}_n$ with $\hat{m}_n \leq \LambdaUV$.

In interpreting the results shown in Fig.~\ref{fig:TimeEvolweffandOmega}, we begin by
observing that while reciprocal rescalings of $\Gamma_0$ and $t$ do not affect
the overall {\it shapes}\/ of the curves representing $\weff$ and 
$\Omegatot/\widetilde{\Omega}_{\rm tot}$ as functions of $\sigma$, such rescalings do
change the value $\sigma_{\rm now} \equiv \Gamma_0 \tnow$ of $\sigma$ which corresponds to 
present time.  In particular, the smaller $\Gamma_0$ is, the smaller the corresponding
value of $\sigma_{\rm now}$.  Consistency with the constraints in 
Eqs.~(\ref{eq:Omegatotbound}) and~(\ref{eq:weffbound}) requires only that these 
constraints be satisfied for $\sigma < \sigma_{\rm now}$.  The results shown in each
row of Fig.~\ref{fig:TimeEvolweffandOmega} therefore suggest that these constraints
can generally be satisfied by choosing a sufficiently small value for $\Gamma_0$ that  
the blue curves for both $\weff$ and $\Omegatot$ never enter the respective gray regions
for all $\sigma$ within the range $\sigma < \sigma_{\rm now}$.  Indeed, we observe that 
consistency with these constraints can always be achieved by taking $\Gamma_0$ to 
be sufficiently small, provided either that the number of $\hat{\chi}_n$ in the ensemble is 
finite and that their lifetimes satisfy $t_0 \ll \tau_0$, or else that the $\hat{\chi}_n$
with lifetimes $\tau_n \lesssim t_0$ collectively contribute only a negligible fraction of
the total abundance of the ensemble at $t = t_0$.  

Thus, when this is the case, we see that the constraints in Eqs.~(\ref{eq:weffbound}) 
and~(\ref{eq:Omegatotbound}) do not simply serve to exclude particular combinations of 
the model parameters $\pi k R$, $m_\phi$, and $\LambdaUV R$ outright, but rather to establish 
an upper bound on $\Gamma_0$ --- or, equivalently, a lower bound on $\tau_0$ --- for any such 
combination of these parameters.  More explicitly, the maximum value $\sigma_{\rm now}^{\rm max}$ 
of $\sigma_{\rm now}$ for which these constraints are simultaneously satisfied determines the 
minimum possible lifetime $\tau_0^{\rm min}$ for the lightest ensemble constituent through 
the relation    
\begin{equation}
  \tau_0^{\rm min} ~=~ \frac{\tnow}{\sigma_{\rm now}^{\rm max}}~.
  \label{eq:tauminbound} 
\end{equation}      
We stress that $\tau_0$ is indeed an independent degree of freedom in this scenario.
Although the overall normalization factors for both the abundances and lifetimes of the 
$\hat{\chi}_n$ both depend on $\hat{f}_X$, the normalization factor for the $\Omega_n$ 
depends not only on additional model parameters, such as the misalignment angle $\theta$, 
but also on the details of the cosmological history at times $t > t_0$.

It is also worth remarking that the results shown in the top two rows of 
Fig.~\ref{fig:TimeEvolweffandOmega} are qualitatively similar to those obtained in the 
$k\rightarrow 0$ limit studied in Ref.~\protect\cite{DDM1}.  The last two rows of the figure 
correspond to cases in which $\pi k R$ is large and therefore represent departures from the 
flat-space case.  We observe that it is when $\pi k R$ and $m_\phi R$ are both large that the 
deviations from the CDM limit are the most dramatic.

\begin{figure*}[t]
\centering
  \includegraphics[width=0.31\linewidth]{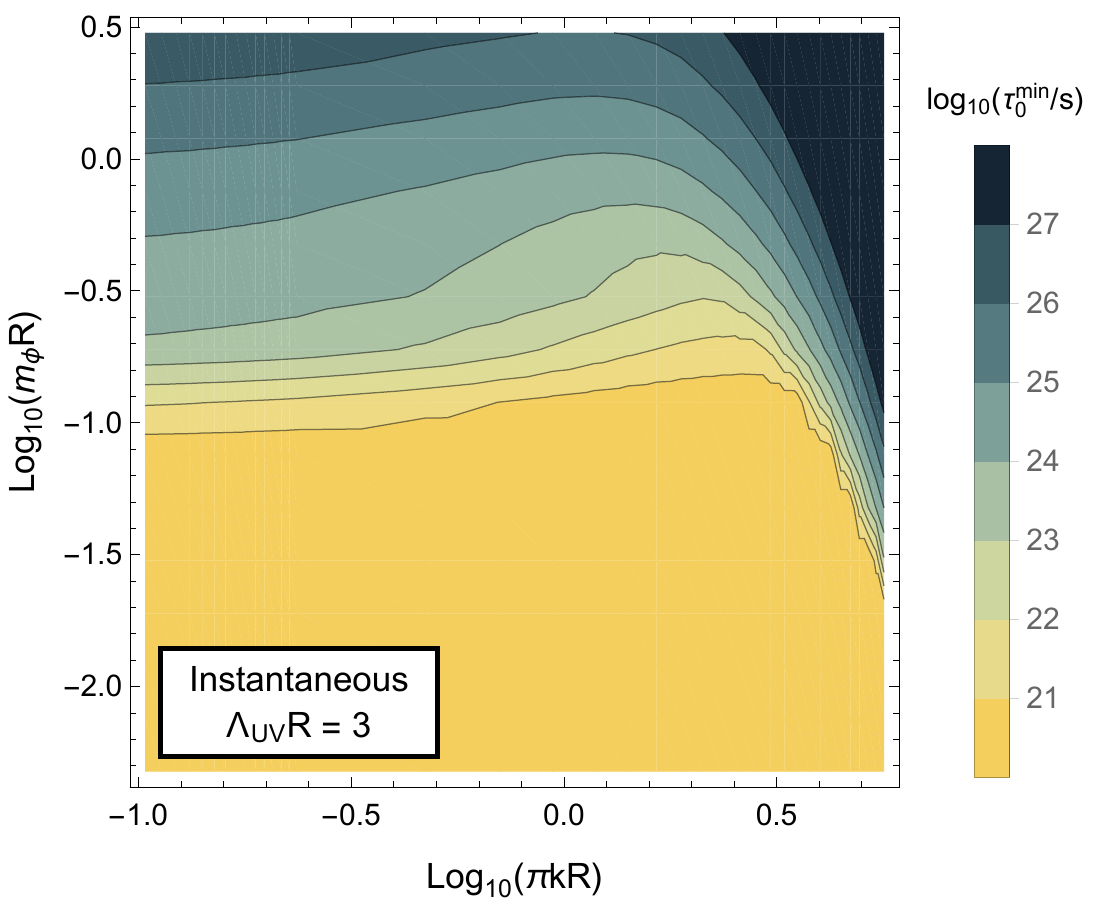}~
  \includegraphics[width=0.31\linewidth]{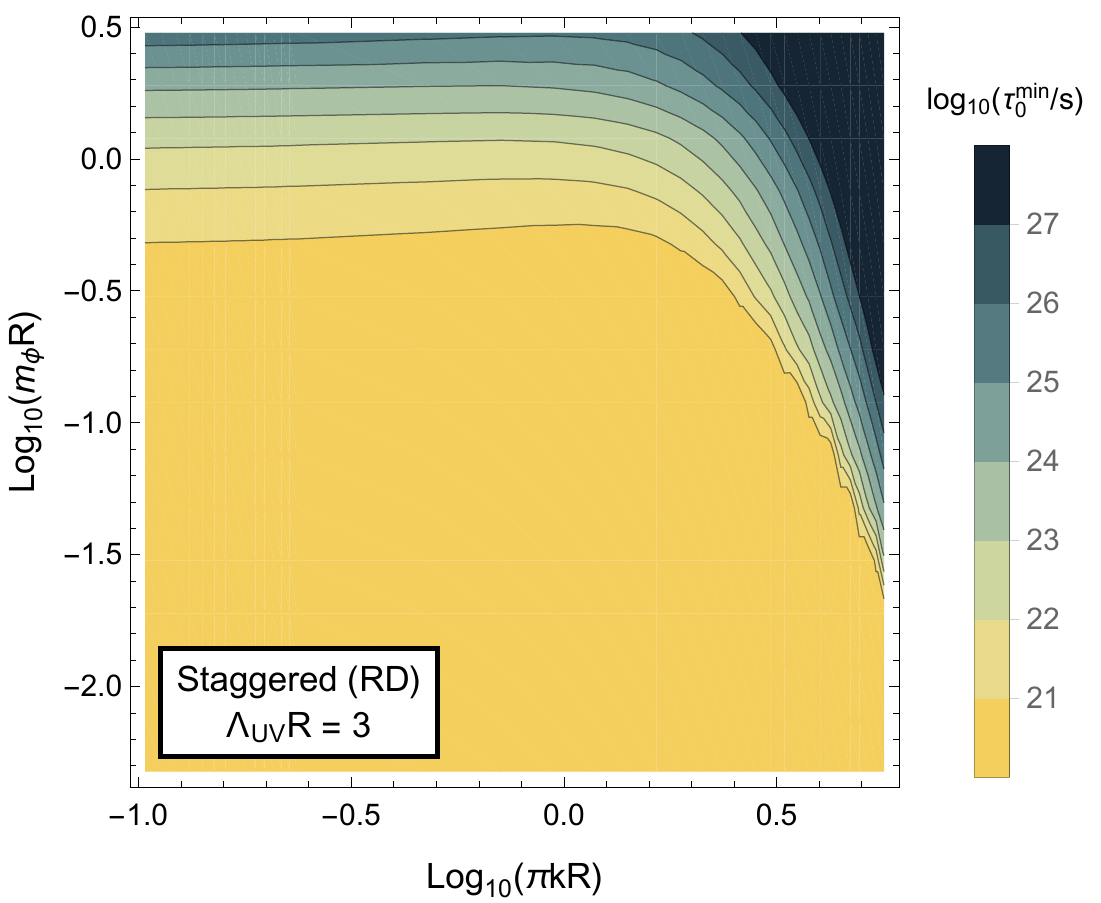}~
  \includegraphics[width=0.31\linewidth]{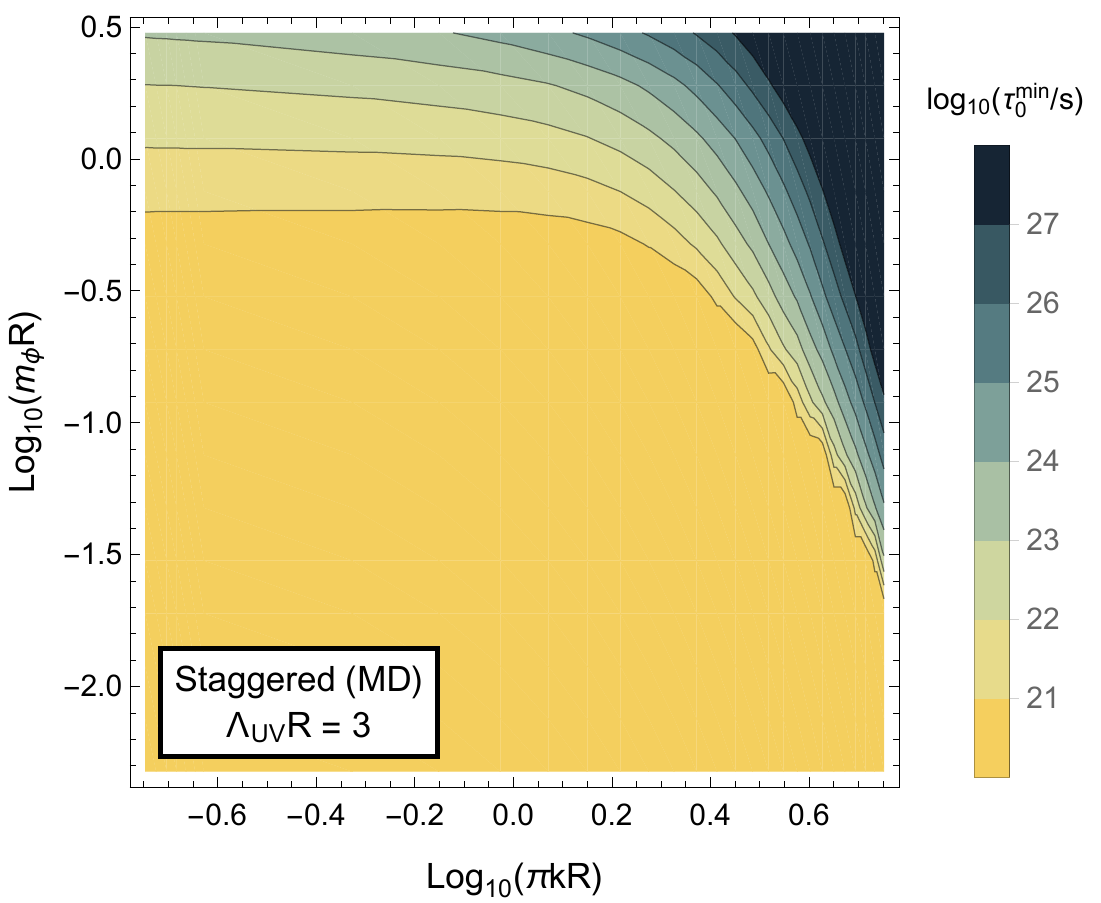}
\caption{Contours of the minimum lifetime $t_0^{\rm min}$ consistent with the 
  constraints in Eqs.~(\protect\ref{eq:weffbound}) 
  and~(\protect\ref{eq:Omegatotbound}), plotted within the $(\pi k R, m_\phi R)$-plane.
  For this plot, we take $\LambdaUV R = 3$.  The left, middle, and right panels 
  respectively correspond to the case of an instantaneous turn-on, a staggered turn-on 
  during a radiation-dominated era, and a staggered turn-on during a matter-dominated era.  
  We see that in general, the bound on $\tau_0$ becomes increasingly stringent as the 
  degree of warping is increased for fixed $m_\phi R$.
  \label{fig:tau0bounds}}
\vskip 0.35cm  
\centering 
  \includegraphics[width=0.31\linewidth]{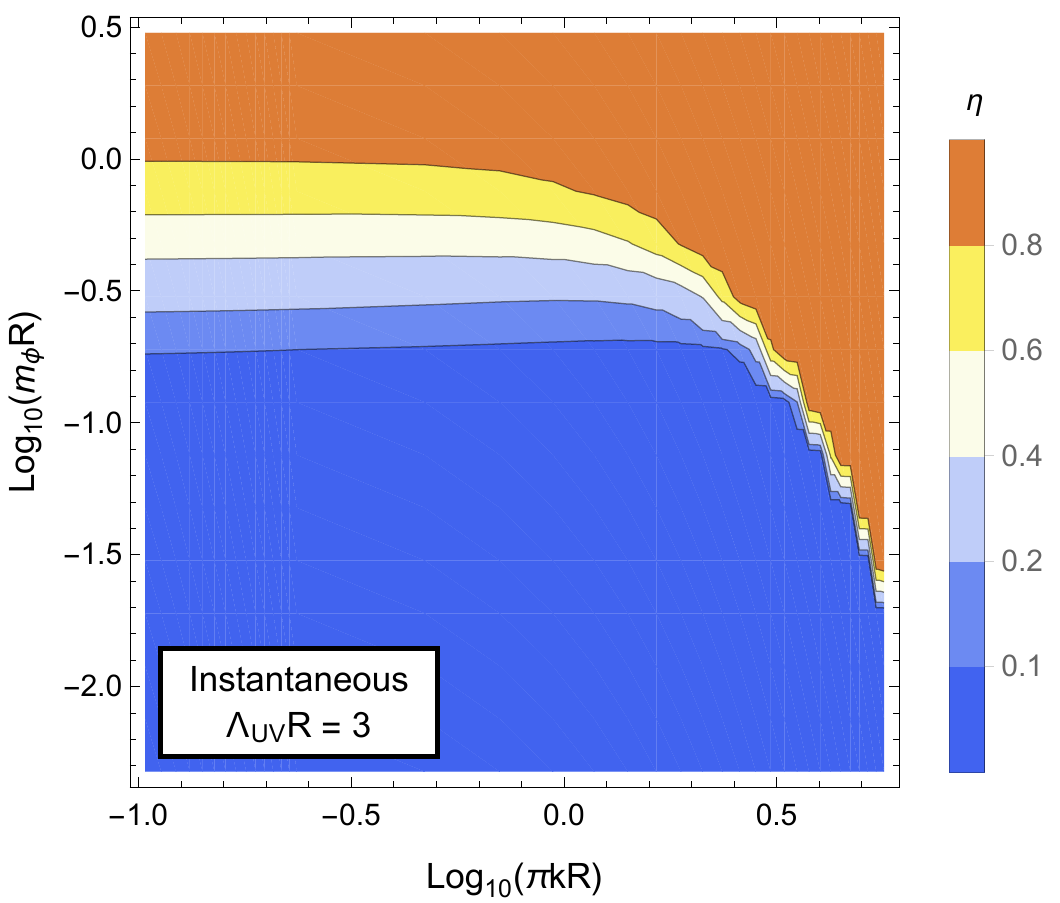}~
  \includegraphics[width=0.31\linewidth]{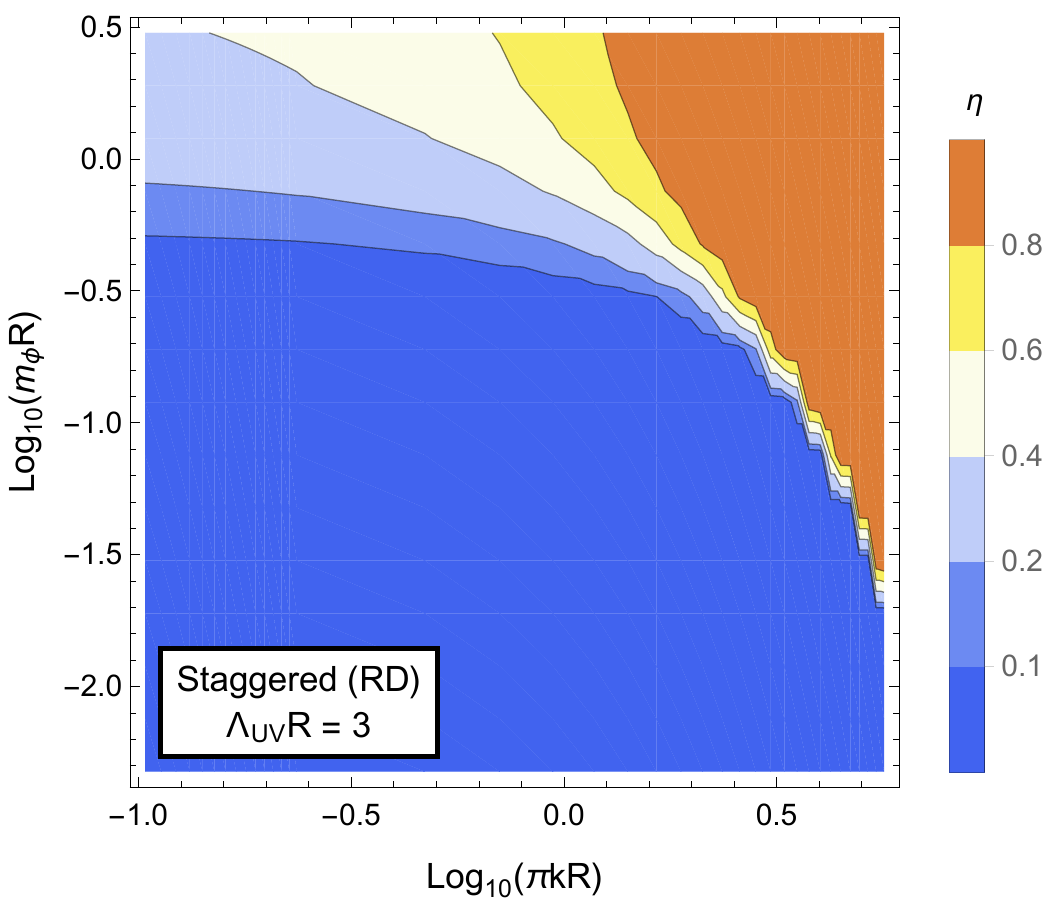}~
  \includegraphics[width=0.31\linewidth]{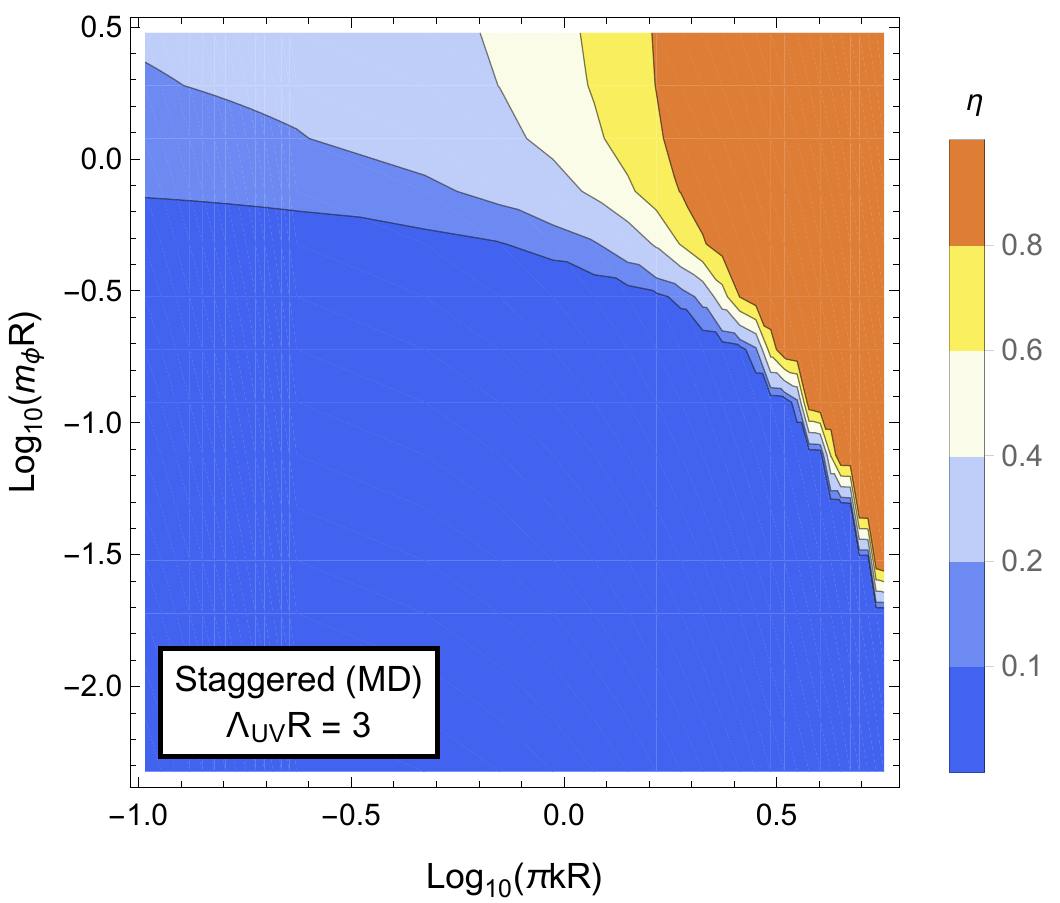}
\caption{Contours of the initial value $\eta(t_0)$ of the DDM tower fraction, plotted 
  within the same $(\pi k R,m_\phi R)$-plane shown in Fig.~\protect\ref{fig:tau0bounds}.  
  Once again, we take $\LambdaUV R = 3$. 
  \label{fig:etas}}
\end{figure*}

We now survey the parameter space of our model, using the criterion in 
Eq.~(\ref{eq:tauminbound}) in order to establish a bound on $\tau_0$ at each point within
that parameter space.  In particular, we hold $\LambdaUV R$ fixed and vary both
$\pi k R$ and $m_\phi R$.  In Fig.~\ref{fig:tau0bounds}, we show contours in 
$(\pi k R,m_\phi R)$-space of $\tau_0^{\rm min}$ for the parameter choice 
$\LambdaUV R = 3$.  The different panels of the figure correspond to the three 
different behaviors for the oscillation-onset times delineated in 
Eq.~(\ref{eq:abundances}).  In particular, the left, middle, and right panels of the 
figure respectively correspond to the case of an instantaneous turn-on, 
a staggered turn-on during a RD era, and a staggered turn-on during an MD era.  

Generally speaking, we observe that in each panel of the figure, the bound on $\tau_0$
tends to become more stringent as $\pi k R$ is increased for a fixed value of $m_\phi R$.
This implies that for a given choice of the parameter $\tau_0$, there is a maximum degree 
of AdS  warping for which a phenomenologically consistent dark sector can emerge for any 
fixed value of $m_\phi R$.  Moreover, we observe that the bound on the AdS curvature scale
generally becomes more and more stringent as $m_\phi R$ is increased, in agreement with 
the results shown in Fig.~\ref{fig:TimeEvolweffandOmega}.  Indeed, the regime in which 
$\pi k R$ and $m_\phi R$ are both large is the regime in which a significant number of 
low-lying states with similar abundances and comparable lifetimes are present within the 
ensemble.  Moreover, comparing results across the three panels of the figure, we see
that the bounds are more stringent for the case of an instantaneous turn-on than they
are for the case of a staggered turn-on during either a RD or MD era.  Indeed,
this is expected, since the $\Omega_n^0$ for the lighter $\hat{\chi}_n$ are enhanced relative 
to the $\Omega_n^0$ for the heavier $\hat{\chi}_n$ in the case of a staggered turn-on.  
These lighter modes, which typically have longer lifetimes, therefore carry a larger
fraction of $\Omegatot$ in this case than in the case of an instantaneous turn-on, 
and as a result the ensemble as a whole is more stable.

While the results in Fig.~\ref{fig:tau0bounds} provide a great deal of information about
the ensembles which arise within the parameter space of our warped-space scenario, 
there are other considerations which we must also take into account in assessing which regions 
of that parameter space are phenomenologically of interest.  In particular, from a DDM
perspective, we are interested in ensembles which are not only consistent with 
observational constraints, but which also represent a significant departure from
traditional dark-matter scenarios --- scenarios in which a single particle species 
contributes essentially the entirety of the dark-matter abundance.  The degree to which 
the contribution from the most abundant individual constituent dominates in 
$\Omegatot$ at any given time can be be parametrized by the ``tower fraction'' $\eta$, 
defined by the relation~\cite{DDM1} 
\begin{equation}
  \eta(t) ~\equiv~ \frac{\Omegatot - \max_n\{\Omega_n\}}{\Omegatot}~,
\end{equation}
the range of which is $0 \leq \eta < 1$.  If the most abundant individual ensemble 
constituent contributes essentially the entirety of $\Omegatot$, with the other 
$\hat{\chi}_n$ contributing negligibly to this total abundance, then $\eta \ll 1$ 
and this individual ensemble constituent is for all intents and purposes a 
single-particle dark-matter candidate.  By contrast, if $\eta \sim \mathcal{O}(1)$, 
multiple $\hat{\chi}_n$ contribute meaningfully to $\Omegatot$ and the ensemble is 
truly DDM-like.     

In Fig.~\ref{fig:etas} we show contours of the initial value $\eta(t_0)$ of the tower 
fraction at the time at which the abundances $\Omega_n$ are effectively established 
within the same region of parameter space as in Fig.~\ref{fig:tau0bounds}, and for 
the same choice of $\LambdaUV R$.  While the present-day tower fraction $\eta(\tnow)$ 
differs from $\eta(t_0)$ as a result of $\hat{\chi}_n$ decays, this difference is 
generally not terribly significant for ensembles which satisfy the constraint on 
$\Omegatot$ in Eq.~(\ref{eq:Omegatotbound}).  

One important feature that emerges upon comparing Figs.~\ref{fig:tau0bounds} 
and~\ref{fig:etas} is that the conditions which make $\eta(t_0)$ large are also those which 
make the bound on $\tau_0$ quite stringent.  In other words, there is an increasing tension 
between these two figures as $\pi k R$ gets large.  Indeed, if we impose an upper bound on 
$\tau_0$ (so that our DDM ensemble continues to be dynamical throughout up to and including the 
present epoch) as well as a lower bound on $\eta(t_0)$ (so that our scenario remains 
``DDM-like,'' with a significant fraction of the total dark-matter abundance shared across 
many ensemble constituents), then for any value of $m_\phi R$ there exists a {\it maximum value 
of warping}\/ which may be tolerated.   
Fortunately, however, we also observe that it is nevertheless possible to achieve a reasonably 
large value of $\eta(t_0)$ without requiring the value of $\tau_0$ to be extreme.

We also note that for $\pi k R \ll 1$, the values of $\eta(t_0)$,
expressed as functions of $m_\phi R$, are in complete agreement with the flat-space
results previously found in Ref.~\cite{DDM1}.  Thus, in this sense, we may view the 
contour plots in Fig.~\ref{fig:etas} as illustrating the structure that emerges
as we move away from the flat-space limit and increase $\pi k R$.


\section{Warped vs.\ Flat from the Dual Perspective}\label{sec:WarpedvsFlat}


Thus far, we have examined a 5D theory involving 
a bulk scalar propagating within a slice of AdS$_5$ and have shown that the mixed KK modes 
of this bulk scalar are capable of satisfying the basic criteria for a phenomenologically 
viable DDM ensemble in which multiple constituents contribute meaningfully to $\Omegatot$.  
This in turn implies that the ensemble of partially composite scalars which arises in the 
4D dual of this warped-space theory can likewise serve as a DDM ensemble as well.  Thus,
we have demonstrated what we set out to demonstrate in this paper --- namely that 
scenarios involving such ensembles are a viable context for model-bulding within the DDM 
framework.  

There are, however, certain aspects of the AdS/CFT dictionary 
that relates the two dual theories which deserve further comment.  Within the regime in 
which the AdS curvature scale is large, this dictionary is reasonably transparent.  
In general, the two dimensionful parameters $k$ and $R$ which characterize the 5D theory at 
times $t \lesssim t_G$ are related to the physical scales $\LambdaUV$ and 
$\LambdaIR$ of the strongly-coupled 4D theory by 
\begin{equation}
  \LambdaIR ~=~ \LambdaUV\, e^{-\pi k R}~.
  \label{eq:UVIRkRDictionary}
\end{equation}
Thus, as briefly mentioned in Sect.~\ref{sec:DynamicalDarkMatter}, the regime in which 
$\pi k R \gg 1$ corresponds to a large hierarchy between $\LambdaIR$ and $\LambdaUV$.  
The lightest mass eigenstate $\hat{\chi}_0$ in the 5D theory corresponds to a state in the 
4D theory which is primarily elementary.  The rest of the low-lying $\hat{\chi}_n$ in the 
5D theory correspond to states in the 4D theory which are primarily composite.        
  
By contrast, within the regime in which $\pi k R \ll 1$ and the theory approaches the 
flat-space limit considered in Ref.~\cite{DDM1}, the relationship between the states
of the 4D and 5D theories is more subtle.  The corresponding regime in the 4D theory is
that in which $\LambdaIR \approx \LambdaUV$.   
The KK eigenstates $\hat{\chi}_n^{(k=0)}$ which emerge in the flat-space limit of 
our warped DDM scenario do not correspond to composite states of the CFT in the 
dual 4D theory.  Rather, these states correspond to a tower of elementary fields
$\phi_n$ with masses $M_n \sim n/R$ which are also generically present in the 
theory and mix with the $\varphi_n$.  Indeed, the elementary scalar $\phi_0$ introduced 
in Sect.~\ref{sec:PartiallyCompositeAxion} may be viewed as the lightest of these fields.  
The $\phi_n$ with $n > 0$ typically do not play a significant role in the phenomenology 
of the partially composite theory when $\LambdaIR \ll \LambdaUV$.
The reason is that within this regime a large number of light states are present in the 
ensemble with masses $\hat{m}_n \ll 1/R$.  These light states have negligible wavefunction 
overlap with any of the $\phi_n$ other than $\phi_0$.  However, in the opposite regime in
which $\LambdaIR \approx \LambdaUV$, no such hierarchy exists between the mass scales of 
the elementary and composite states of the 4D theory.  Within this regime, the $\phi_n$ do 
indeed play an important role in the phenomenology of the model.

In order to understand how the $\phi_n$ affect the properties of the DDM ensemble in the 
$\LambdaIR \approx \LambdaUV$ regime, it is illustrative to compare the structure of the 
mass matrix which emerges in this regime to the structure which emerges in the 
$\LambdaIR \ll \LambdaUV$ regime.  In situations in which the $\varphi_n$ are 
all significantly heavier than at least the lightest several $\phi_n$, the mass eigenstates 
$\chi_n$ of the theory at times $t \lesssim t_G$ 
are simply the $\phi_n$, with the corresponding masses $m_0 = 0$ for $n=0$ and $m_n = M_n$ 
for $n > 0$.  By contrast, at times $t \gtrsim t_G$, the squared-mass matrix in the 
$\phi_n$ basis has the rough overall structure
\begin{equation}
  \mathcal{M}^2 ~=~ 
    \left( \begin{matrix} 
      m_\phi^2 & m_\phi^2         & m_\phi^2       & \ldots \\
      m_\phi^2 & M_1^2 + m_\phi^2 & m_\phi^2       & \ldots \\ 
      m_\phi^2 & m_\phi^2         & M_2^2 + m_\phi^2 & \ldots \\
      \vdots   & \vdots           & \vdots         & \ddots
    \end{matrix} \right)~.
  \label{eq:flatmass}
\end{equation} 
In the regime in which $m_\phi \ll M_n$ for all $n > 0$, the mass eigenstates $\hat{\chi}_n$ are, 
to $\mathcal{O}(m_\phi^2/M_n^2)$, given by   
\begin{equation}
  |\hat{\chi}_n \rangle ~\approx~
    \begin{cases}  
     |\phi_0 \rangle 
       - {\displaystyle\sum_{\ell=1}^\infty}\frac{m_\phi^2}{M_\ell^2} |\phi_\ell\rangle  
       & n = 0 \\  
     \frac{m_\phi^2}{M_n^2} |\phi_0\rangle 
       + | \phi_n \rangle + {\displaystyle\sum_{\ell \neq 0, n}^\infty}
       \frac{m_\phi^2}{M_n^2 - M_\ell^2}|\phi_\ell\rangle  & n > 0~.
  \end{cases}
\end{equation}
To the same order, the corresponding mass eigenvalues are $\hat{m}_0^2 \approx m_\phi^2$
for $n=0$ and $\hat{m}_n^2 \approx M_n^2 + m_\phi^2$ for $n > 0$.

Since all of the $\chi_n$ with $n > 0$ are massive prior to the phase transition, only 
$\chi_0$ can acquire a misaligned vacuum value.  Thus, the mixing coefficients 
$A_n = \langle \chi_0 | \hat{\chi}_n\rangle$ play the same phenomenological role in 
the $\LambdaIR \approx \LambdaUV$ regime as they do in the $\LambdaIR \ll \LambdaUV$ 
regime.  In our truncated theory, these coefficients are given by       
\begin{equation}
  A_n ~=~ \langle \phi_0 | \hat{\chi}_n \rangle ~\approx~ 
    \begin{cases} 
      1 & n=0 \\ 
      \frac{m_\phi^2}{M_n^2} & n > 0~.
  \end{cases}
\end{equation}
The projection coefficients $A_n'$ in this same regime are 
\begin{equation}
  A_n' ~=~ \sum_{\ell=0}^\infty \langle \phi_\ell | \hat{\chi}_n \rangle 
    ~\approx~ \langle \phi_n | \hat{\chi}_n \rangle ~\approx~ 1~, 
\end{equation}
up to corrections of $\mathcal{O}(m_\phi^2/M_n^2)$.  These results agree with those in 
Ref.~\cite{DDM1} for the $m_\phi \ll M_n$ regime, up to $\mathcal{O}(1)$ numerical factors.  
Of course, for $m_\phi$ outside this regime, the full, infinite-dimensional mass matrix is 
required in order to obtain the corresponding expressions for $A_n$ and $A_n'$.

The structure of the mass-squared matrix in Eq.~(\ref{eq:flatmass}) clearly differs 
in several ways from the structure of the mass-squared matrix in Eq.~(\ref{eq:warpedmass})
for the corresponding truncated theory within the $\LambdaIR \ll \LambdaUV$ regime.
However, the mass-squared matrices in Eqs.~(\ref{eq:flatmass}) and~(\ref{eq:warpedmass}) cannot
meaningfully be compared because the former is expressed with respect to the basis of
mass eigenstates prior to the phase transition, whereas the latter is expressed in the 
$\{\phi_0,\varphi_n\}$ basis.  Rather, the mass-squared matrix in Eq.~(\ref{eq:flatmass})
must be compared to the mass-squared matrix $\widetilde{\mathcal{M}}^2$ 
obtained in the $\LambdaIR \ll \LambdaUV$ regime {\it after}\/ the phase transition 
expressed in the basis of the states $\chi_n$ which are mass eigenstates of the 
theory {\it before}\/ the phase transition.  This matrix is given by  
$\widetilde{\mathcal{M}}^2 = \mathcal{U}\mathcal{M}^2\mathcal{U}^\dagger$, where
$\mathcal{M}^2$ is the matrix appearing in Eq.~(\ref{eq:warpedmass}) and 
$\mathcal{U}$ is the unitary matrix which represents the transformation from the 
$\{\phi_0,\varphi_n\}$ basis to the $\chi_n$ basis.  The results in 
Eq.~(\ref{eq:EvecsCompositeHighEnergy}) 
imply that to $\mathcal{O}(\epsilon_n^2)$, this latter matrix is given by
\begin{equation}
  \mathcal{U} ~\approx~ 
    \left(\begin{matrix}
      1- \sum_{m=1}^\infty\frac{\epsilon_m^2}{2g_m^2} & -\frac{\epsilon_1}{g_1} & 
        -\frac{\epsilon_2}{g_2} & \ldots  \\
      \frac{\epsilon_1}{g_1} & 1 - \frac{\epsilon_1^2}{2g_1^2} & 
        \frac{\epsilon_1\epsilon_2 g_2}{g_1(g_1^2-g_2^2)} & \ldots \\
      \frac{\epsilon_2}{g_2} & 
        \frac{\epsilon_2\epsilon_1 g_1}{g_2(g_2^2-g_1^2)} & 
        1 - \frac{\epsilon_2^2}{2g_2^2} & \ldots \\
      \vdots & \vdots & \vdots & \ddots  
    \end{matrix}\right)~.
\end{equation}
As a result, to the same order in $\epsilon_n$, we find that
\begin{widetext}
\begin{equation}
  \widetilde{\mathcal{M}}^2 ~\approx~ 
    \left( \begin{matrix} 
      m_\phi^2\left(1 + \sum_{m=1}^\infty\frac{\epsilon_m^2}{g_m^2}\right) &
      \frac{\epsilon_1}{g_1} m_\phi^2 & 
      \frac{\epsilon_2}{g_2} m_\phi^2 & \ldots \\
      \frac{\epsilon_1}{g_1} m_\phi^2 & 
      (g_1^2+\epsilon_1^2)\LambdaIR^2 + \frac{\epsilon_1^2}{g_1^2}m_\phi^2 & 
      \frac{\epsilon_1\epsilon_2}{g_1g_2}m_\phi^2 &
      \ldots \\
      \frac{\epsilon_2}{g_2} m_\phi^2 & 
      \frac{\epsilon_1\epsilon_2}{g_1g_2}m_\phi^2 &
      (g_2^2+\epsilon_2^2)\LambdaIR^2 + \frac{\epsilon_2^2}{g_2^2}m_\phi^2 & 
      \ldots \\
      \vdots & \vdots & \vdots & \ddots
    \end{matrix} \right)~.
  \label{eq:MassSqdMatTilde}
\end{equation}
\end{widetext}

Comparing the results in Eqs.~(\ref{eq:flatmass}) and~(\ref{eq:MassSqdMatTilde}), we see
that the crucial difference between the structures of these two mass-squared matrices is due
to the factors of $\epsilon_n/g_n$ that appear in both the diagonal and off-diagonal 
contributions to $\widetilde{\mathcal{M}}^2$ which arise a result of the phase transition.
These factors arise in the $\LambdaIR \ll \LambdaUV$ regime as a consequence of the coupling 
between $\phi_0$ and the composite sector engendered by the operator $\mathcal{O}_c$.
The fact that $\epsilon_n/g_n$ varies with $n$ in a non-trivial manner for $\pi k R \gg 1$
accounts for the differences in the resulting mass spectra.

\begin{figure*}[t]
  \centering
  \includegraphics[clip, width=0.4\linewidth]{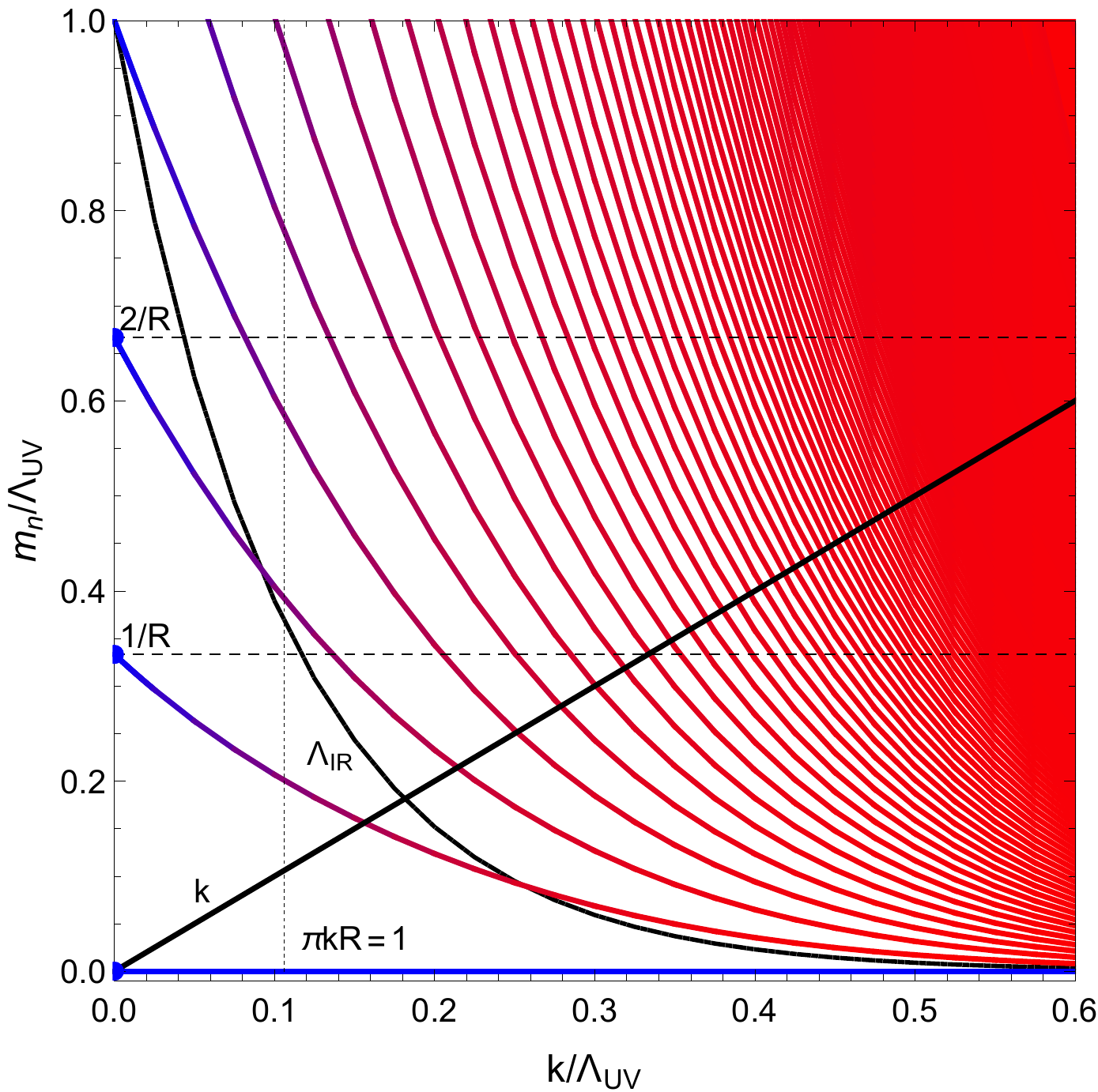} 
    ~~~~~
  \includegraphics[clip, width=0.4\linewidth]{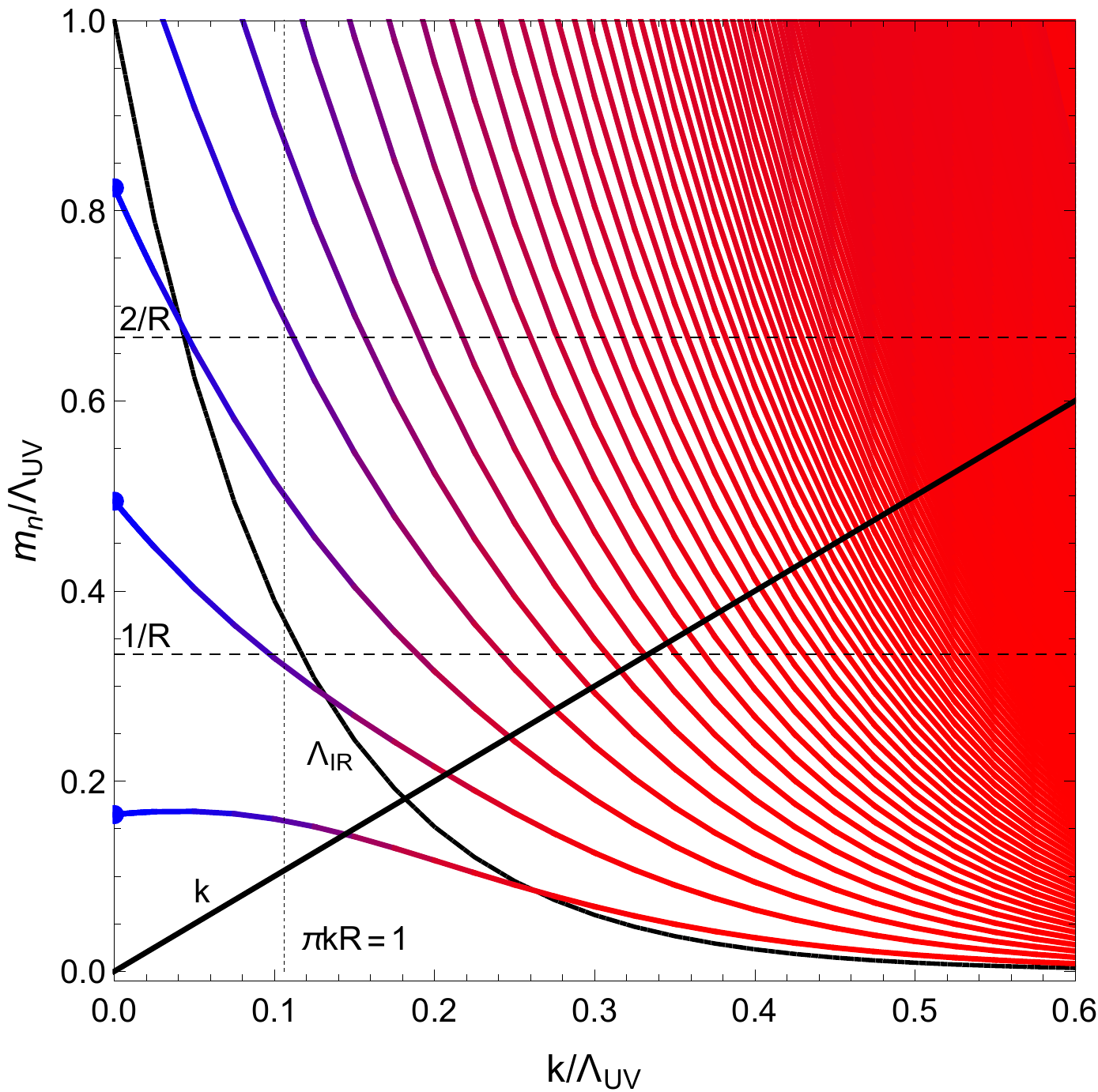}
  \caption{The mass spectrum of the 5D gravity dual of our partially composite DDM theory,
    plotted as a function of the AdS curvature scale $k$ for two representative choices of 
    $m_\phi$.  The results shown in the left panel correspond to the choice of 
    $m_\phi = 10^{-4} \Lambda_{\rm UV}$, while the results shown in the right panel 
    correspond to the choice of $m_\phi = \Lambda_{\rm UV}$.  In both panels, we have taken 
    $R = 3/\LambdaUV$.  Each of the solid curves shown in each panel 
    corresponds to a particular value of the index $n$ and indicates the mass $\hat{m}_n$ of 
    the corresponding ensemble constituent.  Thus, the set of points obtained by taking 
    a vertical ``slice'' through either panel collectively represent the mass 
    spectrum of the theory for the corresponding value of $k$.  The color at any given 
    point along each curve provides information about the extent to which the
    corresponding state in the partially composite theory in 4D is primarily elementary 
    or composite.  In particular, the color indicates the absolute value of the projection 
    coefficient $A'_n$ at that point, normalized to the absolute value of
    the projection coefficient $A_n^{\prime (k=0)}$ obtained for the same choice of 
    $m_\phi$ and $R$, but with $k = 0$.  A value near 
    $|A'_n/A_n^{\prime (k=0)}| = 0$ (red) suggests that the state is 
    primarily composite, while a value near $|A'_n/A_n^{\prime (k=0)}| = 1$ 
    (blue) suggests that the state is primarily elementary.  Curves indicating the
    value of $k$ (solid black line with unit slope), $1/R$ and $2/R$ (dashed black 
    horizontal lines), and $\Lambda_{\rm IR}$ (dot-dashed black curve) are also provided 
    for reference.         
  \label{fig:dictionarycartoon}}
\end{figure*}

We now turn to examine how the structural differences between the matrices in 
Eqs.~(\ref{eq:flatmass}) and~(\ref{eq:MassSqdMatTilde}) affect the actual mass spectra of
the theory.  In Fig.~\ref{fig:dictionarycartoon}, we show how the mass spectrum of the 
5D gravity dual of our partially composite DDM theory varies as a function of $k$ for 
two representative choices of $m_\phi$.  The results shown in the left panel correspond to 
the choice of $m_\phi = 10^{-4} \Lambda_{\rm UV}$, while the results shown in the the 
right panel correspond to the choice of $m_\phi = \Lambda_{\rm UV}$.  In both panels, we
have taken $R = 3/\Lambda_{\rm UV}$.  Each of the solid curves shown in each 
panel corresponds to a particular value of the index $n$ and indicates the mass $\hat{m}_n$
of the corresponding ensemble constituent.  Thus, the set of points obtained by taking a
vertical ``slice'' through either panel collectively represent the mass spectrum of the
theory for the corresponding value of $k$.  The color at any given point along each curve
provides information about the degree to which the corresponding state in the
partially composite theory is elementary or composite.  In particular the color indicates 
the absolute value of the projection coefficient $A'_n$ at that point, normalized to the
absolute value of the projection coefficient $A_n^{\prime (k=0)}$ obtained for the same 
choice of $m_\phi$ and $R$, but with $k = 0$.  

In order to motivate why this quantity is a useful proxy for compositeness, we note
once again that the flat-space limit of the 5D dual theory corresponds to the 
limit in which all of the states of the corresponding 4D theory are purely elementary.
As discussed in Appendix~\ref{app:FlatLimit}, the bulk profile of each state in 
this limit reduces to
\begin{equation}
  \hat{\zeta}_n^{(k=0)}(y) ~=~ \frac{r_n}{\sqrt{\pi R}} \cos\left(\frac{n\pi y}{R}\right)~,
\end{equation}
where we have defined
\begin{equation}
  r_n ~\equiv~ \begin{cases} 1 & n=0 \\ \sqrt{2} & n > 0~. 
    \end{cases}
\end{equation}
Using the completeness relation in Eq.~(\ref{eq:CompletenessRel}) for these 
flat-space bulk profiles with $y'=0$, we may express $A'_n$ for 
general $k$ in the more revealing form
\begin{eqnarray}
  A'_n &=& \sqrt{\frac{1-e^{-2\pi k R}}{2k}}\int_0^{\pi R} \hat{\zeta}_n(y)\delta(y) dy 
      \nonumber \\
    &=& \sqrt{\frac{1-e^{-2\pi k R}}{2k}}\int_0^{\pi R} \hat{\zeta}_n(y) \sum_{m=0}^{\infty} 
       \frac{r_m^2}{\pi R}\cos\left(\frac{m\pi y}{R}\right) dy \nonumber \\
   &=& \sqrt{\frac{1-e^{-2\pi k R}}{2\pi k R}}\sum_{m=0}^\infty r_m \!\!
       \int_0^{\pi R}\!\! \hat{\zeta}_m^{(k=0)}(y) \hat{\zeta}_n(y) dy ~, \nonumber \\
\end{eqnarray}
where in going from the first to the second line, we have used the completeness relation
\begin{equation}
  \sum_{n=0}^\infty \hat{\zeta}_n^{(k=0)}(y)\hat{\zeta}_n^{(k=0)}(y') ~=~ \delta(y-y')~,
\end{equation}
with $y'=0$.
Thus, up to an overall normalization coefficient and an additional factor of $r_n$ which 
appears in each term of the sum, $A'_n$ can be viewed as a sum of the overlap integrals between 
the state $\hat{\chi}_n$ within the ensemble and the individual mass-eigenstate fields
$\hat{\chi}_m^{(k=0)}$ of a theory with $k = 0$ and the same values of $m_\phi$ and $R$.
We choose to normalize this quantity to $A_n^{\prime (k=0)}$ because 
$0 \leq |A'_n/A_n^{\prime (k=0)}| \leq 1$, with $|A'_n/A_n^{\prime (k=0)}| = 1$ occurring in
the $k = 0$ limit.  A value of $|A'_n/A_n^{\prime (k=0)}|$ near unity therefore suggests 
that the degree of overlap between $\hat{\chi}_n$ and the $\hat{\chi}_m^{(k=0)}$ is large 
and that the corresponding state in the partially composite theory is mostly elementary.  
By contrast, a value near zero suggests that the degree of overlap is small and that the
corresponding state is mostly composite.  

The results shown in the left panel of the Fig.~\ref{fig:dictionarycartoon} are 
characteristic of the regime in which $m_\phi$ is considerably smaller than all of the
other relevant scales in the problem.  In this regime, for $k=0$, the spectrum consists 
of one light state $\hat{\chi}_0$ with a mass 
$\hat{m}_0 \ll 1/R$ and several additional states with masses $\hat{m}_n \approx n/R$, 
all of which are elementary.  As $k$ is increased, $\hat{m}_0$ remains approximately
constant and $\hat{\chi}_0$ remains approximately elementary.  By contrast, the masses
of the additional $\hat{\chi}_n$ decrease while the degree of compositeness for each of 
these states increases.  Furthermore, additional $\hat{\chi}_n$ whose masses descend 
from infinity successively appear in the spectrum of the theory below 
$\LambdaUV$ as $k$ increases.  The process continues as $k$ is further increased until
we enter the $\pi k R \gg 1$ regime in which the spectrum includes a large number of
low-lying states with masses in the range $k \gg \hat{m}_n \gtrsim \mKK$, all of which
exhibit a high degree of compositeness, as expected.

By contrast, the results shown in the right panel of Fig.~\ref{fig:dictionarycartoon}, which 
are characteristic of the regime in which $m_\phi$ is significantly larger than both $k$ and 
$1/R$, differ from those in the left panel primarily for small $k$.  Most notably, the masses 
of the states obtained for $k=0$ are not given by $\hat{m}_n \approx n/R$ as they are in 
the left panel, but rather by $\hat{m}_n \approx ( n + \frac{1}{2})/R$.  Once again, this
accords with the expected behavior of the $\hat{m}_n$ in the flat-space limit~\cite{DDM1}.  
For larger $k$, the only qualitative difference between the mass spectra obtained in the 
small-$m_\phi$ and large-$m_\phi$ regimes is that the spectrum in the latter regime lacks 
the single, primarily elementary state with $\hat{m}_0 \ll \mKK$ present in the former
regime.  Indeed, for large $m_\phi$, we see that all of the low-lying states within the
ensemble are primarily composite when $k$ is large.


\section{Conclusions\label{sec:Conclusions}}


In this paper, we have investigated a novel realization of the DDM framework
within the context of a strongly-coupled CFT.  In this scenario, the constituent 
particles of the DDM ensemble are the composite states which emerge in the spectrum 
of the theory below the scale at which conformal invariance is spontaneously broken.  
Abundances and decay widths for these ensemble constituents can be generated through 
mixing between these composite states and an additional, elementary scalar $\phi_0$, 
yielding a spectrum of partially composite mass-eigenstates whose degree of compositeness 
varies across the ensemble.  Informed by the AdS/CFT correspondence, we have derived 
the masses, decay widths, and cosmological abundance for these partially composite states 
within the context of the gravity dual of this scenario --- a theory involving a scalar 
field propagating in the bulk of a slice of AdS$_5$.  We have investigated the extent 
to which model-independent bounds on the total abundance and the equation of state for the 
ensemble constrain the parameter space of this scenario, and we have shown that indeed a 
balancing between decay widths and abundances appropriate for a DDM ensemble arises 
within large regions of that parameter space, even within the regime wherein the degree of 
warping in the dual theory is significant --- a regime which corresponds to the regime in 
which there exists a significant hierarchy of scales $\LambdaIR \ll \LambdaUV$ in the partially 
composite theory.  However, we have also shown that constraints on the ensemble become 
increasingly stringent as the degree of warping increases.  Moreover, we have shown that
interesting qualitative features, such as non-monotonicities in the spectrum of decay widths, 
can develop in the highly-warped regime of the dual theory which do not arise in the flat-space 
limit.

A few comments are in order.  First of all, because our primary focus in this paper has 
been the 4D partially composite DDM scenario, we have regarded the 5D gravity dual 
of this theory primarily as a calculational tool for obtaining information about the 
properties of the ensemble in the 4D theory.  However, the fact that a viable DDM ensemble 
can emerge in the context of a scenario involving a warped extra dimension is interesting in 
its own right.  Indeed from this perspective, we may regard the results in 
Sects.~\ref{sec:DualMisalignmentMechanism} and~\ref{sec:DynamicalDarkMatter} as generalizations
of the flat-space results derived in Refs.~\cite{DDM1,DDM2} to warped space.    

On a final note, in constraining the parameter space of our scenario, we have focused 
on considerations such as limits on $\weff$ and $\Omegatot$ in bounding 
the parameter space of our scenario --- considerations which do not depend sensitively 
on the identities of the final-state particles into which the ensemble constituents decay.
If the $\hat{\chi}_n$ decay solely into other, lighter particles which reside within
the dark sector but are external to the ensemble (\eg, particles which behave as dark 
radiation rather than dark matter), these constraints are typically the leading ones.  
By contrast, if the $\hat{\chi}_n$ decay into final states involving visible-sector 
particles, additional constraints apply.  It would be interesting to consider how such 
constraints further restrict the parameter space of our ensemble for certain well-motivated 
decay scenarios in which decays to SM particles dominate the width of each $\hat{\chi}_n$.

\begin{acknowledgments}

The research activities of YB and TG are supported in part by the Department of 
Energy under Grant DE-SC0011842.  The research activities of KRD are supported in 
part by the Department of Energy under Grant DE-FG02-13ER41976 (DE-SC0009913) and 
by the National Science Foundation through its employee IR/D program.
The research activities of BT are supported in part by the National Science 
Foundation under Grant PHY-1720430.  The opinions and conclusions expressed 
herein are those of the authors, and do not represent any funding agencies. 

\end{acknowledgments}

\appendix


\section{Mixing and Projection Coefficients in the Flat-Space Limit}\label{app:FlatLimit}


In Sect.~\ref{sec:DualMisalignmentMechanism}, we derived analytic approximations
for $A_n$ and $A'_n$ valid within the regime in which $\pi k R \gg 1$ and 
$ k \gg \hat{m}_n \gg \mKK$.  In this Appendix, in order to make contact with the results
obtained in Refs.~\cite{DDGAxion,DDM1} for the case of a flat extra dimension, we 
derive the corresponding expressions valid in the regime in which $\hat{m}_n \gg k$ and 
then demonstrate that these expressions reduce to the expected results in the 
$k\rightarrow 0$ limit. 

We begin by considering the mass spectrum of the theory at early times 
$t \lesssim t_G$, before the phase transition occurs.  The mass spectrum of the 
$\chi_n$ in this phase of the theory consists of the solutions to
Eq.~(\ref{eq:MassSpecEqnNomphi}).  In the regime in which $\pi k R \ll 1$, this 
equation reduces to 
\begin{equation}
  \sin \left(\pi m_n R \right) ~\approx ~ 0~,
\end{equation}
which implies that $m_n \approx n/R$, in accord with the expected flat-space result.
    
We now consider the mass spectrum of the theory at times $t \gtrsim t_G$, after 
the brane mass has been generated.  Since the action in the flat-space limit 
is symmetric under the coordinate transformation $y \rightarrow \pi R - y$, the mass 
spectrum of the $\hat{\chi}_n$ in this limit is the same regardless of whether the 
dynamics that generates $m_\phi$ is localized on the UV or IR brane.  We therefore 
focus on the case in which this dynamics is localized on the UV brane.  The mass 
spectrum of the $\hat{\chi}_n$ in this phase of the theory consists of the solutions 
to Eq.~(\ref{eq:MassSpecEqnmphi}).  In the regime in which $\hat{m}_n \gg k$, regardless
of the value of $\pi k R$, the Bessel functions in this equation are well approximated
by
\begin{eqnarray}
  J_\alpha(x) &\approx& \sqrt{\frac{2}{\pi x}}
    \cos\left(x - \frac{\alpha \pi}{2} - \frac{\pi}{4}\right) \nonumber \\
  Y_\alpha(x) &\approx& \sqrt{\frac{2}{\pi x}}
    \sin\left(x - \frac{\alpha \pi}{2} - \frac{\pi}{4}\right)~. 
\end{eqnarray}
One therefore finds that, in this regime, Eq.~(\ref{eq:MassSpecEqnmphi}) reduces to  
\begin{equation}
  \frac{m_\phi^2}{2k}\left(1 - e^{-2\pi k R}\right)
    \cot\left[\frac{\hat{m}_n}{k}\left(e^{\pi k R} - 1\right)\right] ~\approx~ \hat{m}_n~.
  \label{eq:MassEquationmggkFull}
\end{equation}
In the regime in which $\pi k R \ll 1$, this equation further reduces to
\begin{equation}
  \pi m_\phi^2 R \cot (\pi \hat{m}_n R) ~\approx~ \hat{m}_n~.
  \label{eq:boundarymass}
\end{equation}
This result --- and therefore the mass spectrum of the $\hat{\chi}_n$ obtained in this
regime --- agrees with the corresponding flat-space expression in Ref.~\cite{DDM1,DDGAxion}.
The solutions for $\hat{m}_n$ are given by $\hat{m}_n \approx (n + \frac{1}{2})/R$ for 
$n \ll \pi m_\phi^2 R^2$ and $\hat{m}_n \approx n/R$ for $n \gg \pi m_\phi^2 R^2$ and
interpolate smoothly between these asymptotic expressions. 

In order to derive the corresponding analytic approximations for $A_n$ and $A'_n$,
we begin by noting that for $m_n \gg k$, the expression for the bulk profile $\zeta_n(y)$ 
of the early-time mass eigenstate $\chi_n$ in Eq.~(\ref{eq:profile}) reduces to
\begin{equation}
  \zeta_n(y) ~\approx~ \frac{r_n}{\sqrt{\pi R}} e^{3 k y /2} 
    \cos\left[\frac{m_n}{k}(e^{k y} - e^{\pi k R})\right]~,
  \label{eq:BulkProfilemnggk}
\end{equation}
where we have defined
\begin{equation}
    r_n ~\equiv~ \sqrt{\frac{4\pi m_n R}
      {\frac{2m_n}{k}(e^{\pi k R} - 1)+
      \sin\left[\frac{2m_n}{k}(e^{\pi k R} - 1)\right]}}~.
  \label{eq:rnCoeff}
\end{equation}
The expression for the bulk profile $\hat{\zeta}_n$ of the late-time
mass eigenstate $\hat{\chi}_n$ is identical in form to the expression for
$\zeta_n(y)$ in Eq.~(\ref{eq:BulkProfilemnggk}), but with $\hat{m}_n$ in place
of $m_n$.

In the regime in which $\pi k R \ll 1$, Eq.~(\ref{eq:BulkProfilemnggk}) reduces to 
\begin{equation}
  \zeta_n (y) ~\approx ~ \frac{r_n}{\sqrt{\pi R}} 
    \cos \big[m_n (y-\pi R)\big]~,
  \label{eq:ZetaApproxInFlatSpaceLimit}
\end{equation}
where
\begin{equation}
  r_n ~\approx~ \sqrt{
    \frac{2}{1 + \frac{\sin (2 \pi m_n R)}{2 \pi m_n R}}}~.
\end{equation}
We note that for either $m_n \approx n/R$ or $m_n \approx (n+\frac{1}{2})/R$ 
with $n \in \mathbb{Z}$, this quantity is well approximated by
\begin{equation}
  r_n ~\approx~ 
    \begin{cases} 
      1 & n ~=~ 0 \\
      \sqrt{2} & n ~>~ 0~.
    \end{cases}
\end{equation}
Taking into account the difference in normalization conventions, these results agree
with those derived in Ref.~\cite{DDM1}.  Since $A_n$ and $A'_n$ are derived directly
from $\zeta_0(y)$ the corresponding bulk profile $\hat{\zeta}_n(y)$, it therefore follows 
that the mixing and projection coefficients obtained in the $k\rightarrow 0$ limit of 
our warped-space scenario reproduce those obtained in Refs.~\cite{DDGAxion,DDM1} as well. 

Substituting our analytic approximation for $\hat{\zeta}_n(y)$ 
into Eq.~(\ref{eq:Unl}), we find that the mixing-matrix elements are 
\begin{eqnarray}
  A_n &\approx& \sqrt{\frac{2k}{1-e^{-2\pi k R}}}\frac{\hat{r}_n}{\sqrt{\pi R}} 
    \nonumber \\ &  & \times
    \int^{\pi R}_0 e^{-ky/2}\cos\left[\frac{\hat{m}_n}{k}(e^{k y} - e^{\pi k R})\right] dy~,
    \nonumber \\
  \label{eq:AnWithCosIntegral}
\end{eqnarray}
where $\hat{r}_n$ is given by Eq.~(\ref{eq:rnCoeff}), but with $\hat{m}_n$ in
place of $m_n$.  In order to simplify this expression further, we observe that 
the integral over $y$ can be written in terms of the Fresnel integrals
\begin{eqnarray}
  C(x) &\equiv& \int_0^x \cos\left(\frac{\pi t^2}{2} \right) dt \nonumber \\
  S(x) &\equiv& \int_0^x \sin\left(\frac{\pi t^2}{2} \right) dt~.
\end{eqnarray}
In particular, we find that
\begin{eqnarray}
  A_n &\approx& \frac{1}{\sqrt{\hat{m}_n R(1-e^{-2\pi k R})}}
    \left(\frac{4\hat{m}_n \hat{r}_n}{k}\right) 
    \nonumber \\ &  & \times
    \Bigg\{ \sin\left(\textstyle\frac{\hat{m}_n}{\mKK}\right)
      \bigg[C\left(\textstyle\sqrt{\frac{2\hat{m}_n}{\pi \mKK}}\right) 
        - C\left(\textstyle\sqrt{\frac{2\hat{m}_n}{\pi k}}\right) \bigg] 
      \nonumber \\ & &
      - \cos\left(\textstyle\frac{\hat{m}_n}{\mKK}\right)
      \bigg[S\left(\textstyle\sqrt{\frac{2\hat{m}_n}{\pi \mKK}}\right) 
        - S\left(\textstyle\sqrt{\frac{2\hat{m}_n}{\pi k}}\right) \bigg] 
      \nonumber \\ &  &
      + \sqrt{\frac{k}{2\pi \hat{m}_n}}
        \cos\left(\textstyle\frac{\hat{m}_n}{\mKK} - \frac{\hat{m}_n}{k}\right)
      - \sqrt{\frac{\mKK}{2\pi \hat{m}_n}} \Bigg\}~.
    \nonumber \\
    \label{eq:AnFresnalExpansion}
\end{eqnarray}
Since we are working within the regime in which $\hat{m}_n \gg k$, we may simplify
this expression by making use of the well-known asymptotic expansions for 
$C(x)$ and $S(x)$.  In particular, for large arguments $x \gg 1$, these 
integrals are well approximated by
\begin{eqnarray}
  C(x) &\approx& \frac{1}{2} +\frac{1}{\pi x}\sin\left(\frac{\pi x^2}{2}\right)
    - \frac{1}{\pi^2 x^3}\cos\left(\frac{\pi x^2}{2}\right) \nonumber \\
  S(x) &\approx& \frac{1}{2} -\frac{1}{\pi x}\cos\left(\frac{\pi x^2}{2}\right)
    - \frac{1}{\pi^2 x^3}\sin\left(\frac{\pi x^2}{2}\right)~. \nonumber \\
\end{eqnarray}
With these approximations, we find that Eq.~(\ref{eq:AnFresnalExpansion}) reduces to
\begin{equation}
  A_n ~\approx~ \sqrt{\frac{2k}{\pi \hat{m}_n^2 R(1-e^{-2 \pi k R})}} \hat{r}_n
    \sin\left[\frac{\hat{m}_n}{k}(e^{\pi k R} - 1)\right]~.
\end{equation}
Using Eq.~(\ref{eq:MassEquationmggkFull}) in order to eliminate the trigonometric 
functions, we arrive at our final expression for $A_n$ in the $\hat{m}_n \gg k$ regime.
After some algebra, we find that
\begin{equation}
  A_n ~\approx~ \sqrt{\frac{\frac{m_\phi^2}{\hat{m}_n^2}
    \left(\frac{1-e^{-2\pi k R}}{e^{\pi k R} - 1}\right)}
    {\frac{\hat{m}_n^2}{m_\phi^2} + \frac{m_\phi^2}{4k^2}(1-e^{-2\pi k R})^2+
    \frac{1 - e^{-2\pi k R}}{2(e^{\pi k R} - 1)}}}~.
  \label{eq:AnFinalmnggk}
\end{equation}
We note that for $\pi k R \ll 1$, this expression reduces to
\begin{equation}
    A_n ~\approx~ \frac{\sqrt{2}m_\phi}{\hat{m}_n}
      \frac{1}{\sqrt{\frac{\hat{m}_n^2}{m_\phi^2} 
      + \pi^2 m_\phi^2 R^2 + 1}}~,
\end{equation}
which once again agrees with the corresponding result in Refs.~\cite{DDGAxion,DDM1}.

The analytic approximation for $A'_n$ in the $\hat{m}_n \gg k$ regime, 
obtained by substituting Eq.~(\ref{eq:ZetaApproxInFlatSpaceLimit}) into 
Eq.~(\ref{eq:projectionUV}), is
\begin{equation}
  A_n' ~\approx~ \sqrt{\frac{1-e^{-2\pi k R}}{2\pi k R}} \hat{r}_n 
    \cos \left[\frac{\hat{m}_n}{k}(e^{\pi k R} - 1)\right]~.
\end{equation}
Using Eq.~(\ref{eq:MassEquationmggkFull}) in order to eliminate trigonometric 
functions, we find that this expression simplifies to 
\begin{equation}
    A'_n ~\approx~ \sqrt{\frac{\frac{\hat{m}_n^2}{m_\phi^2}
    \left(\frac{1-e^{-2\pi k R}}{e^{\pi k R} - 1}\right)}
    {\frac{\hat{m}_n^2}{m_\phi^2} + \frac{m_\phi^2}{4k^2}(1-e^{-2\pi k R})^2+
    \frac{1 - e^{-2\pi k R}}{2(e^{\pi k R} - 1)}}}~.
  \label{eq:AprimeGenmnggk}
\end{equation}

Comparing this result with Eq.~(\ref{eq:AnFinalmnggk}), we observe that 
$A_n' \approx (\hat{m}_n/m_\phi)^2 A_n$ within this regime, in accord with 
the relationship between the mixing and projection coefficients obtained  
in Refs.~\cite{DDGAxion,DDM1}.  Moreover, we observe that the expression
in Eq.~(\ref{eq:AprimeGenmnggk}) increases monotonically with $\hat{m}_n$.
Thus, in the regime in which the AdS curvature is sufficiently small that the 
criterion $\hat{m}_n \gg k$ is satisfied for all $\hat{\chi}_n$ within the ensemble, the 
$A_n'$ --- and therefore also the decay widths $\Gamma_n$ --- do not exhibit 
the non-monotonicities discussed in Sect.~\ref{sec:DynamicalDarkMatter}, which 
can arise when the ensemble includes states with masses $\hat{m}_n \lesssim k$.


\section{Alternative Brane-Localization Scenarios}\label{app:DifferentScenarios}


In Sect.~\ref{sec:DualMisalignmentMechanism} we derived expressions for the 
mixing and projection coefficients $A_n$ and $A_n'$ for the ensemble 
constituents for the case in which the dynamics which generates the mass term 
$m_\phi$ and the SM particles into which these ensemble constituents decay are 
both localized on the UV brane.  In this Appendix, we derive the 
corresponding expression for $A_n$ for the case in which the dynamics that 
generates $m_\phi$ is localized on the IR brane and the corresponding 
expression for $A_n'$ for the case in which the SM is localized on the IR brane.  
From these results and those appearing in Eqs.~(\ref{eq:mixing}) 
and~(\ref{eq:projectionUV}), the scaling exponents $\alpha$ and $\beta$ for
all possible combinations of locations for the mass-generating dynamics 
and the SM can be determined in a straightforward manner.     

\subsection{Mass-Generating Dynamics on the IR Brane}

We begin by deriving the mixing coefficients $A_n$ for the case in which
the dynamics that generates $m_\phi$ is localized on the IR brane.  At 
times $t \lesssim t_G$ before the scale at which the mass-generating 
phase transition occurs, the action is essentially the same as it is in the case 
in which $m_\phi$ is localized on the UV brane.  The lightest state is likewise
massless, with a profile given by Eq.~(\ref{eq:profilezeromode}), while the 
remaining states have masses given by the solutions of Eq.~(\ref{eq:MassSpecEqnNomphi})
and profiles given by Eqs.~(\ref{eq:profile}).  However, at times $t \gtrsim t_G$, 
the action in this case is given not by Eq.~(\ref{eq:SwithaUVBraneMass}), 
but rather by
\begin{equation}
  \mathcal{S}_{\chi} ~=~ - \int d^5 x \sqrt{-g} 
    \Bigg[ \frac{1}{2} \partial_M \chi \partial^M \chi  
    - m_B \chi^2 \delta (y - \pi R) \Bigg]~.
  \label{eq:SwithanIRBraneMass}
\end{equation}

The masses and bulk profiles of the mass eigenstates $\hat{\chi}_n$ can be 
determined by solving the equation of motion derived from
Eq.~(\ref{eq:SwithanIRBraneMass}) with the boundary conditions 
\begin{eqnarray}
  \partial_y \chi(x, y) \big|_{y = 0} &=& 0 \nonumber \\ 
  \left( \partial_y - m_B \right) 
  \chi(x, y) \big|_{y = \pi R} &=& 0~.
  \label{eq:LowEnergyBCsIRBraneMass}
\end{eqnarray}
In particular, the masses $\hat{m}_n$ are the solutions to the equation 
\begin{multline}
  J_1\left(\frac{\hat{m}_n}{k}\right)\left[ 
    \frac{m_B}{\hat{m}_n e^{\pi kR}} 
    Y_2\left(\frac{\hat{m}_n}{\mKK}\right) 
    -  Y_1\left(\frac{\hat{m}_n}{\mKK}\right)\right] \\ ~=~
  Y_1\left(\frac{\hat{m}_n}{k}\right)
    \left[ \frac{m_B}{\hat{m}_n e^{\pi kR}} 
    J_2\left(\frac{\hat{m}_n}{\mKK}\right) 
    - J_1\left(\frac{\hat{m}_n}{\mKK}\right)\right]~,
  \label{eq:MassSpecEqnmphiIRMass}
\end{multline}
while the bulk profiles once again take the form
\begin{equation}
  \hat{\zeta}_n (y) ~=~ \hat{\mathcal{N}}_n e^{2ky} 
    \left[ J_2 \left(\frac{\hat{m}_n}{ke^{-ky}}  \right) + 
    \hat{b}_n Y_2 \left( \frac{\hat{m}_n}{ke^{-ky}} \right) \right]~,
  \label{eq:LowEnergyfnIRBraneMass}
\end{equation}
where $\hat{\mathcal{N}}_n$ is a normalization coefficient.  However, due to the 
difference in boundary conditions in this case relative to the case in which $m_\phi$ 
is localized on the UV brane, the constant $\hat{b}_n$ is given not by Eq.~(\ref{eq:bn}), 
but rather by 
\begin{equation}
    \hat{b}_n ~=~ -\frac{J_1\left(\frac{\hat{m}_n}{k}\right)}
      {Y_1\left(\frac{\hat{m}_n}{k}\right)}~.
\end{equation}
  
For any given choice of $m_\phi$, $k$, and $R$, evaluating $A_n = U_{n 0}$ for the 
case in which $m_\phi$ is localized on the IR brane is simply a matter of substituting the 
expression for $\hat{\zeta}_n (y)$ in Eq.~(\ref{eq:LowEnergyfnIRBraneMass}) into 
Eq.~(\ref{eq:Unl}).  However, we note that simple analytic expressions for the 
$A_n$ can be derived within the regime in which $m_\phi \ll \mKK$.  As in the case in
which $m_\phi$ is localized on the UV brane, we find that $A_0$ is approximately unity.
Moreover, within the same regime, we find that the mixing coefficients for all $\hat{\chi}_n$ 
with masses within the regime $k \gg \hat{m}_n \gg \mKK$ are well approximated by  
\begin{equation}
    A_n ~ \approx ~ 
      e^{-3\pi kR} \left( \frac{m_\phi}{m_{\text{KK}}} \right)^2 
      \left( \frac{m_{\text{KK}}}{\hat{m}_n} \right)^2~,
  \label{eq:mixing2}
\end{equation} 
while the $\hat{m}_n$ themselves are once again well approximated by the expression
in Eq~(\ref{eq:mnApproxLowmphi}).

The initial abundances $\Omega_n^0$ for the ensemble constituents with masses 
within the regime $k \gg \hat{m}_n \gg \mKK$ in the case in which the
mass-generating dynamics is localized on the IR brane may be obtained simply
by substituting our result for $A_n$ in Eq.~(\ref{eq:mixing2}) into
Eq.~(\ref{eq:abundances}).  This yields 
\begin{equation}
  \Omega_n^0 ~\sim~ 
    \begin{cases} 
      \hat{m}_n^{-2} & \text{instantaneous} \\ 
      \hat{m}_n^{-\frac{7}{2}} & \text{staggered (RD era)} \\ 
      \hat{m}_n^{-4} & \text{staggered (MD era)}~. \\
  \end{cases}
\end{equation} 

\subsection{SM on the IR Brane}

We have seen in cases in which the SM fields into which the $\hat{\chi}_n$ decay are 
localized on the UV brane, the quantity $A_n'$ plays a crucial role in determining how 
$\Gamma_n$ scales with $\hat{m}_n$ across the ensemble.  By contrast, in cases in
which the SM fields are localized on the IR brane, it is the quantity 
$A_n''$, which describes the projection of $\hat{\chi}_n$ onto the IR brane at 
$y = \pi R$, which plays this same role.  In general, these projection coefficients 
are given by 
\begin{eqnarray}
  A_n'' &\equiv & e^{-4\pi kR}\sqrt{\frac{1-e^{-2\pi k R}}{2k}} \nonumber \\
    & & \times \sum_{\ell=0}^{\infty} \zeta_\ell ( \pi R ) \int_{0}^{\pi R}  
    dy~e^{-2ky} \zeta_\ell (y) \hat{\zeta}_n (y) \nonumber \\ 
  & = & e^{-4\pi kR}\sqrt{\frac{1-e^{-2\pi k R}}{2k}}\hat{\zeta}_n(\pi R)~,    
  \label{eq:projectionIR}
\end{eqnarray}
where in going from the first to the second line, we have once again used the 
completeness relation in Eq.~(\ref{eq:CompletenessRel}).  

As was the case with our expression for $A'_n$ in Eq.~(\ref{eq:projectionUV}), the 
expression for $A_n''$ in Eq.~(\ref{eq:projectionIR}) turns out to have a simple 
analytic form within the regime in which 
$m_\phi \ll \mKK$ and $k \gg \hat{m}_n \gtrsim \mKK$.  In particular, we find that     
\begin{equation}
  A_n'' ~\approx ~ e^{-3\pi k R}~.
\end{equation}
The scaling relation for the decay widths in this regime may be obtained by substituting this 
result for $A''_n$ into Eq.~(\ref{eq:decaywidth}), which yields   
\begin{equation}
  \Gamma_n ~\propto~ \hat{m}_n^3~.    
\end{equation}
The density of states per unit $\Gamma$ in this case is therefore
\begin{equation}
  n_{\Gamma} ~\sim~ n_{m}(\Gamma) \left( \frac{d\Gamma}{dm}\right)^{-1} 
    ~\sim~ \Gamma^{-2/3}. 
\end{equation} 

Given these results above and the results in Sect.~\ref{sec:DynamicalDarkMatter}, it is 
straightforward to evaluate the values of $\alpha$ and $\beta$ obtained in the 
$\hat{m}_n \ll \mKK$ regime for any of the four possible configurations for the brane mass
and the SM fields.  These values are tabulated in Table~\ref{tab:scalingtable}.


\end{document}